\documentclass[12pt]{article}
\usepackage{amscd, amsmath}
\usepackage{amsthm, amssymb}
\usepackage{mathtools}
\usepackage{scalerel}
\usepackage[bottom,flushmargin]{footmisc}
\usepackage{mathrsfs}
\usepackage[format=plain, labelfont=it]{caption}
\usepackage{float}
\usepackage{array}
\usepackage{tabularx}
\usepackage{slashed}
\usepackage{accents}
\usepackage[toc, page]{appendix}

\newcommand{\linebr}{\\[0.2em]}

\DeclareMathAlphabet{\mathpzc}{OT1}{pzc}{m}{it}

\newtheorem*{assumption(A)}{Condition (A)}

\theoremstyle{remark}

\theoremstyle{remark}

\numberwithin{equation}{section}

\newcommand{\CC}{\mathbb{C}}

\newcommand{\Lie}{\mathcal L}

\newcommand{\VV}{\mathscr V}
\newcommand{\HH}{\mathcal H}
\newcommand{\mph}{|\phi|}
\newcommand{\tphi}{\tilde{\phi}}
\newcommand{\FF}{\mathcal F}

\newcommand{\tbeta}{\tilde{\beta}}

\newcommand{\overbar}[1]{\mkern 1.5mu\overline{\mkern-1.5mu#1\mkern-1.5mu}\mkern 1.5mu}
\newcommand{\hb}{\overbar{h}}
\newcommand{\sbb}{\overbar{s}}
\newcommand{\vb}{\overbar{v}}
\newcommand{\ub}{\overbar{u}}

\newcommand{\SU}{\mathrm{SU}(3)}
\newcommand{\su}{\mathfrak{su}(3)}
\newcommand{\Utwo}{\mathrm{U}(2)}
\newcommand{\utwo}{\mathfrak{u}(2)}
\newcommand{\SUtwo}{\mathrm{SU}(2)}
\newcommand{\sutwo}{\mathfrak{su}(2)}
\newcommand{\kk}{\mathfrak{k}}

\newcommand{\dirac}{\slashed{D}}
\newcommand{\dd}{{\mathrm d}}
\newcommand{\vol}{\mathrm{vol}}
\newcommand{\Vol}{\mathrm{Vol}}
\newcommand{\diag}{\mathrm{diag}}

\DeclareMathOperator{\Tr}{Tr}
\DeclareMathOperator{\Ad}{Ad}
\DeclareMathOperator{\ad}{ad}
\DeclareMathOperator{\Real}{Re}
\DeclareMathOperator{\Imaginary}{Im}
\DeclareMathOperator{\grad}{grad}
\DeclareMathOperator{\divergence}{div}

\newcommand{\beq}{\begin{equation}}
\newcommand{\eeq}{\end{equation}}
\newcommand{\bal}{\begin{align}}
\newcommand{\eal}{\end{align}}

\newcommand{\bmatr}{\begin{bmatrix}}
\newcommand{\ematr}{\end{bmatrix}}

\newcommand*\conj[1]{\overbar{#1}}

\newcommand{\LL}{{\scaleto{L}{4.8pt}}}
\newcommand{\RR}{{\scaleto{R}{4.8pt}}}
\newcommand{\LLL}{{\scaleto{L}{3.8pt}}}
\newcommand{\RRR}{{\scaleto{R}{3.8pt}}}

\newcommand{\HHH}{{\scaleto{\mathcal H}{3.1pt}}}

\makeatletter

\renewenvironment{bmatrix}
  {\left[\mkern3.5mu\env@matrix}
  {\endmatrix\mkern3.5mu\right]}

\def\blfootnote{\xdef\@thefnmark{}\@footnotetext}
\makeatother

\begin{document}

\begin{titlepage}
\title{\bf Higher-dimensional routes \\ to the Standard Model fermions}

\author{{Jo\~ao Baptista}}  
\date{May 2021}

\maketitle

\thispagestyle{empty}
\vspace{.5cm}
\vskip 20pt
{\centerline{{\large \bf{Abstract}}}}
\noindent
In the old spirit of Kaluza-Klein, we consider a spacetime of the form $P = M_4 \times K$, where $K$ is the Lie group $\SU$ equipped with a left-invariant metric that is not fully right-invariant. We observe that a complete generation of fermionic fields can be encoded in the 64 components of a single spinor over the 12-dimensional spacetime. The behaviour of the spinorial function along the internal space $K$ can be chosen so that, after pairing and fibre-integration over $K$, the resulting Dirac kinetic terms in four dimensions couple to the $\mathfrak{u}(1) \oplus \sutwo \oplus \su$ gauge fields in the exact chiral representations present in the Standard Model. Although we describe the action of the internal Dirac operator on the 12-dimensional spinor, the full calculation of the fermionic mass terms produced by the model is longer and is not carried out here. We calculate instead the action of the internal Laplace operator on the spinor components.

\end{titlepage}

\tableofcontents

\newpage

\section{Introduction}
  
Traditional Kaluza-Klein theories propose to replace four-dimensional Minkowski spacetime $M_4$ with a higher-dimensional product manifold $P \!=\! M_4 \times K$, where the internal space $K$ is a Lie group or a homogeneous space with very small volume. The proposed Lorentzian metric on P is not the simple product of the metrics on $M_4$ and $K$, but has non-diagonal terms that can be interpreted as the observed gauge fields on $M_4$. Geometrically, the projection $P \to M_4$ should be a Riemannian submersion with fibre $K$. 

The original Kaluza-Klein choice $K = \mathrm{U}(1)$ has the remarkable feature that geodesics on $P$ project down to paths on $M_4$ satisfying the Lorentz law for moving electric charges. For general choices of $K$, it can be shown that a natural quantity on $P$, namely its scalar curvature $R_P$, can be written as a sum of components that include the individual scalar curvatures of $M_4$ and $K$ and, more remarkably, the norm $|F_A|^2$ of the gauge field strength. Since the scalar curvature is also the Lagrangian density for general relativity, it follows that the Einstein-Hilbert action on the higher-dimensional $P$ produces, after projection down to $M_4$, two of the essential ingredients of physical field theories in four dimensions: the Einstein-Hilbert and the Yang-Mills Lagrangians on $M_4$.

Kaluza-Klein theories, however, do present challenging difficulties when interpreted simply as higher-dimensional versions of general relativity, i.e. as dynamical field theories for a metric tensor on $P$ that satisfies the full Einstein equations on the higher-dimensional space. Although unifying and appealing, the direct extension of general relativity to higher dimensions seems to imply the existence of many unobserved scalar fields satisfying complicated equations of motion with few physically reasonable solutions. The new fields generally do not bear much resemblance to the well-known field content of the Standard Model. Moreover, following the interpretation of fermions in Kaluza-Klein theory as zero modes of the Dirac operator on the internal space K, there does not seem to be a good choice of Riemannian manifold $K$ able to deliver the necessary zero modes in the chiral representations appearing in the Standard Model. For reviews and discussions of Kaluza-Klein theory from different viewpoints, see for instance \cite{Bailin, Duff, Witten81, Witten83, WessonOverduin, Coq, Hogan, Bleecker}. Some of the early original references are \cite{Original}, with much more complete lists given in the mentioned reviews.

The plan of the present investigation is to dig deeper into some of the geometrical aspects of the Kaluza-Klein framework. The companion study \cite{Baptista} suggest that, besides the curvature $|F_A|^2$ of the gauge fields, there are other natural objets in a Riemannian submersion that resemble the field content of the Standard Model. When the fibre $K$ is a Lie group equipped with a left-invariant metric, the second fundamental form of the fibres generates terms in the four-dimensional Lagrangian sharing notable similarities with the covariant derivative of a Higgs field, including quadratic terms in the gauge fields that generate the gauge bosons' mass. When $K$ is chosen to be the group $\SU$ equipped with a specific family of left-invariant metrics, denoted by $g_\phi$, then the terms generated by the second fundamental form contain the precise covariant derivative $\dd^A \phi$ that appears in the Standard Model, namely a $\CC^2$-valued Higgs field coupled to the electroweak gauge fields through the correct representation.

In the present study we suggest ways to integrate fermions into this picture. For an internal space $K = \SU$, we regard fermions as spinorial functions on the 12-dimensional spacetime $M_4 \times K$ with a prescribed behaviour along $K$. A complete generation of fermionic fields can then be encoded in the 64 complex components of a single higher-dimensional spinor. Moreover, the vertical behaviour of this spinor can be chosen so that, after fibre-integration over K, the resulting Dirac kinetic terms in four dimensions couple to the $\mathfrak{u}(1) \oplus \sutwo \oplus \su$ gauge fields in the exact chiral representations present in the Standard Model. Perhaps one could think of the prescribed vertical behaviour as a sort of elementary, spinorial oscillation along the compact direction $K$.

\subsubsection*{Spinorial functions and fermionic gauge couplings} 

This second part of the Introduction gives a brief description of the calculations underlying the present study. Spinorial functions on the 12-dimensional spacetime $P = M_4 \times K$ can be regarded as functions $\Psi$ with values on a 64-dimensional complex vector space $\Delta_{12}$ equipped with a hermitian pairing $\langle \Psi_1, \Psi_2 \rangle \simeq \Tr (\Psi_1^\dag \Psi_2)$. This identification uses the fact that Lie groups, such as $K$ and $M_4\times K$, have trivial tangent and spin bundles. In addition, spin structures are unique on simply connected spaces.

In the usual Kaluza-Klein setting, one considers the action of the higher-dimensional Dirac operator $\dirac$ on spinorial functions. If $\{  X_a \} = \{  X_\mu^\HH , X_j \}$ denotes an orthonormal basis of the tangent space to $P$, where the $X_j$ are vectors tangent to $K$ and the $X_\mu^\HH$ are the horizontal lifts of vectors $X_\mu$ tangent to $M_4$, then we can schematically write the Dirac operator on $P$ as
\beq
\dirac \Psi \ := \ \sum_{a= 0}^{11}\, \Gamma^a \, \nabla_a \Psi \ = \ \sum_{\mu= 0}^{3}\, \Gamma^\mu \, \Lie_{X_\mu^\HHH} \Psi \ + \ \sum_{j= 4}^{11}\, \Gamma^j \, \nabla_{X_j} \Psi \ .
\eeq
Here the $\Gamma^a$ are a set of $(11,1)$-dimensional gamma matrices and we have used the triviality of the Levi-Civita connection on Minkowski space to replace $\nabla_{X_\mu^\HHH}$ with the Lie derivative $\Lie_{X_\mu^\HHH}$. A scalar density on $P$ can be obtained through the Dirac pairing of spinors,
\beq
\big\langle \Gamma^0 \Psi, \, \dirac \Psi \big\rangle \ = \ \sum_{\mu} \big\langle \Gamma^0 \Gamma^\mu \Psi, \,  \Lie_{X_\mu^\HHH} \Psi \big\rangle \, + \, \sum_{j} \big\langle \Gamma^0 \Gamma^j  \Psi, \, \nabla_{X_j} \Psi \big\rangle  , 
\eeq
where we have used standard hermiticity properties of the gamma matrices (signs may depend on the conventions). Now, by definition of the higher-dimensional metric $g_P$ (see section 3 of \cite{Baptista}), the horizontal lift to $P$ of a vector $X_\mu$ tangent to $M_4$ is given by
\beq \label{HorizontalLift}
X_\mu^\HH  \ := \  X_\mu \ + \  A^j_\LL (X_\mu) \, e_j^\LL   \ - \ A_\RR^j (X_\mu) \, e_j^\RR  \ ,
\eeq
where $A_\LL$ and $A_\RR$ are one-forms on $M_4$ with values in the Lie algebra $\su$; the vectors $\{ e_j\}$ form a basis of $\su$; the symbols $e_j^\LL$ and $e_j^\RR$ denote the extensions of $e_j$ as a left or right-invariant vector field on $K = \SU$, respectively. In particular, the kinetic terms that appear in the first sum of the scalar density have the form
\beq \label{KineticTerms1}
 \big\langle \, \Gamma^0 \Gamma^\mu \Psi, \,  \Lie_{X_\mu^\HHH} \Psi \, \big\rangle \ = \ \big\langle \, \Gamma^0 \Gamma^\mu \Psi, \ \,  \partial_\mu \Psi \, + \,  A^j_\LL (X_\mu) \, \Lie_{e_j^\LLL} \Psi  \, - \, A_\RR^j (X_\mu) \, \Lie_{e_j^\RRR} \Psi \, \big\rangle \ .
\eeq
Thus, the coupling of the 12-dimensional spinor $\Psi$ to the gauge fields on $M_4$ is determined by the Lie derivatives of $\Psi$ along the left and right-invariant vector fields on $K$.

Although simple enough, the previous reasoning seems headed towards an apparent contradiction with a basic property of the Standard Model, namely the fact that leptons do not couple to the strong force, mediated by the gauge fields that here are called $A_\RR^j (X_\mu)$. Having \eqref{KineticTerms1} in mind, this property should mean that the Lie derivatives $\Lie_{e_j^\RRR} \Psi$ of the leptonic components of $\Psi$ are identically zero. Since this happens for all right-invariant fields $e_j^\RR$, it follows that these components are invariant functions under left-multiplication on the group $K$, i.e. they are simply constant on $K$. But constant functions on $K$ also have vanishing derivatives along all left-invariant vector fields $e_j^\LL$, and hence the leptonic components of $\Psi$ should not couple to the electroweak gauge fields $A_\LL^j (X_\mu)$ either, which is not true.

The gap in the previous argument comes from looking at expression \eqref{KineticTerms1} and assuming that the Lie derivatives $\Lie_{e_j^\RRR} \Psi$ must all vanish when the leptonic component of $\Psi$ do not couple to the fields $A_\RR^j (X_\mu)$. In reality, these derivatives may perfectly well have plenty of non-zero terms, as long as they integrate to zero along the fibres $K$, since the fermionic gauge representations in the four-dimensional Lagrangian are obtained only after projecting the scalar density \eqref{KineticTerms1} from $M_4\times K$ down to $M_4$. Schematically, down in four dimensions the couplings of fermions to the fields $A_\RR^j (X_\mu)$ should be determined by the integrals
\beq \label{GaugeCouplings1}
\int_K \, \big\langle \, \Gamma^0 \Gamma^\mu \Psi, \  \Lie_{e_j^\RRR} \Psi \, \big\rangle \, \vol_K \ ,
\eeq
and similarly for the couplings to the $A_\LL^j (X_\mu)$ gauge fields.

The main purpose of this study is to search for the appropriate spinorial functions $\Psi^P (x, h)$, defined on the higher-dimensional spacetime $M_4\times \SU$, such that the scalar integrals \eqref{GaugeCouplings1} produce the exact fermionic gauge representations present in the Standard Model. More precisely, given a function $\Psi (x)$ on $M_4$ with values in $\Delta_{12}$, we can extend it to a function $\Psi^P (x, h)$ on the whole $P$ through a simple ``separation of variables'' prescription:
\beq
\Psi^P (x, h) \ := \ S(h)\, \Psi (x) \ ,
\eeq
for all points $(x,h)$ in $M_4\times K$, where $S(h): \Delta_{12} \to \Delta_{12}$ is a linear transformation that depends smoothly on the point $h \in K$. For spinorial functions of this sort, expansion \eqref{HorizontalLift} implies that
\beq
\Lie_{X_\mu^\HHH} \, \Psi^P  \ = \   S \, \partial_\mu \Psi \, + \,  A^j_\LL (X_\mu) \, \big(\Lie_{e_j^\LLL} S \big) \Psi  \, - \, A_\RR^j (X_\mu) \, \big(\Lie_{e_j^\RRR} S \big) \Psi \ ,
\eeq
which is again a function on $M_4 \times K$ with values on $\Delta_{12}$. Remarkably, one can then show that for a relatively simple choice of vertical transformations $S(h)$, the scalar density in four dimensions obtained after fibre-integration,
\beq \label{GaugeCouplings2}
\int_K \ \sum_{\mu = 0}^3  \big\langle \, \big(\Gamma^0 \Gamma^\mu \Psi \big)^P , \  \Lie_{X_\mu^\HHH} \big( \Psi^P \big) \, \big\rangle \ \vol_K \ ,
\eeq
contains all the kinetic terms present in the Standard Model Lagrangian for one generation of fermions. This includes the correct hypercharges and gauge representations for all the leptons and quarks, right-handed and left-handed, particles and anti-particles. All this information is encoded in the single 12-dimensional spinor $\Psi^P (x, h)$.

The next section of this study is dedicated to the definition of the spinor's vertical transformation $S(h)$ and to the calculation of the gauge representations induced in four dimensions, after the fibre-integrals \eqref{GaugeCouplings2}. It is the main section of this study. In section 3 we discuss the action of the internal Dirac and Laplace operators on the spinor $\Psi^P (x, h)$, as they are a natural source of fermionic mass terms in the model. The four-dimensional mass terms induced by this action are explicitly calculated only in the case of the Laplace operator. In the more pertinent case of the internal Dirac operator, the calculations are significantly more evolved and are not carried out here. The discussion in this study also does not encompass the fundamental quantum aspects of the Standard Model.

\newpage

\section{Spinorial functions on $M_4 \times K$}

\subsection*{The 12-dimensional gamma matrices} 
\addcontentsline{toc}{subsection}{12-dimensional gamma matrices}

To compare the fibre-integrals \eqref{GaugeCouplings2} with the usual kinetic terms of four-dimensional fermions, we want to find a representation of the 12-dimensional gamma matrices $\Gamma^\mu$ in terms of the more familiar Pauli matrices $\sigma^\mu$ and four-dimensional gamma matrices $\gamma^\mu$. Start by identifying the 64-dimensional spinor space $\Delta_{12}$ with the space of $8\times8$ complex matrices,
\beq
\Delta_{12} \ \simeq \  M_{8\times8} (\CC)  \ . 
\eeq
Now consider the three Pauli matrices:
\beq \label{PauliMatrices}
\sigma^1 \, = \, \bmatr 0  &  \ 1 \\ 1  & \ 0 \ematr \qquad  \sigma^2 \, = \, \bmatr 0  &  \ -i \\ i  & \ 0 \ematr \qquad \sigma^3 \, = \, \bmatr 1  &  \ 0 \\ 0 & \ -1 \ematr \ ,
\eeq
satisfying the relations $\sigma^a \sigma^b = \delta^{ab} I_2 + i \sum_c \varepsilon^{abc} \sigma^c$, and choose a set of four-dimensional gamma matrices $\gamma^\mu$ on Minkowski space equipped with the metric $\eta = \diag (-1, 1, 1, 1)$. We use the mathematical convention where gamma matrices in space dimensions are anti-hermitian and correspond to square roots of $- 1$, so that
\bal
\big\{ \gamma^\mu,\, \gamma^\nu \big\} \ = \ - 2 \, \eta^{\mu \nu} \, I_4  \qquad \quad (\gamma^0)^\dag \ &= \ \gamma^0  \qquad \quad (\gamma^l)^\dag \ = \ - \gamma^l  \ ,
\end{align}
for $l = 1, 2, 3$. The corresponding gamma matrices in four-dimensional Euclidian space are defined by $\gamma_E^l = \gamma^l $, for $l = 1, 2, 3$, and by $\gamma_E^4 = i \gamma^0$. It follows that all the $\gamma_E^k$ are anti-hermitian and satisfy
\beq
\big\{ \gamma^k_E,\,  \gamma^l_E \big\} \ = \ - 2 \, \delta^{k l} \, I_4  \ .
\eeq
Then the Euclidian operator $\gamma^k_E\, \partial_k$ is formally self-adjoint with respect to the pairing of spinors $\langle \psi, \, \psi \rangle = \psi^\dag \psi$, while the operator $i \gamma^\mu\, \partial_\mu$ on Minkowski space is formally self-adjoint with respect to the pairing $\langle \gamma^0 \psi, \, \psi \rangle$. The chiral operator in four dimensions is defined by
\beq
\gamma^5 \ := \ i\, \gamma^0 \,\gamma^1\, \gamma^2\, \gamma^3 \ = \ -\, \gamma_E^1 \,\gamma_E^2 \, \gamma_E^3\, \gamma_E^4 \ =: \ \gamma^5_E \ .
\eeq
Since it coincides in Euclidian and Lorentzian signatures, we will refer to it simply as $\gamma^5$. It is represented by a hermitian matrix that anti-commutes with all the $\gamma^\mu$ and $\gamma_E^k$ and satisfies $\gamma^5 \gamma^5 = I_4$.

To build the 12-dimensional gamma matrices $\Gamma^a$ out of four-dimensional blocks, we use the standard Kronecker (tensor) product of matrices. It can be applied to matrices of any dimensions and is defined by
\beq   \label{KroneckerProduct}
A \otimes B \, := \, \bmatr a_{11}B  &  \cdots  &  a_{1m'} B  \linebr \vdots  &   \ddots  &   \vdots  \linebr   a_{1m}B  &  \cdots  &  a_{mm'} B   \ematr \qquad \quad \in \ M_{mn \times m'n'} (\CC)  \ ,
\eeq
where $A$ and $B$ have dimensions $m\times m'$ and $n \times n'$, respectively. It is an associative product satisfying
\bal 
(A \otimes B) \, (C \otimes D) \  &= \  (A\,C) \otimes (B\,D)  \linebr
(A \otimes B)^\dag \ &= \ A^\dag \otimes B^\dag    \nonumber  \linebr 
\det (A \otimes B) \ &= \ (\det A)^n\, (\det B)^m \ ,   \nonumber
\end{align}
where the last property makes sense only when $A$ and $B$ are square matrices.

With these pieces in place, we can write down the following set of $(11,1)$-dimensional gamma matrices:
\bal     \label{GammaMatrices}
\Gamma^0 \, \Psi \ &= \  \big(\gamma^0 \otimes I_2\big) \, \Psi  \linebr
\Gamma^b \, \Psi \ &= \  \big(\gamma^0 \gamma^5 \otimes \sigma^b\big) \, \Psi    \qquad    & b&=1,2,3     \nonumber \linebr
\Gamma^{3+k} \, \Psi \ &= \  \big(\gamma^5 \otimes  I_2 \big) \, \Psi  \,  \big( I_2 \otimes  \gamma^k_E \big)   \qquad    & k&=1,2,3,4    \nonumber \linebr
\Gamma^{7+l} \, \Psi \ &= \  \big(\gamma^l_E \gamma^4_E \otimes  I_2 \big) \, \Psi  \,  \big( \sigma^3 \otimes  \gamma^5 \big)   \qquad    & l&=1,2,3    \nonumber \linebr
\Gamma^{11} \, \Psi \ &= \  \big(\gamma^5 \otimes  I_2 \big) \, \Psi  \,  \big(  i\sigma^2 \otimes  \gamma^5 \big)    \ .  \nonumber
\end{align}
These formulae define the action of $\Gamma^a$ on the $8\times8$ matrix $\Psi$ through right and left-multiplication of $\Psi$ by matrices of the same dimension, which themselves are built out of 2 and 4-dimensional blocks using the Kronecker product. Using the properties of the $\gamma$, $\gamma_E$ and $\sigma$ matrices one can readily check that $\Gamma^0$ is a hermitian operator with respect to the pairing $\langle \Psi_1, \Psi_2 \rangle \simeq \Tr (\Psi_1^\dag \Psi_2)$ in spinor space $\Delta_{12}$, while all the other $\Gamma^a$ are anti-hermitian. Furthermore, they satisfy the Clifford relation
\beq
\big\{ \Gamma^a,\,  \Gamma^b \big\} \ = \ - 2 \, \eta^{ab} \, I   \qquad \qquad \qquad  a, b \in \{ 0,1, \ldots, 11\} \ ,
\eeq
as operators on $\Delta_{12}$, where $\eta$ is the Lorentzian metric with signature $(- + \cdots +)$. Thus, the gamma matrices define unitary transformations on spinor space. An application of the properties of the Kronecker product shows that all the $\Gamma^a$ have determinant equal to one, and therefore are elements of ${\mathrm{SU}} ( \Delta_{12})$. Composing the $\Gamma^a$ in sequence, one can also check that
\bal \label{ChiralOperator}
i\, \Gamma^{0}\, \Gamma^{1}\, \Gamma^{2}\, \Gamma^{3} \, \Psi \ &= \ \big(\gamma^5 \otimes I_2\big) \, \Psi  \linebr
\Gamma^{4}\, \Gamma^{5}\,   \cdots \, \Gamma^{11} \, \Psi \ &= \  -\, \Psi \, \big(\sigma^1 \otimes \gamma^5 \big) \,     \nonumber   \linebr
\hat{\Gamma} \, \Psi \ := \ i \, \Gamma^{0}\, \Gamma^{1}\,   \cdots \, \Gamma^{11} \, \Psi \ &= \  -\, \big(\gamma^5 \otimes I_2\big) \, \Psi \, \big(\sigma^1 \otimes \gamma^5 \big) \ ,    \nonumber
\end{align}
where $\hat{\Gamma}$ is the chiral operator in $(11, 1)$ dimensions. Decomposing the $8 \times8$ matrix $\Psi$ into four blocks,
\beq   
\Psi \ = \ \bmatr B_1  &  B_2  \linebr B_3  &   B_4   \ematr  \ ,
\eeq
each of them a $4 \times4$ matrix, the first relation in \eqref{ChiralOperator} implies that blocks $B_1$ and $B_2$ have ``four-dimensional chirality'' with opposite sign to that of the bottom blocks $B_3$ and $B_4$, as determined by the four-dimensional operator $\gamma^5 = \diag (I_2, - I_2)$. The 12-dimensional chiral operator $\hat{\Gamma}$, on the other hand, acts on $\Delta_{12}$ as
\beq \label{ActionChiralOperator}  
\hat{\Gamma} \, \Psi \ = \ \bmatr -B_2\, \gamma^5  &  - B_1\,\gamma^5  \linebr B_4\,\gamma^5  &   B_3\, \gamma^5   \ematr  \ .
\eeq
It is clear that $\hat{\Gamma}\, \hat{\Gamma} = I$ and that $\hat{\Gamma}$ is a hermitian operator with respect to the pairing $\Tr (\Psi_1^\dag \Psi_2)$ on $\Delta_{12}$. Further ahead, we will establish a correspondence between the entries of $\Psi$ and the different fermions in the Standard Model. The action of $\hat{\Gamma}$ on $\Psi$ will then correspond to an exchange of particles and anti-particles.

The explicit formulae for the 12-dimensional gamma matrices allow us to write a formula for the product $\Gamma^0 \, \Gamma^\mu$, appearing in \eqref{GaugeCouplings2}, in terms of the lower-dimensional gamma matrices. From definition \eqref{GammaMatrices} it is clear that $\Gamma^0 \, \Gamma^0 = I_8$ while $\Gamma^0 \, \Gamma^b = \gamma^5 \otimes \sigma^b$ for $b=1, 2, 3$. Abbreviating $\tilde{\Gamma}^\mu  := \Gamma^0 \, \Gamma^\mu$, we therefore have a four-vector of $8\times8$ matrices,
\beq \label{TildeGammaFormula}
(\tilde{\Gamma}^\mu )_{\, 0 \leq \mu \leq 3} \ = \ (\Gamma^0 \, \Gamma^\mu )_{\, 0 \leq \mu \leq 3} \ = \ \big(\, I_8, \ \gamma^5 \otimes \sigma^1,  \ \gamma^5 \otimes \sigma^2, \ \gamma^5 \otimes \sigma^2 \, \big) \ .
\eeq
When applied to a spinor $\Psi \in \Delta_{12}$, as in the Lagrangian density \eqref{GaugeCouplings2}, this four-vector describes how the Pauli matrices $\sigma^b$ act differently on the components of $\Psi$ corresponding to different eigenspaces of $\gamma^5$, i.e. how the $\sigma^b$ act differently on left-handed and right-handed Weyl spinors.

\subsection*{Spinor behaviour along the internal space $K$} 
\addcontentsline{toc}{subsection}{Spinor behaviour along the internal space $K$}

Having laid out some algebraic requisites to manipulate spinors $\Psi \in \Delta_{12}$, now we want to define the prescription for the $K$-dependence of the functions $\Psi^P (x, h)$ that produces the correct gauge representations for four-dimensional fermions. To begin with, write the $8 \times8$ matrix $\Psi$ as a juxtaposition of two $8\times4$ complex matrices $\psi_\pm$,
\beq
\Psi  \ = \ \big[\ \psi_+   \ \ \,  \psi_{-} \ \big]  \ .
\eeq
Each of these two components will have a different (in fact, complex conjugate) dependence on the $K$-coordinate $h$. They will correspond to the particle and anti-particle fermionic representations, respectively. Now consider the map $S: \SU \to M_{4\times 4} (\CC)$ defined by
\beq \label{DefinitionS}
S(h) \ := \ \bmatr \ \conj{s(h)} \  &   \quad   \linebr   &   h \ \    \ematr  \ ,
\eeq 
where $h$ is in $\SU$ and the complex number $s(h)$ is determined by the scalar function on the group
\beq \label{Definition_s}
s(h) \ := \    \sqrt{2} \, \left[ \, (h_{11} )^2 \, + \, (h_{21})^2 \, + \,  (h_{31} )^2  \, \right] \ = \  \sqrt{2} \, \big(h^T\, h \big)_{11} \  \ .
\eeq
Here we have labelled as $h_{kj}$ the entries of the square matrix $h$. Then, given any spinorial function $\Psi (x)$ on $M_4$, we extend it to the full spacetime $P$ by
\beq \label{VerticalTransformationSpinors}
\Psi^P (x, h) \ := \  \Big[\  \ \Big(\, S (h) \otimes I_2 \,\Big)\, \psi_+(x) \ \conj{S(h)}  \ \qquad \ \, \Big(\, \conj{S(h)} \otimes I_2  \,\Big)\, \psi_-(x) \ S(h)  \ \ \Big]  \ .
\eeq
Since $S(h)$ is a $4\times4$ matrix, the product $S (h) \otimes I_2$ is $8 \times8$ and $\psi_\pm$ are $8 \times4$ matrices, all multiplications above are well defined. The symbol $\conj{S(h)}$ denotes the complex conjugate of $S(h)$, not the hermitian conjugate nor the inverse matrix; therefore, the vertical transformation of $\psi_\pm$ does not follow a $\SU$-representation as $h$ varies.

To descend into more detail without overburdening the notation, we will write the $8\times4$ matrices $\psi_\pm$ as
\beq
\psi_\pm \ = \ \bmatr \ a_\pm \  &   c_\pm^T   \linebr    b_\pm &   D_\pm \  \ematr  \ 
\eeq
and regard them as $4\times4$ matrices whose entries are two-component Weyl spinors. Thus, $a_\pm$ are simple Weyl spinors; $b_\pm$ and $c_\pm$ are 3-vectors of Weyl spinors; and $D_\pm$ are $3\times3$ matrices of such spinors. This notation is analogous to the usual way of writing a Dirac spinor in four dimensions as $[ \psi_\RR  \ \ \psi_\LL]^T$, i.e. as a 2-vector whose entries are Weyl spinors. With these conventions, the vertical transformations of the components of $\Psi^P (x, h)$ along $K$, as determined by rule \eqref{VerticalTransformationSpinors}, can be written as
\bal \label{VerticalTransformationSpinors2}
a^P_\pm (x,h) \ &= \  \big| s(h) \big|^2 \, a_\pm (x)    &    \linebr
b^P_+ (x,h) \ &= \   s(h)\ h \ b_+ (x)     &     b^P_- (x,h) \ &= \   \conj{s(h)}\ \conj{h} \ b_- (x)   \nonumber \linebr
c^P_+ (x,h) \ &= \   \conj{s(h)}\ h^\dag \ c_+ (x)     &     c^P_- (x,h) \ &= \   s(h)\ h^T \ c_- (x)   \nonumber \linebr
D^P_+ (x,h) \ &= \   h \ D_+ (x) \ \conj{h}     &     D^P_- (x,h) \ &= \   \conj{h} \ D_- (x)\ h   \nonumber  \ ,
\end{align}
where $h^T$ and $\conj{h}$ denote the transpose and complex conjugate matrices of $h$, respectively. It is implicit in these formulae that when a matrix $h$ of $\SU$ multiplies the vectors $b_\pm$ and $c_\pm$, or the matrices $D_\pm$, it produces a linear mix of their entries (which are Weyl spinors) but does not act inside the two-component Weyl spinors themselves. A more complete notation would be to write $(h \otimes I_2) \, b_+$ instead of $h\, b_+$, for instance, but this would make the formulae ahead somewhat less transparent.

Consider again the main scalar Lagrangian density \eqref{GaugeCouplings2}. Since the pairing $\langle \Psi_1, \Psi_2 \rangle$ is just the matrix trace $ \Tr (\Psi_1^\dag \Psi_2)$, it is clear that the decomposition $\Psi_P = \big[   \psi_+^P \ \ \psi_-^P  \big]$ leads to a simple sum of traces
\beq \label{BigTraces}
\int_K \, \left( \sum_{\mu = 0}^3   \Tr  \Big\{  \Big[\big(\Gamma^0 \Gamma^\mu \psi_+ \big)^P \Big]^\dag  \,  \Lie_{X_\mu^\HHH} \big( \psi_+^P \big)  \Big\}  \, + \,  \Tr  \Big\{  \Big[\big(\Gamma^0 \Gamma^\mu \psi_- \big)^P \Big]^\dag  \,  \Lie_{X_\mu^\HHH} \big( \psi_-^P \big)  \Big\} \right)  \vol_K \, .
\eeq
The decomposition of $\psi_\pm$ into components $a$, $b$, $c$ and $D$ will further break these two big traces into a sum of traces of the smaller components, leading to integrals of the general form
\beq \label{SmallTraces}
\int_K \   \Tr  \Big[  \big( D_2^P \big)^\dag  \,  \Lie_{X_\mu^\HHH} \big( D_1^P \big)  \Big] \  \vol_K \ , 
\eeq
and analogous integrals for the remaining components $a$, $b$ and $c$. Here we have used two different generic matrices $D_1$ and $D_2$ to account for the fact that, inside the traces in \eqref{BigTraces}, one of the $\psi_\pm$ appears multiplied by $\Gamma^0 \Gamma^\mu$ while the other does not. It is also implicit that both matrices $D_j^P (x, h)$ transform along $K$ according to the rules \eqref{VerticalTransformationSpinors2}.

Recalling expression \eqref{HorizontalLift} for the horizontal vector fields $X_\mu^\HHH$ in terms of the gauge fields $A_\LL^j(X_\mu)$ and $A_\RR^j(X_\mu)$ on $M_4$, the overall conclusion of the previous observations is that, in order to obtain the couplings of the gauge fields to the different components of the spinor $\Psi \in \Delta_{12}$, we will have to calculate several fibre-integrals of the form
\beq \label{SmallTraces2}
\int_K \   \Tr  \Big[  \big( D_2^P \big)^\dag  \,  \Lie_{v^\LLL} \big( D_1^P \big)  \Big] \  \vol_K \   \qquad  \qquad  \int_K \   \Tr  \Big[  \big( D_2^P \big)^\dag  \,  \Lie_{v^\RRR} \big( D_1^P \big)  \Big] \  \vol_K \ , \nonumber
\eeq
where $v$ is any vector in the Lie algebra $\su$; $v^\LL$ and $v^\RR$ denote the extension of $v$ to a left and right-invariant vector field on $\SU$; and the matrix functions $D_j^P (x, h)$ behave along $K$ according to \eqref{VerticalTransformationSpinors2}. Analogous integrals should be calculated for the other components $a^P$, $b^P$ and $c^P$ of $\psi_\pm^P$. These calculations will reveal the correspondence between the components of $\Psi^P$ and the fermions of the Standard Model. It will be the work of the next section.

\subsection*{Gauge representations induced in four dimensions} 
\addcontentsline{toc}{subsection}{Gauge representations induced in four dimensions}

\subsubsection*{Components $D_\pm (x,h)$} 
\addcontentsline{toc}{subsubsection}{Components $D_\pm (x,h)$}

We start by describing the procedure in detail in the simplest example: the matrix components $D_\pm$ of $\psi_\pm$. Simplifying the notation of \eqref{VerticalTransformationSpinors2} for the case $D_+$, we write
\beq
D^P (h) \ = \ h\, D \, \conj{h} \ ,
\eeq
where it is implicit that the $3\times 3$ matrix $D$ may depend on the coordinate $x$. Let $v$ denote any vector in $\su$ and let $v^\LL$ and $v^\RR$ denote its extension as left and right-invariant vector fields on $\SU$. Then the Lie derivatives of $D^P$ along these fields are simple enough:
\bal  \label{LieDerivativesD}
\big( \Lie_{v^\LLL} D^P \big) \, (h) \ &= \ \frac{\dd}{\dd t}\,  D^P \big[h \cdot \exp(tv) \big] \ |_{t=0} \ = \ h\, v\, D \, \hb \ + \ h \, D \, \hb \, \vb  \linebr
\big( \Lie_{v^\RRR} D^P \big) \, (h) \ &= \ \frac{\dd}{\dd t}\,  D^P \big[\exp(tv) \cdot h \big] \ |_{t=0} \ = \ v\, h\, D \, \hb \ + \ h \, D \, \overline{v} \, \hb  \nonumber \ .
\end{align}
As usual, one should be aware that the flux associated to a left-invariant field $v^\LL$ on a group $K$ is given by right-multiplication by $\exp(tv)$, and vice-versa for the fields $v^\RR$. If we have two matrix functions $D_1^P$ and $D_2^P$, it is possible to construct the pairing
\bal
\Tr \Big[\,  \big(D_2^P \big)^\dag \,  \big(\Lie_{v^\LLL} D_1^P \big) \, \Big] \ &= \ \Tr \Big[\, D_2^\dag \,v \, D_1 \, + \,  D_2^\dag \, D_1 \, \hb \, \vb \, h^T  \, \Big] \linebr
\Tr \Big[\,  \big(D_2^P \big)^\dag \,  \big(\Lie_{v^\RRR} D_1^P \big) \, \Big] \ &= \ \Tr \Big[\, D_2^\dag \, h^\dag \,v \, h\,  D_1 \, + \,  D_2^\dag \, D_1 \, \vb  \, \Big]  \nonumber \ ,
\end{align}
where $h^T = \hb^\dag$ denotes the transpose matrix and we used the cyclic properties of the trace. Integrating along the fibre $K \! = \SU$ equipped with a bi-invariant volume form $\vol_K$ leads to a significant simplification, since it can be shown that (see \eqref{Baux1} and subsequent comments)
\beq
\int_{h\in K} \, \hb \ \vb \ h^T \ \vol_K  \ = \  \int_{h\in K}\, h^\dag \, v \ h \ \vol_K  \ = \ 0 
\eeq
for any traceless matrix $v \in \su$. Thus, we get the simpler expressions
\bal
\int_K \, \Tr \Big[  \big(D_2^P \big)^\dag   \big(\Lie_{v^\LLL} D_1^P \big)  \Big] \, \vol_K \, &= \, \Tr \big[\, D_2^\dag \,v \, D_1\, \big] \, (\Vol\, K) \, = \,   \int_K \, \Tr \Big[  \big(D_2^P \big)^\dag \,   \big(v\, D_1\big)^P   \Big] \, \vol_K  \nonumber  \linebr
\int_K \, \Tr \Big[  \big(D_2^P \big)^\dag   \big(\Lie_{v^\RRR} D_1^P \big)  \Big]  \, \vol_K \, &= \, \Tr \big[\, D_2^\dag \, D_1\, \vb \, \big] \, (\Vol\, K) \, = \,   \int_K \, \Tr \Big[  \big(D_2^P \big)^\dag \,   \big(D_1 \, \vb \big)^P   \Big] \, \vol_K . \nonumber 
\end{align}
It therefore seems that, after pairing and fibre-integration, taking the Lie derivative $\Lie_{v^\LLL} D_1^P$ looks just like multiplying $D_1$ on the left by the matrix $v$, while the derivative $\Lie_{v^\RRR} D_1^P$ looks like $(D_1 \, \vb \big)^P$. Combining these results with decomposition \eqref{HorizontalLift} of horizontal fields $X_\mu^\HH$ on $P$, we get the nice expression
\beq
\int_K \, \Tr \Big[\,  \big(D_2^P \big)^\dag \,  \big(\Lie_{X_\mu^\HHH} D_1^P \big) \, \Big] \, \vol_K \ = \ \Tr \Big[\, D_2^\dag \, \big(\nabla^A_\mu D_1   \big)\, \Big] \, (\Vol\, K) \ ,
\eeq
where we have abbreviated
\beq \label{CovariantDerivativeD}
\nabla^A_\mu D_1 \ := \ \partial_\mu D_1 \ + \ \sum_{j = 1}^4 \, A_\LL^j(X_\mu) \, e_j \, D_1 \ -  \sum_{j = 1}^8 \, A_\RR^j(X_\mu) \, D_1 \, \conj{e_j}  \ .
\eeq
Recall from section 3 of \cite{Baptista} that $\{ e_j\}$ denotes a basis of matrices of $\su$ such that the subset $\{ e_j: 1 \leq j \leq 4\}$ spans the subalgebra $\iota(\utwo)$ inside $\su$. Therefore, we conclude that the gauge fields on $M_4$ couple to the $3\times3$ matrix function $D_1 (x)$ in a rather simple form, after the fibre-integrals kill half of the terms coming from \eqref{LieDerivativesD}. Let us look at this coupling in more detail by decomposing the matrix $D_1$ into its three lines:
\beq
D_1 \ = \ \bmatr u_1^T \  \linebr u_2^T \linebr \ u_3^T \  \ematr \, 
\eeq
with $u_a \in \CC^3$. Since the matrices $e_j$ are anti-hermitian, we have that
\beq
u_a^T \, \conj{e_j} \ = \ \big(  e_j^\dag \, u_a \big)^T \ = \ - \, \big(   e_j \, u_a \big)^T \ .
\eeq
Thus, the gauge fields $A_\RR^j(X_\mu)$ in \eqref{CovariantDerivativeD}, which according to \cite{Baptista} can be identified with the strong force fields, act on the vector $u_a$ through simple left-multiplication $u_a \mapsto e_j u_a$ by the corresponding matrix in the $\su$ basis. At the Lie algebra level, this is the action corresponding to the fundamental representation of $\SU$ on $\CC^3$. Therefore, the three lines of $D_1 (x)$ couple to the fields $A_\RR^j(X_\mu)$ just as quarks couple to gluons in the Standard Model.

To analyze the coupling to $D_1(x)$ of the fields $A_\LL^j(X_\mu)$, which according to \cite{Baptista} can be identified with the electroweak fields, recall that the subspace $\iota(\utwo)$ of $\su$ spanned the basis subset $\{ e_j: 1 \leq j \leq 4\}$ consists of matrices of the form
\beq
\iota(a) \ = \ \bmatr - \Tr (a) &  \  \linebr  & a \ \ematr \ ,
\eeq
with $a \in \utwo$. The terms associated with $A_\LL^j(X_\mu)$ in the covariant derivative \eqref{CovariantDerivativeD} say that these matrices act on $D_1 (x)$ simply by left-multiplication $\iota(a) \, D_1 (x)$. In particular, when $a$ is in the smaller subspace $\sutwo \subset \utwo$, i.e. when $\Tr(a) = 0$, we recognize that the corresponding gauge field $A_\LL^j(X_\mu)$ does not couple to the first line of $D_1(x)$, while it mixes the second and third lines as in the fundamental $\SUtwo$-action on this doublet of 3-vectors.
Finally, when $a$ is the diagonal generator of the subspace $\mathfrak{u}(1) \subset \utwo$, that is when
\beq
\iota (a) \ = \ \bmatr - 2\alpha\, i &  \  \linebr  & \alpha \, i I_2 \ \ematr \ ,
\eeq
with $\alpha \in \mathbb{R}$, formula \eqref{CovariantDerivativeD} for the covariant derivative tells us that the hypercharge associated to the first line of $D_1 (x)$ is minus twice the hypercharge associated to the second and third lines.

Comparing with a table of fermionic representations in the Standard Model \cite{Baez, Hamilton}, we recognize that the gauge couplings of $D_1 (x)$ fit very well if we identify its first line with the right-handed down quark $d_\RR$ and its second and third lines with the doublet of left-handed up and down quarks, to exemplify with the first fermionic generation only. Thus, the behaviour of $D_+^P (x,h)$ along $K$ prescribed by rule \eqref{VerticalTransformationSpinors} originates the correct strong and electroweak gauge representations if we identify the lines of this $3\times3$ matrix with quark fields according to
\beq \label{IdentificationD}
D_+ (x) \ = \ \bmatr d_\RR (x)^{T} \  \linebr u_\LL(x)^T \linebr \ d_\LL(x)^T \  \ematr \ .
\eeq
A similar procedure can be used to analyze the behaviour along $K$ of the components $D_{-} (x)$ of $\psi_{-}$, as determined by the transformation rule \eqref{VerticalTransformationSpinors2}. It leads to an identification analogous to \eqref{IdentificationD}, but with the particle fields on the right-hand side exchanged by the corresponding antiparticle fields.

\subsubsection*{Components $b_\pm (x,h)$} 
\addcontentsline{toc}{subsubsection}{Components $b_\pm (x,h)$}

We will analyze the behaviour along $K$ of the components $b_+^P (x,h)$ of the matrix $\psi_+^P$, as defined in \eqref{VerticalTransformationSpinors2}. In simplified notation, this behaviour is
\beq
b_+^P (x,h) \ = \ s(h) \, h \ b \ ,
\eeq
where $h \in \SU$ and it is implicit that the vector $b \in \CC^3$ may depend on the coordinate $x$ of Minkowski space. The Lie derivatives of $b^P$ along left and right-invariant vector fields on $K$ are
\bal
\Lie_{v^\LLL} \, b^P \ &= \  \big(\Lie_{v^\LLL} \, s\big)\, h \, b \ + \ s(h)\, h\, v\, b \linebr
\Lie_{v^\RRR} \, b^P \ &= \  \big(\Lie_{v^\RRR} \, s\big)\, h \, b \ + \ s(h)\, v\, h\, b  \ , \nonumber
\end{align}
for any matrix $v$ in $\su$. If we have two vectorial functions $b_1^P$ and $b_2^P$, it is possible to take the hermitian products
\bal
\big( b_2^P \big)^\dag\, \big( \Lie_{v^\LLL}\,  b_1^P \big) \ &= \  \sbb \, \big(\Lie_{v^\LLL} \, s\big)\, b_2^\dag \, b_1 \ + \ |s|^2\,  b_2^\dag \, v\, b_1 \linebr
\big( b_2^P \big)^\dag\, \big( \Lie_{v^\RRR}\,  b_1^P \big) \ &= \  \sbb \, \big(\Lie_{v^\RRR} \, s\big)\, b_2^\dag \, b_1 \ + \ |s|^2\,  b_2^\dag \, h^\dag \,  v\, h\,  b_1  \ . \nonumber
\end{align}
Using the explicit form of the scalar function $s(h) = \sqrt{2} \, (h^T h)_{11}$ defined in \eqref{Definition_s}, one can readily calculate that
\beq
\big(\Lie_{v^\LLL} \, s\big) (h)\ = \ 2\, \sqrt{2}\, (h^T h\, v)_{11}  \qquad \qquad  \big(\Lie_{v^\RRR} \, s\big) (h)\ = \ 2\, \sqrt{2}\, (h^T v \,  h)_{11} \ .
\eeq
But the integrals \eqref{A4Integral1}, \eqref{A4Int2} and \eqref{A4Int3} calculated in appendix A.4 say that
\bal  \label{SpecialIntegralsb}
\int_K \, \sbb \, \big(\Lie_{v^\RRR} \, s\big) \ \vol_K \ &= \ 0 \ = \ \int_K\, |s|^2 \, h^\dag\, v\, h \ \vol_K \linebr
\int_K \, \sbb \, \big(\Lie_{v^\LLL} \, s\big) \ \vol_K \ &= \ 2\, v_{11}  \int_K \, |s|^2 \ \vol_K \ , \nonumber
\end{align}
for any $v \in \su$ and any bi-invariant volume form $\vol_K$ on the group $K = \SU$. This produces a significant simplification of the integrals of the hermitian products,
\bal \label{Couplings_b}
\int_K \, \big( b_2^P \big)^\dag\, \big( \Lie_{v^\LLL}\,  b_1^P \big) \ \vol_K \ &= \  b_2^\dag\, (2\, v_{11} I_3 \, + \, v )\, b_1     \int_K |s|^2 \ \vol_K   \nonumber \linebr
\int_K \, \big( b_2^P \big)^\dag\, \big( \Lie_{v^\RRR}\,  b_1^P \big) \ \vol_K \ &= \ 0 \ .
\end{align}
Having in mind decomposition \eqref{HorizontalLift}, expressing the horizontal lift of $X_\mu$ in terms of the gauge fields $A^j(X_\mu)$ on $M_4$, we get that
\beq
\int_K \,  \big(b_2^P \big)^\dag \,  \big(\Lie_{X_\mu^\HHH}\, b_1^P \big)  \, \vol_K \ = \  b_2^\dag \, \big(\nabla^A_\mu \, b_1 \big) \,  \int_K |s|^2 \ \vol_K  \ ,
\eeq
where we have abbreviated
\beq \label{CovariantDerivativeb}
\nabla^A_\mu\, b_1 \ := \ \partial_\mu b_1 \ + \ \sum_{j = 1}^4 \, A_\LL^j(X_\mu) \, \left[ 2\, \big(e_j)_{11} \,  I_3 \, + \, e_j   \right] \, b_1   \ ,
\eeq
which is a function on $M_4$ with values in $\CC^3$. The first conclusion is that the gauge fields $A_\RR^j(X_\mu)$, representing the strong force, do not couple to the components of the 3-vector $b_1(x)$. So these components must represent leptonic fields.

To analyze the coupling of the fields $A_\LL^j(X_\mu)$ to $b_1(x)$, recall that the vectors $\{ e_j: 1 \leq j \leq 4\}$ span the subspace $\iota(\utwo)$ of $\su$ consisting of block-diagonal matrices of the form $\iota(a) = \diag \big(-\Tr(a), \, a \big)$, with $a \in \utwo$. For matrices with this structure we have
\beq
2\, \big[ \, \iota(a) \,]_{11} \,  I_3 \, + \, \iota(a) \ = \ \bmatr  -  3\Tr(a) &   \linebr   &  a \, - \, 2 \Tr(a) I_2 \ematr \ .
\eeq
In particular, when $a$ is in the subspace $\sutwo \subset \utwo$, i.e. when $\Tr(a) = 0$, we see that the gauge field $A_\LL^j(X_\mu)$ corresponding to $e_j = \iota(a)$ does not couple to the top scalar component of $b_1(x) \in \CC^3$, while it mixes the second and third components as in the fundamental $\SUtwo$-action on this doublet. On the other hand, when  $a = \diag (i \alpha, i \alpha)$ is a generator of the diagonal subspace $\mathfrak{u}(1) \subset \utwo$, we have that $\iota(a) = e_1$ is just the diagonal matrix $\diag(-2i \alpha, i \alpha, i \alpha)$ in $\su$, and thus
\beq
2\, \big( \, e_1 \,)_{11} \,  I_3 \, + \, e_1 \ = \ \bmatr  -  6 \, \alpha\, i &   \linebr   &  -  3 \, \alpha\, i \,  I_2 \ematr \ .
\eeq
Inserting this result into formula \eqref{CovariantDerivativeb} for the covariant derivative and comparing with a table of fermionic representations in the Standard Model \cite{Baez, Hamilton}, we see that the gauge field $A_\LL^1(X_\mu)$ corresponding to the basis vector $e_1$ couples to the top component of $b(x)$ just like the Standard Model's subgroup $U(1)_Y$ couples to the right-handed electron field $e^-_R$, whereas the coupling of $A_\LL^1(X_\mu)$ to the second and third components of $b(x)$ is like the coupling of $U(1)_Y$ to the doublet formed by the left-handed electron and electron-neutrino. The conclusion is that the behaviour of the components $b_+^P (x,h)$ along $K$ prescribed by \eqref{VerticalTransformationSpinors} originates the correct strong and electroweak gauge representations if we identify
\beq \label{IdentificationB}
b_+ (x) \ = \ \bmatr e^-_\RR (x)  \  \linebr \nu_\LL(x) \linebr \ e^-_\LL(x) \  \ematr \ .
\eeq
A similar analysis applied to the component $b_{-} (x)$ of $\psi_{-}$ leads to an identification analogous to \eqref{IdentificationB}, but with the particle fields on the right-hand side exchanged by the corresponding antiparticles.

\subsubsection*{Components $c_\pm (x,h)$} 
\addcontentsline{toc}{subsubsection}{Components $c_\pm (x,h)$}

The vertical behaviour along $K$ of the components $c_\pm (x,h)$ of $\psi^P_+$, as defined in \eqref{VerticalTransformationSpinors2}, can be written in simplified notation as 
\beq
c_+^P (x,h) \ = \ \conj{s(h)} \, h^\dag \ c \ ,
\eeq
being implicit that the vector $c \in \CC^3$ may depend on the coordinate $x$ of $M_4$. The Lie derivatives of $c^P$ along left and right-invariant vector fields on $K$ are
\bal
\Lie_{v^\LLL} \, c^P \ &= \  \big(\Lie_{v^\LLL} \, \sbb \big)\, h^\dag \, c \ - \ \sbb\, v\, h^\dag\, c \linebr
\Lie_{v^\RRR} \, c^P \ &= \  \big(\Lie_{v^\RRR} \,  \sbb \big)\, h^\dag \, c \ - \ \sbb \, h^\dag \, v \, c  \ , \nonumber
\end{align}
for any matrix $v$ in $\su$. If we have two vectorial functions $c_1^P$ and $c_2^P$, it is possible to take the hermitian products
\bal
\big( c_2^P \big)^\dag\, \big( \Lie_{v^\LLL}\,  c_1^P \big) \ &= \  s \, \big(\Lie_{v^\LLL} \, \sbb \big)\, c_2^\dag \, c_1 \ - \ |s|^2\,  c_2^\dag \, h\, v\, h^\dag \, c_1 \linebr
\big( c_2^P \big)^\dag\, \big( \Lie_{v^\RRR}\,  c_1^P \big) \ &= \  s \, \big(\Lie_{v^\RRR} \, \sbb \big)\, c_2^\dag \, c_1 \ - \ |s|^2\,  c_2^\dag \, v\,  c_1  \ . \nonumber
\end{align}
Complex conjugation of functions commutes with Lie derivatives and integration over $K$, so the integrals in \eqref{SpecialIntegralsb} imply that
\bal  \label{SpecialIntegralsc}
\int_K \, s \, \big(\Lie_{v^\RRR} \, \sbb \big) \ \vol_K \ &= \ 0       \linebr
\int_K \, s \, \big(\Lie_{v^\LLL} \, \sbb \big) \ \vol_K \ &= \ -2\, v_{11}  \int_K \, |s|^2 \ \vol_K \ , \nonumber
\end{align}
where we have used that $\conj{v_{11}} = - v_{11}$ because $v \in \su$. Furthermore, for traceless $v$ the integrals \eqref{A4Integral1} calculated in the appendix say that
\beq
 \int_K \, \big| s(h) \big|^2 \ h \, v\, h^\dag \, \vol_K \ = \ 0 
\eeq
for any bi-invariant volume form $\vol_K$. Therefore, we get the simplified results:
\bal
\int_K \, \big( c_2^P \big)^\dag\, \big( \Lie_{v^\LLL}\,  c_1^P \big) \ \vol_K \ &= \  c_2^\dag\, (- 2\, v_{11}\, c_1)     \int_K |s|^2 \ \vol_K   \nonumber \linebr
\int_K \, \big( c_2^P \big)^\dag\, \big( \Lie_{v^\RRR}\,  c_1^P \big) \ \vol_K \ &= \  c_2^\dag\, (- \, v \, c_1)     \int_K |s|^2 \ \vol_K  \ 
\end{align}
for all matrices $v$ in $\su$. Combining this last calculation with decomposition \eqref{HorizontalLift}, where the $X_\mu^\HH$ are written in terms of the gauge fields $A^j(X_\mu)$ on $M_4$, we get that
\beq \label{Coupling_c}
\int_K \,  \big(c_2^P \big)^\dag \,  \big(\Lie_{X_\mu^\HHH}\, c_1^P \big)  \, \vol_K \ = \  c_2^\dag \, \big(\nabla^A_\mu \, c_1 \big) \,  \int_K |s|^2 \ \vol_K  \ ,
\eeq
together with the definition
\beq \label{CovariantDerivativec}
\nabla^A_\mu \,c_1 \ := \ \partial_\mu c_1 \ - \ 2\, \sum_{j = 1}^4 \, A_\LL^j(X_\mu) \, (e_j)_{11} \, c_1   \ + \  \sum_{j = 1}^8 \, A_\RR^j(X_\mu) \, e_j  \, c_1        \ .
\eeq
This covariant derivative of $c_1$ is a function on $M_4$ with values in $\CC^3$. As before, we will analyze its couplings to the gauge fields $A^j(X_\mu)$ and try to identify the components of $c_1 (x)$ with standard fermionic fields.

The strong force gauge fields $A_\RR^j(X_\mu)$ act on $c_1$ through simple multiplication $c_1 \mapsto e_j \, c_1$ by the generators of $\su$. At the Lie algebra level, this is the action corresponding to the fundamental representation of $\SU$ on $\CC^3$, so the three components of $c_1(x)$ should be interpreted as quark fields.

To analyze the coupling of the $A_\LL^j(X_\mu)$ to $c_1(x)$, recall that the subset of generators $\{ e_j: 1 \leq j \leq 4\}$ spans the subspace of $\su$ consisting of block-diagonal matrices of the form $\iota(a) = \diag \big(-\Tr(a), \, a \big)$, with $a$ in $\utwo$. Thus, when the matrix $a$ is in the subspace $\sutwo \subset \utwo$, i.e. when $\Tr(a) = 0$, the gauge fields $A^j_\LL$ corresponding to the generator $e_j = \iota(a)$ do not couple to $c_1(x)$, since $(e_j)_{11} = 0$ in this case. This is the same as saying that the components of $c_1(x)$ do not couple to the weak force. On the other hand, when  $a = \diag (i \alpha, i \alpha)$ is a generator of the diagonal subspace $\mathfrak{u}(1) \subset \utwo$, the corresponding generator $e_1 = \iota(a)$ in $\su$ is just the diagonal matrix $\diag(-2i \alpha, i \alpha, i \alpha)$. So we have that
\beq 
-2\, A_\LL^1(X_\mu) \, (e_1)_{11} \, c_1 \ = \ A_\LL^1(X_\mu) \, ( 4\, \alpha \, i \, c_1) \ .
\eeq
Comparing with a table of fermionic representations in the Standard Model and its hypercharge assignment \cite{Baez, Hamilton}, we conclude that the fields $A_\LL^j(X_\mu)$ couple to $c_1(x)$ just as the electroweak fields couple to the right-handed up quark $u_\RR$ with its three colour components. Thus, we identify
\beq \label{AuxIdentification3}
c_+ (x) \ = \ u_\RR (x) \ .
\eeq
A similar analysis applied to the component $c_{-} (x)$ of $\psi_-$ leads to an identification of this vector with the antiparticle of $u_\RR$, namely $\conj{u}_\LL$ with its three colour components.

\subsubsection*{Components $a_\pm (x,h)$} 
\addcontentsline{toc}{subsubsection}{Components $a_\pm (x,h)$}

Finally, the behaviour along $K$ of the components $a_+^P (x,h)$ of $\psi_+^P$, as defined in \eqref{VerticalTransformationSpinors2}, is just
\beq
a^P (x,h) \ = \ | s(h) |^2 \ a \ ,
\eeq
being implicit that the component $a$ may depend on the coordinate $x$ of $M_4$. The Lie derivatives of $a^P$ along left and right-invariant vector fields on $K$ are simply
\bal
\Lie_{v^\LLL} \, a^P \ &= \  \big(\Lie_{v^\LLL} \, | s(h)  |^2 \big) \, a     \linebr
\Lie_{v^\RRR} \, a^P \ &= \  \big(\Lie_{v^\RRR} \, | s(h) |^2 \big) \, a \ . \nonumber
\end{align}
Given two scalar functions $a_1^P$ and $a_2^P$ we can take the hermitian products
\beq
\big( a_2^P \big)^\dag\, \big( \Lie_{v^{\LLL / \RRR}}\,  a_1^P \big) \ = \  |s|^2  \left(\Lie_{v^{\LLL / \RRR}} \, |s|^2 \right)\, a_2^\dag \, a_1 \   = \  \Lie_{v^{\LLL / \RRR}} \left( \frac{1}{2} \, |s|^4 \,  a_2^\dag \, a_1 \right) \ .
\eeq
This is a total derivative, and therefore for any bi-invariant volume form $\vol_K$ on the group $K$ we have
\beq
\int_K \, \big( a_2^P \big)^\dag\, \big( \Lie_{v^{\LLL / \RRR}}\,  a_1^P \big) \, \vol_K \ = \ 0 \ ,
\eeq
both for right-invariant and left-invariant vector fields on $K$. Combining with decomposition \eqref{HorizontalLift} of horizontal vector fields $X_\mu^\HH$ on $P$, it follows that
\beq \label{Coupling_a}
\int_K \, \big( a_2^P \big)^\dag\, \big( \Lie_{X_\mu^\HHH}\,  a_1^P \big) \ \vol_K \ = \   a_2^\dag \, (\partial_\mu a_1) \, \int_K |s|^4 \ \vol_K  \ .
\eeq
Since none of the gauge fields $A_\LL^j(X_\mu)$ and $A_\RR^j(X_\mu)$ couple to the function $a_1(x)$, we can identify $a_+ (x)$ with the right-handed neutrino field $\nu_\RR (x)$. Similarly, $a_- (x)$ can be identified with its antiparticle $\conj{\nu}_\LL$.

\subsubsection*{Bringing the components together} 
\addcontentsline{toc}{subsubsection}{Bringing the components together}

All in all, the preceding calculations show that if we take a spinor $\Psi^P (x,h)$ on the product manifold $P = M_4 \times K$ and endow it with the vertical behaviour along $K$ prescribed by rule \eqref{VerticalTransformationSpinors}, namely
\beq
\Psi^P (x, h) \ := \  \Big[\  \ \Big(\, S (h) \otimes I_2 \,\Big)\, \psi_+(x) \, \conj{S(h)}  \ \qquad \ \, \Big(\, \conj{S(h)} \otimes I_2  \,\Big)\, \psi_-(x) \, S(h)  \ \ \Big]  \ ,
\eeq
then the fibre-integral in \eqref{GaugeCouplings2} produces kinetic terms on $M_4$ of the desired form:
\bal \label{ScalarDensity2}
\int_K \, \big\langle \, \big(\Gamma^0 \Gamma^\mu \Psi \big)^P , \  \Lie_{X_\mu^\HHH} \big( \Psi^P \big) \, \big\rangle \, \vol_K \ &= \ \Tr \Big[   \big(\Gamma^0 \Gamma^\mu \Psi \big)^\dag\, \nabla_\mu^A \Psi    \Big] \, (\Vol\, K) \linebr
&= \ \Tr \Big[  \Psi^\dag \, \big(\gamma^0 \gamma^\mu \otimes I_2 \big)^\dag\, \nabla_\mu^A \Psi    \Big] \, (\Vol\, K)  \nonumber \ ,
\end{align}
where $\gamma^\mu$ are the gamma matrices on $M_4$ and the covariant derivative of the matrix $\Psi = [\, \psi_+ \ \ \psi_- \, ]$ can be written as
\beq \label{CovariantDerivative}
\nabla^A \,\Psi \ := \ \dd \Psi \ + \  \sum_{j} \, A_\LL^j \ \big[\, \rho_{e_j}^\LL (\psi_+)   \ \ \    \rho_{\conj{e_j}}^\LL (\psi_-) \, \big]   \ + \  A_\RR^j \ \big[\, \rho_{e_j}^\RR (\psi_+)   \ \ \    \rho_{\conj{e_j}}^\RR (\psi_-) \, \big]      \ .
\eeq
Here $\{e_j\}$ denotes a basis of $\su$; the actions $\rho^\LL$ and $\rho^\RR$ are determined by the formulae
\bal \label{RhoAction}
\rho^\LL_v (\psi_\pm) \ &= \  \rho^\LL_v  \bmatr  a & c^T \linebr b & D  \ematr \ = \ \bmatr  0 & -2\, v_{11} \, c^T \linebr \big(  2\, v_{11}\, I_3 \, + \, v \big)\, b & v\, D  \ematr  \nonumber \\[0.6em]
\rho^\RR_v (\psi_\pm) \ &= \  \rho^\RR_v  \bmatr  a & c^T \linebr b & D  \ematr \ = \ \bmatr \ 0 \ & ( v\, c)^T \linebr \  0\  & - D \conj{v}\  \ematr 
\end{align}
for all matrices $v$ in $\su$; and the one-form $\dd \Psi$ is obtained by differentiation along the coordinates $x^\mu$ of $M_4$,
\beq
\dd \psi_\pm \ := \  \bmatr  \zeta\, \partial_\mu a & \partial_\mu c^T \linebr \partial_\mu  b & \partial_\mu D  \ematr \ \dd x^\mu \ .
\eeq
The awkward factor $\zeta$ appearing in front of the derivatives of $a_\pm$ comes from \eqref{Coupling_a}. Following the calculation in \eqref{A4I2}, it is equal to
\beq
\zeta \ := \ (\Vol\, K)^{-1} \, \int_K |s|^4 \ \vol_K \ = \ \frac{4}{3} \ .
\eeq
This factor cannot be made to disappear by rescaling $s(h)$ because we have already normalized this function so that the integral of $|s|^2$ over $K$ equals $\Vol \, K$. This normalization choice ensures that no similarly awkward factors appear in front of the components $\partial_\mu b$ and $\partial_\mu c$ (see \eqref{Couplings_b} and  \eqref{Couplings_b}), since such factors would affect the classical electromagnetic charges associated to those fields. The fermionic fields $a_\pm (x)$, on the other hand, do not couple to any gauge fields, so an overall factor $\zeta$ in front of its kinetic terms does not affect the equations of motion and seems comparatively harmless. 

Let us now make some general remarks about the actions $\rho^\LL$ and $\rho^\RR$ that determine the fermionic gauge couplings. Since the matrices $v \in \su$ are traceless and anti-hermitian, it is not difficult do check from definition \eqref{RhoAction} that the linear transformations $\rho^\LL_v$ and $\rho^\RR_v$ on $M_{4\times 4} (\CC)$ are also traceless and anti-hermitian with respect to the product of matrices $\Tr (A^\dag B)$. Therefore, regarding $\rho^\LL_v$ and $\rho^\RR_v$ as transformations of the larger space $M_{4\times 4} (\CC) \simeq \Delta_{12}$, they are elements in $\mathfrak{su} (\Delta_{12} )$. Moreover, from definition \eqref{RhoAction} one can readily calculate that the commutators of $\rho^\LL$ and $\rho^\RR$ satisfy
\bal \label{Commutators_rho}
\Big( \, \big[\rho^\RR_u , \,  \rho^\RR_v \big] \, - \, \rho^\RR_{[u,v]} \, \Big) (\psi_\pm) \ &= \ 0  \linebr
  \big[\rho^\RR_u , \,  \rho^\LL_v \big] \,  (\psi_\pm) \ &= \ 0     \nonumber \linebr
\Big( \, \big[\rho^\LL_u , \,  \rho^\LL_v \big] \, - \, \rho^\LL_{[u,v]} \, \Big)  \bmatr  a & c^T \linebr b & D  \ematr \ &= \ \bmatr  0 & 2\, [u,v]_{11} \, c^T  \linebr -2\, [u,v]_{11}\, b & 0 \ematr  \nonumber 
\end{align}
for all matrices $u$ and $v$ in $\su$. These relations say that $\rho^\RR$ defines a Lie algebra homomorphism from $\su$ to $\mathfrak{su} (M_{4\times 4})$, whereas $\rho^\LL$ does not. In particular, starting from two copies of $\su$, the direct sum $\rho^\LL +  \rho^\RR$ does not define a homomorphism from $\su \oplus \su$ to $\mathfrak{su} (M_{4\times 4})$, and hence to the larger $\mathfrak{su} (\Delta_{12} )$. This fact is in agreement with the observation in \cite{GG} saying that the fermionic gauge representations of the Standard Model cannot be obtained simply by restricting to $\mathrm{U}(1) \times \SUtwo \times \SU$ a well-chosen representation of the larger group $\SU \times \SU$. Notice, however, that when $u$ and $v$ are in the subspace $\mathfrak{u} (1) \oplus \sutwo = \utwo$ of the original $\su$, then the number $[u,v]_{11}$ is always zero, so it follows from \eqref{Commutators_rho} that $\rho^\LL +  \rho^\RR$ does define a Lie algebra homomorphism from $\utwo \oplus \su$ to the algebras $\mathfrak{su} (M_{4\times 4})$ and $\mathfrak{su} (\Delta_{12} )$. In fact, $\utwo \oplus \su$ is the largest subalgebra of $\su \oplus \su$ such that the map $\rho^\LL +  \rho^\RR$ is a homomorphism.

These observations fit well with the discussion in the preceding subsections, where we showed that when $A_\LL$ has values in the subspace $\utwo$ of $\su$ and $A_\RR$ has values on the whole $\su$, the associated action $\rho^\LL +  \rho^\RR$ on the components of $\psi_+ (x)$ coincides with the gauge representations of the different Standard Model fermions. In fact, the discussion in those sections, together with \eqref{CovariantDerivative}, \eqref{RhoAction} and \eqref{TildeGammaFormula}, allows us to conclude that the scalar density \eqref{ScalarDensity2} produces kinetic terms on $M_4$ that coincide with those of a fermionic generation of fields, provided that we identify the components of the spinor matrix $\Psi(x) = [\, \psi_+(x) \ \ \psi_-(x) \, ]$ as
\beq \label{FermionicIdentification1}
\Psi (x) \ = \ 
\bmatr \nu_\RR  & \, u_\RR^r & \, u_\RR^g & \, u_\RR^b & \, \conj{\nu}_\LL & \, \conj{u}_\LL^{\conj{r}} & \, \conj{u}_\LL^{\conj{g}} & \, \conj{u}_\LL^{\conj{b}} \   \\[0.4em]
 e_\RR^{-} & \, d_\RR^r & \, d_\RR^g & \, d_\RR^b & \, e_\LL^{+} & \, \conj{d}_\LL^{\conj{r}} & \, \conj{d}_\LL^{\conj{g}} & \, \conj{d}_\LL^{\conj{b}} \,  \\[0.4em]
 \nu_\LL  & \, u_\LL^r & \, u_\LL^g & \, u_\LL^b & \, \conj{\nu}_\RR & \, \conj{u}_\RR^{\conj{r}} & \, \conj{u}_\RR^{\conj{g}} & \, \conj{u}_\RR^{\conj{b}} \,  \\[0.4em]
 e_\LL^{-} & \, d_\LL^r & \, d_\LL^g & \, d_\LL^b & \, e_\RR^{+} & \, \conj{d}_\RR^{\conj{r}} & \, \conj{d}_\RR^{\conj{g}} & \, \conj{d}_\RR^{\conj{b}} \quad 
\ematr \ .
\eeq
This should be regarded as an $8\times8$ matrix field on Minkowski space $M_4$, since each of its entries is a Weyl spinor with two complex components. The conventions for naming the particles and anti-particles are as in \cite{Baez}, with $\{r, g, b \}$ being the colour indices of quarks. Observe how the 12-dimensional chiral operator $\hat{\Gamma}$, described in \eqref{ActionChiralOperator}, acts on the matrix $\Psi$ by exchanging each particle with the corresponding anti-particle and changing half of the signs:
\beq
\hat{\Gamma}\, \Psi \ = \ 
\bmatr  - \conj{\nu}_\LL &  -\conj{u}_\LL^{\conj{r}} &  \conj{u}_\LL^{\conj{g}} &  \conj{u}_\LL^{\conj{b}}  &  - \nu_\RR  & - u_\RR^r &  u_\RR^g &  u_\RR^b \,   \\[0.4em]
 - e_\LL^{+} &  - \conj{d}_\LL^{\conj{r}} &  \conj{d}_\LL^{\conj{g}} &  \conj{d}_\LL^{\conj{b}}  & - e_\RR^{-} & - d_\RR^r &  d_\RR^g &  d_\RR^b    \\[0.4em]
 \conj{\nu}_\RR &  \conj{u}_\RR^{\conj{r}} &  - \conj{u}_\RR^{\conj{g}} &  - \conj{u}_\RR^{\conj{b}} &   \nu_\LL  &  u_\LL^r &  -u_\LL^g &  -u_\LL^b   \\[0.4em]
 e_\RR^{+} &  \conj{d}_\RR^{\conj{r}} & - \conj{d}_\RR^{\conj{g}} &  - \conj{d}_\RR^{\conj{b}}  &  e_\LL^{-} &  d_\LL^r &  - d_\LL^g & - d_\LL^b  
\ematr \ .
\eeq
In particular, this implies that the eigenvectors of $\hat{\Gamma}$ are elementary linear combinations of the particle and anti-particle fields.

It seems remarkable that all the fields of a fermionic generation can be orderly fit into an $8\times8$ matrix representing a single spinor over the 12-dimensional spacetime $M_4 \times \SU$. In other words, the plethora of degrees of freedom of elementary fermions in one generation --- spin, chirality, weak isospin, lepton or quark, colour, particle or anti-particle --- can perhaps be regarded as a veiled reflection of the 64 degrees of freedom of a spinor in 12 dimensions. They look very different from each other because the individual components of the spinor are associated either to the $M_4$ or the $\SU$ factors of spacetime and couple to the gauge fields according to different representations.

\subsection*{Charges and coupling constants} 
\addcontentsline{toc}{subsection}{Charges and coupling constants}

\subsubsection*{Electromagnetic charges} 
\addcontentsline{toc}{subsubsection}{Electromagnetic charges}

The discussion in the previous section was slightly rushed when we identified the components of the matrix $\psi_+(x)$ with the usual fermionic fields. There was a small gap in the argument for associating the lower components of $b_+(x)$ and $D_+(x)$ with the left-handed particles. As a result, identification \eqref{FermionicIdentification1} is strictly valid only in the special case where the ``Higgs vector'' $\phi \in \CC^2$, as used in the definition of the ``vacuum'' metric $g_\phi$ on the internal space, is chosen to be proportional to $[0 \ \ 1]^T$ in the unitary gauge, which is the traditional choice in the literature. In the next few paragraphs we want to clarify how this small detail comes about. In the process, and more importantly, we want to verify explicitly how the photon gauge field $A^\gamma$ couples to the different components of $\Psi$, i.e. we want to calculate the electromagnetic charges implicit in the model.

Let us consider the case of $b_+ (x)$, for instance, which we identified before as
\beq
b_+ (x) \ \simeq \ \bmatr e^-_\RR (x)  \  \linebr \nu_\LL(x) \linebr \ e^-_\LL(x) \  \ematr \ .
\eeq
In the previous section we concluded that none of these components couples to the strong force fields $A_\RR$, while only the bottom two components couple to the weak fields $A_\LL$ with values in the subspace $\iota(\sutwo)$ of $\su$. From this observation and the consistent assignment of hypercharges, we concluded that the bottom two components of $b_+$ must correspond to the left-handed leptons. However, nothing in the reasoning showed that the middle component corresponds to $\nu_\LL$ and the bottom one to $e_\LL^-$, and not the other way around. In general, the bottom and middle components of $b_+$ may correspond to independent linear combinations of $\nu_\LL$ and $e_\LL^-$, and to distinguish them one should verify how they couple to the photon field $A^\gamma_\LL$, since electrons and neutrinos have different charges. 

Recall from section 2 of \cite{Baptista} that, among the left-invariant vector fields on $K = \SU$, there is a special one that is a Killing field of the metric $g_\phi$. It is generated by a vector $\gamma_\phi$ in the subalgebra $\iota(\utwo)$ of $\su$ that satisfies $[\gamma_\phi, \, \iota(\phi)] = 0$. Explicitly and up to a normalization factor, it is given by
\beq \label{GammaVector}
\gamma_\phi \ := \  \frac{i}{\sqrt{3}}  \bmatr -1 &    \linebr     & 2 I_2 -  3 \, |\phi|^{-2}\, \phi\, \phi^\dag  \ematr \quad   \in \ \su ,
\eeq
where $\phi \in \CC^2$ is the parameter in the definition of the metric $g_\phi$. Decomposing $\su$ into the sum of the span of $\gamma_\phi$ and its orthogonal complement, the photon field $A_\LL^\gamma$ is defined as the component of $A_\LL$ with values in the line of $\su$ generated by $\gamma_\phi$. 

The fermionic couplings of the photon field $A_\LL^\gamma$ are obtained simply by substituting $v = \gamma_\phi$ in formula \eqref{RhoAction} for the $\rho_v^\LL$ actions (the couplings of the $Z$ and $W$ gauge fields can be similarly obtained from the respective generators). Thus, we get that
\bal \label{RhoPhoton}
\rho^\LL_{\gamma_\phi} (b_+) \ &= \  i\, \sqrt{3} \,  \bmatr  -1 &   \linebr  &  - |\phi|^2\, \phi \phi^\dag  \ematr \, b_+      \qquad \qquad \quad      \rho^\LL_{\gamma_\phi} (c_+) \ = \ -\, \frac{2\, i}{\sqrt{3}} \ c_+      \nonumber \linebr
\rho^\LL_{\gamma_\phi} (D_+) \ &= \  \frac{i}{\sqrt{3}} \,  \bmatr  -1 &   \linebr   &   2 \, I_2 \, - \, 3 |\phi|^2\, \phi \phi^\dag  \ematr \, D_+  \ .
\end{align}
With the traditional choice for Higgs field in the unitary gauge $\phi  \propto [0 \ \ 1]^T$, the action of $\rho^\LL_{\gamma_\phi}$ is represented by the simpler diagonal matrices
\bal \label{RhoPhoton2}
\rho^\LL_{\gamma_\phi} (b_+) \ &= \  i\, \sqrt{3} \  \diag\left(-1,\, 0,\, -1\right) \, b_+      \qquad \qquad     \rho^\LL_{\gamma_\phi} (c_+) \ = \ i\, \sqrt{3} \  \diag\left(\frac{-2}{3},\, \frac{-2}{3},\, \frac{-2}{3}\right) \, c_+    \nonumber \linebr
\rho^\LL_{\gamma_\phi} (D_+) \ &= \ i\, \sqrt{3} \  \diag \left(\frac{-1}{3}, \,\frac{2}{3}, \,\frac{-1}{3}\right)  \, D_+  \ .
\end{align}
Thus, we obtain the correct relative assignment of electromagnetic charges if the components of $b_+$, $c_+$ and $D_+$ are identified with the fermions $e^-$, $\nu$, $u$ and $d$ through \eqref{FermionicIdentification1}, as in the preceding section. It is also clear, however, that identification \eqref{FermionicIdentification1} produces the correct charges only when $\phi  \propto [0 \ \ 1]^T$. In the case of a general non-zero $\phi \in \CC^2$, the doublets of left-handed particles inside the $8\times8$ matrix \eqref{FermionicIdentification1} should be modified according to the rule
\bal
\bmatr  \nu_\LL \linebr e_\LL^-  \ematr \ &\longmapsto \ \bmatr \,  e_\LL^- \, |\phi|^{-1} \, \phi \ + \ \nu_\LL \, |\phi|^{-1} \, \tphi \,  \ematr   \qquad \in \ \CC^2 \linebr
\bmatr  u_\LL \linebr d_\LL  \ematr \ &\longmapsto \ \bmatr \,  d_\LL \, |\phi|^{-1} \, \phi \ + \ u_\LL \, |\phi|^{-1} \, \tphi \,  \ematr \qquad \in \ \CC^2   \ , \nonumber
\end{align}
where we define the vector $\tphi  :=  [ \conj{\phi_2} \ \, - \conj{\phi_1}]^T$ whenever $\phi  =  [\phi_1 \ \ \phi_2]^T$. Using \eqref{RhoPhoton}, one can check that with these choices the fields  $e_\LL^-$, $\nu_\LL$, $u_\LL$ and $d_\LL$ will again couple to $A_\LL^\gamma$ with the correct relative charges. The modifications of \eqref{FermionicIdentification1}  for the anti-particle fields are analogous.

So far in this section we have discussed only the {\it relative} electromagnetic charges of the components of $\Psi (x)$, as induced by the Lagrangian density \eqref{GaugeCouplings2}. We have not discussed the absolute value of the charges implicit in the model. To address this topic, one should go back to the calculations of \cite{Baptista} and combine the Yang-Mills density on $M_4$ obtained there with the fermionic kinetic terms proposed here. The classical fermionic charges can then be read from the coupling of the Maxwell term $|F_{A_\LLL}^\gamma|^2$ with the Dirac terms $(A_\LL^\gamma)_\mu \, \big\langle \psi, \, \gamma^0 \gamma^\mu \, \rho_{\gamma_\phi}^\LL (\psi) \big\rangle$.

More precisely, recall that the Maxwell density $|F_{A_\LLL}^\gamma|^2$ for the photon field on $M_4$ was derived in \cite{Baptista} from the scalar curvature $R_P$ of the higher-dimensional metric $g_P$ on $P$ using fibre-integration over $K$, as in the standard Kaluza-Klein calculation. So we can combine the bosonic Lagrangian density of \cite{Baptista} with the fermionic density \eqref{GaugeCouplings2} in a single fibre-integral:
\beq \label{GaugeCouplings3}
\frac{1}{2\, \kappa_P}\  \int_K \, \left\{ R_P \ - \ 2\, \Lambda_P \ + \ i \xi \, \sum_{\mu = 0}^3  \big\langle \, \big(\Gamma^0 \Gamma^\mu \Psi \big)^P , \  \Lie_{X_\mu^\HHH} \big( \Psi^P \big) \, \big\rangle  \right\} \, \vol_{K} \ ,
\eeq
where $\xi$ is a positive constant. All the terms of this integral involving the photon field $A_\LL^\gamma$ have been calculated in the previous section and in sections 3 and 5 of \cite{Baptista}. A careful collection of these terms yields an expression of the form
\beq \label{AuxDensity1}
\left\{ - \frac{1}{4}\,  \beta(e_k, e_j) \ (F_{A_\LLL}^k)^{\mu \nu}  (F_{A_\LLL}^j)_{\mu \nu}  \ + \  i \xi \, (A_\LL^\gamma)_\mu \, \Tr \Big[ \Psi^\dag \, (\gamma^0 \gamma^\mu  \otimes I_2)\, \rho_{\gamma_\phi}^\LL (\Psi) \Big]  \right\} \frac{\Vol\, K}{2\, \kappa_P} \,   .
\eeq
There are no additional terms coming from the covariant derivative $\dd^A\phi$ because the commutator $\big[\gamma_\phi, \, \iota(\phi) \big]$ vanishes in $\su$, by definition of $\gamma_\phi$. In this formula, the symbol $\beta$ may stand either for the $\Ad_{\SU}$-invariant inner-product on $\su$, as in section 3 of \cite{Baptista}, or for the more general $\Ad_{\Utwo}$-invariant product $\tbeta$ described in section 5.

Now, working near the vacuum configuration where $\phi = {\rm{constant}} = \phi_0$, we have already identified the normalization $\accentset{\circ}{\gamma}_\phi$ of the vector $\gamma_\phi$ that is necessary for the curvature $F_{A_\LLL}^\gamma$ to appear in the four-dimensional Lagrangian ${\mathscr L}_M$ as it does in the traditional Einstein-Maxwell Lagrangian. These normalization conditions are standard in the Kaluza-Klein approach and were discussed in section 3.7 of \cite{Baptista} as necessary for both Lagrangians to produce the same linearized equations of motion. It was observed that we should have
\[
\kappa_P \ = \  \kappa_M \, \Vol(K, \, g_{\phi}) 
\]
and that the vector $\accentset{\circ}{\gamma}_\phi$ that leads to the canonical normalization of the photon field $A_\LL^\gamma$ should satisfy
\[
\beta(\accentset{\circ}{\gamma}_{\phi}, \accentset{\circ}{\gamma}_{\phi})  \ = \ 2\,  \kappa_M \ . 
\]
Therefore, if we choose the coefficient of the fermionic kinetic term to have the same value, $\xi = 2\, \kappa_M$, it will also cancel the $(\Vol\, K)\, (2 \, \kappa_P)^{-1}$ factor and produce canonically normalized terms in the approximation where $\phi$ is nearly constant:
\beq \label{AuxDensity2}
- \frac{1}{4}\,  \ (F_{A_\LLL}^\gamma)^{\mu \nu}  (F_{A_\LLL}^\gamma)_{\mu \nu}  \ + \ i \, (A_\LL^\gamma)_\mu \, \Tr \Big[ \Psi^\dag \, (\gamma^0 \gamma^\mu  \otimes I_2)\, \rho_{\accentset{\circ}{\gamma}_\phi}^\LL (\Psi) \Big]  \ .
\eeq
Hence the electromagnetic charges of the components of $\Psi$ are just the eigenvalues of the matrix $-i \rho_{\accentset{\circ}{\gamma}_\phi}^\LL$ \cite{Hamilton, Weinberg}. But the relation between the normalized and non-normalized vectors is simply
\beq
\accentset{\circ}{\gamma}_\phi \ = \ \sqrt{\frac{2\,  \kappa_M}{\beta(\gamma_\phi,\, \gamma_\phi)}} \ \gamma_\phi  \ = \   2\, \sqrt{\frac{\kappa_M}{\lambda_1 + 3\, \lambda_2}} \ \gamma_\phi    \ .
\eeq
Thus, working with the non-normalized vector \eqref{GammaVector}, the electromagnetic charges are the eigenvalues of the matrix
\beq
- i \, \sqrt{\frac{2\,  \kappa_M}{\beta(\gamma_\phi,\, \gamma_\phi)}} \ \rho_{\gamma_\phi}^\LL \ , 
\eeq
which are invariant under rescaling of $\gamma_\phi$ by any positive factor. Finally, the calculation in \eqref{RhoPhoton2} showed that $\rho_{\gamma_\phi}^\LL$ has an eigenvalue of $- \sqrt{3} \, i$ for the electron fields $e_\LL^-$ and $e_\RR^-$, so we conclude that the electromagnetic charge of the positron in the model determined by Lagrangian \eqref{GaugeCouplings3} is given simply by
\beq
e \ = \ \sqrt{\frac{6\, \kappa_M}{\beta(\gamma_{\phi},\, \gamma_{\phi})}}  \ \ .
\eeq
Recall again that we are working in the vacuum approximation and that in these formulae the parameter $\phi$ stands for the vacuum value $\phi_0$.

The formula for the charge of the positron can be recast using the relation between the norm $\beta(\gamma_{\phi},\, \gamma_{\phi})$ and the Riemannian length of the circle inside $\SU$ defined by the one-parameter subgroup generated by $\gamma_\phi$. To obtain this relation, start with definition \eqref{GammaVector} of $\gamma_\phi$ and observe that, since $\phi\, \phi^\dag$ commutes with $I_2$,
\beq
\exp\left[t\, \frac{i}{\sqrt{3}}\, \Big(2 I_2 -  3 \, |\phi|^{-2}\, \phi\, \phi^\dag \Big) \right] \ = \exp\left(t \, \frac{2i}{\sqrt{3}} \, I_2 \right) \, \exp\left(- it \sqrt{3} \, |\phi|^{-2}\, \phi\, \phi^\dag \right) \nonumber
\eeq
as $2\times2$ matrices. Moreover, we have that $(\phi\, \phi^\dag )^n = |\phi|^{2n-2} \phi\, \phi^\dag$ for any positive integer $n$, so the second exponential is just
\beq
 \exp\left(- it \sqrt{3}  |\phi|^{-2}\, \phi\, \phi^\dag \right)  \ = \ I_2 + \Big(  e^{- it \sqrt{3}} -1 \Big) \, |\phi|^{-2}\, \phi\, \phi^\dag  \nonumber \ .
\eeq
It follows that the one-parameter subgroup of $\SU$ generated by $\gamma_\phi$ is
\beq \label{ExponentialGammaVector}
\exp(t\, \gamma_\phi) \ = \  e^{-it / \sqrt{3} } \, \bmatr \ 1 &    \linebr     &\ e^{i \sqrt{3} \, t}\, I_2 \,+\,  \Big( 1 - e^{i \sqrt{3}\, t}  \Big)  |\phi|^{-2}\, \phi\, \phi^\dag \ematr \ .
\eeq
This subgroup defines a circle inside $\SU$ as the parameter $t$ ranges from 0 to $2 \sqrt{3}\, \pi$. Since the metric $g_\phi$ is left-invariant and coincides with the product $\beta$ when restricted to the subalgebra $\utwo$ of $\su$, the length $\ell_\gamma$ of this circle is simply
\beq
\ell_\gamma \ = \ 2 \, \pi  \sqrt{3 \, g_\phi (\gamma_\phi, \gamma_\phi)} \ = 2 \, \pi  \sqrt{3\, \beta (\gamma_\phi, \gamma_\phi)} \ .
\eeq
So we conclude that
\beq
e \ = \ \frac{6 \, \pi \sqrt{\, 2\, \kappa_M} }{ \ell_\gamma} \ .
\eeq
This formula agrees with the result obtained by Weinberg in \cite{WeinbergCharges} on more general grounds\footnote{Provided that one makes the following adjustments: 1) the gravitational constant $\kappa$ of \cite{WeinbergCharges} is related to the definition used here by $\kappa^2 =  2\, \kappa_M$; 2) formula 16 of \cite{WeinbergCharges} represents the value of the positive elementary charge of which all other charges are integer multiples, so it is the antiquark charge $e/3$; 3) for an isometry subgroup ${\rm U}(1)_L$ acting on the internal space $K = \SU$ one should take $N_e = 1$ in the cited formula.}. In Lorentz-Heaviside-Planck units with $c= \hbar = \varepsilon_0 = \mu_0 = 8\pi G = 1$,  we have $\kappa_M = 1$ and $e = \sqrt{4\pi \alpha}$, so the formula implies a length of the internal ``electromagnetic circle'' of
\[
\ell_\gamma \ \simeq\  88.0\ \ell_P \ \simeq \  7.13 \times 10^{-33}\  \rm{m} \ , 
\]
where $\ell_P$ is the rationalized Planck length $\sqrt{8\pi G \hbar c^{-3}}$. It is similar to the estimates obtained in 5D Kaluza-Klein theory.

\subsubsection*{Gauge coupling constants} 
\addcontentsline{toc}{subsubsection}{Gauge coupling constants}

The purpose of the next paragraphs is to read from the four-dimensional Lagrangian the value of the classical electroweak and strong coupling constants of the model. Their values will be expressed in terms of the three positive constants --- $\lambda_1$, $\lambda_2$ and $\lambda_3$ --- that were used in section 5 of \cite{Baptista} to write the general $\Ad_{\Utwo}$-invariant metric on $\su$ that appears in the Lagrangian density.

Consider the Yang-Mills and Dirac terms coming from Lagrangian \eqref{GaugeCouplings3} that involve the electroweak fields $A_\LL$, which are taken with values in the subspace $\utwo$ of $\su$:
\beq \label{AuxDensity4}
\left\{ - \frac{1}{4}\,  \beta(e_k, e_j) \ (F_{A_\LLL}^k)^{\mu \nu}  (F_{A_\LLL}^j)_{\mu \nu}  \ + \  i \xi \, (A_\LL^j)_\mu \, \Tr \Big[ \Psi^\dag \, (\gamma^0 \gamma^\mu  \otimes I_2)\, \rho_{e_j}^\LL (\Psi) \Big]  \right\} \frac{\Vol\, K}{2\, \kappa_P} \,   .
\eeq
Consider first the case of the diagonal matrix $Y := \iota(i I_2) = \diag(-2i, i, i)$ in $\su$. Just as in the case of the electromagnetic charge, the coupling constant $g'/2$ associated to the subgroup $U(1)_Y$  can be read from Lagrangian \eqref{AuxDensity4} by looking at the eigenvalues of the matrix $-i \, \rho^\LL_{\accentset{\circ}{Y}}$ using the normalization $\accentset{\circ}{Y}$ of $Y$ such that
\beq
  \beta\big(\accentset{\circ}{Y},\,  \accentset{\circ}{Y} \big) \ = \  \xi \ =\  2\,  \kappa_M \ = \ 2\,  \kappa_P  \, (\Vol\, K)^{-1} .
\eeq
Indeed, with this normalization the components of the gauge fields associated to the direction $\accentset{\circ}{Y} \in \su$ define a canonically normalized Lagrangian inside \eqref{AuxDensity4}, so the coupling constant $g'/2$ is minus the eigenvalue of the matrix $-i \, \rho^\LL_{\accentset{\circ}{Y}}$ when applied to a fermionic field of hypercharge $-1$, say the electron field $e_\LL^-$ \cite{Hamilton, Weinberg}.
But $\accentset{\circ}{Y}$ and $Y$ are related by the factor $\sqrt{2\,  \kappa_M \,/\, \beta(Y, Y) }$. Thus, working with the non-normalized vector, $g'/2$ is minus the eigenvalue of the matrix
\beq
- i \ \sqrt{\frac{2\,  \kappa_M}{\beta(Y,\, Y)}} \ \, \rho_{Y}^\LL \  
\eeq
acting on the left-handed electron field. The definition of $\rho$ in \eqref{RhoAction} says that $-i \,  \rho_{Y}^\LL$ has an eigenvalue of $-3$ when acting on $e_\LL^-$. Hence we conclude that the $U(1)_Y$ coupling constant of the model is given by
\beq
\frac{g'}{2} \ = \ 3 \ \sqrt{\frac{2\,  \kappa_M}{\beta\big( Y,\,  Y \big)}}   \ .
\eeq
The coupling constant of the weak $\SUtwo$ subgroup of the the Standard Model group can be obtained in a similar way. Take the generator $i \sigma^3$ of $\sutwo$ and consider the corresponding matrix $T_3:= \iota(i \sigma^3)$ regarded as an element of $\su$. Take the normalization $\accentset{\circ}{T}_3$ of $T_3$ such that
\beq
 \beta\big(\accentset{\circ}{T}_3,\, \accentset{\circ}{T}_3 \big) \ = \ \xi  \ = \  2\,  \kappa_M \ .
\eeq
Then the components of the gauge fields associated to the direction $\accentset{\circ}{T}_3 \in \su$ define a canonically normalized Lagrangian inside \eqref{AuxDensity4}, so the coupling constant $g/2$ is the eigenvalue of the matrix $-i \, \rho^\LL_{\accentset{\circ}{T}_3}$ when applied to a fermionic field of weak isospin $1/2$, say the neutrino field $\nu_\LL$ \cite{Hamilton, Weinberg}.
The matrices $\accentset{\circ}{T}_3$ and $T_3$  are related by the factor $\sqrt{2\,  \kappa_M \,/\, \beta(T_3, T_3) }$. Thus, working with the non-normalized vector, $g/2$ is the eigenvalue of the matrix
\beq
- i \ \sqrt{\frac{2\,  \kappa_M}{\beta(T_3,\, T_3)}} \ \, \rho_{T_3}^\LL \  
\eeq
acting on the left-handed neutrino field. The definition of $\rho$ in \eqref{RhoAction} says that $-i \,  \rho_{T_3}^\LL$ has an eigenvalue of $1$ when acting on $\nu_\LL$, so we conclude that the coupling constant is just
\beq
\frac{g}{2} \ = \ \sqrt{\frac{2\,  \kappa_M}{\beta\big(T_3,\, T_3 \big)}}  \ .
\eeq
Finally, to obtain the strong coupling constant, denoted by $g_s$, start by considering the four-dimensional density \eqref{GaugeCouplings3} and collect all the terms involving the gluon fields $A_\RR^j$. These terms have been calculated in the previous section and in section 5 of \cite{Baptista}, yielding an expression of the form
\beq \label{AuxDensity3}
\left\{ - \frac{1}{4}\,  \tilde{\lambda} \, \Tr(e_k^\dag \, e_j) \ (F_{A_\RRR}^k)^{\mu \nu}  (F_{A_\RRR}^j)_{\mu \nu}  \, + \, i \xi \, (A_\RR^j)_\mu \, \Tr \Big[ \Psi^\dag \, (\gamma^0 \gamma^\mu  \otimes I_2)\, \rho_{e_j}^\RR (\Psi) \Big]  \right\} \frac{\Vol\, K}{2\, \kappa_P} \,   .
\eeq
Now consider the generator $t_3 := i \, \diag(1, -1, 0)$ of $\su$, which is just the anti-hermitian version of the third Gell-Mann matrix. For a canonically normalized Lagrangian, the coupling constant $g_s/2$ can be defined as the eigenvalue of the matrix $-i \, \rho^\RR_{t_3 }$ when applied to a red quark field, say the field $u_\RR^r$ \cite{Hamilton}. For the Lagrangian \eqref{AuxDensity3} we should look at the eigenvalues of the matrix $-i \, \rho^\LL_{\accentset{\circ}{t_3}}$ using the normalization $\accentset{\circ}{t}_3$ of $t_3$ such that
\beq
 \tilde{\lambda} \, \Tr \big(\accentset{\circ}{t}_3^{\,\dag} \ \accentset{\circ}{t}_3 \big) \ = \ \xi  \ = \  2\,  \kappa_M \ .
\eeq
The matrices $\accentset{\circ}{t}_3$ and $t_3$  are related by
\[
\accentset{\circ}{t}_3 \ = \ \sqrt{ \frac{2\,  \kappa_M}{\tilde{\lambda} \, \Tr (t_3^{\, \dag} \, t_3) } } \ t_3 \ = \ \sqrt{ \frac{\, \kappa_M}{\tilde{\lambda} } } \ t_3 \ .
\]
Thus, working with the non-normalized vector, $g_s/2$ is the eigenvalue of the matrix
\beq
- i \ \sqrt{\frac{\,\kappa_M}{ \tilde{\lambda} } } \ \, \rho_{t_3}^\RR \ 
\eeq
acting on the quark field $u_\RR^r$. The definition of $\rho$ in \eqref{RhoAction}, together with the fermion identifications \eqref{AuxIdentification3} and \eqref{FermionicIdentification1},  say that $-i \,  \rho_{t_3}^\RR$ has an eigenvalue of $1$ when acting on $u_\RR^r$, so we conclude that the strong coupling constant is
\beq
\frac{g_s}{2} \ = \ \sqrt{\frac{\,\kappa_M}{ \tilde{\lambda} }} \ .
\eeq
These results for the gauge coupling constants can be recast in terms of the three constants $\lambda_1$, $\lambda_2$ and $\lambda_3$ that were used in section 5 of \cite{Baptista} to define the $\Ad_{\Utwo}$-invariant metric on $\su$, which was denoted there by $\tbeta$ but is called here simply $\beta$. A straightforward calculation using the definition of that metric yields
\[
\beta\big( Y,\,  Y \big) \ = \ 6\, \lambda_1 \qquad \qquad   \beta\big( T_3,\,  T_3 \big) \ = \ 2\,  \lambda_2  \qquad \qquad  \beta\big( \gamma_\phi,\,  \gamma_\phi \big) \ = \ (  \lambda_1 + 3\,  \lambda_2) /2 \ .
\]
Combining with the definition of the constant $\tilde{\lambda}$ as the weighted average $(  \lambda_1 + 3 \lambda_2 + 4 \lambda_3 ) /8$, and keeping in mind that in Lorentz-Heaviside-Planck units $\kappa_M =1$, we finally obtain the relations
\bal \label{RelationsCoupling constants}
\frac{g'}{2} \ &= \ \sqrt{\frac{ 3}{ \, \lambda_1}}   &   e\  &= \  \frac{2\, \sqrt{3}}{\sqrt{ \lambda_1 +  3\, \lambda_2}}    \linebr   
\frac{g}{2} \ &= \    \frac{1}{\sqrt{\lambda_2}}      &  \frac{g_s}{2} \ &= \ \frac{2\,\sqrt{2}}{\sqrt{ \lambda_1 +  3\, \lambda_2 + 4\, \lambda_3 }} \ . \nonumber
\end{align}

\subsection*{Uniqueness of the spinor vertical behaviour} 
\addcontentsline{toc}{subsection}{Uniqueness of the spinor vertical behaviour}

A natural question to ask about the model described in the last few sections is whether the vertical transformation rule \eqref{VerticalTransformationSpinors} for the spinor components $\Psi^P (x, h)$ is unique, in the sense that it is the only possible behaviour of $\Psi$ along $K = \SU$ that projects down, after fibre-integration, to the correct fermionic representations in the four-dimensional Lagrangian. The answer is no, it is not unique. Even leaving alone the non-abelian $3\times3$ components of the transformation matrix $S(h)$, as in \eqref{DefinitionS}, one can check that there exists more than one scalar function $s(h)$ that produces the same representations in $M_4$. To recognize this fact, start by observing that the calculations of section 2.3 only require that the complex function $s: K \to \CC$ satisfies the following integral identities:
\bal \label{sIdentities}
\int_K \, \sbb \, \big(\Lie_{v^\RRR} \, s\big) \ \vol_K \ &= \ 0     \linebr
\int_K \, \sbb \, \big(\Lie_{v^\LLL} \, s\big) \ \vol_K \ &= \ 2\, v_{11}  \int_K \, |s|^2 \ \vol_K \    \nonumber \linebr
\int_K\, |s|^2 \, h^\dag\, v\, h \ \vol_K \ &= \ \int_K\, |s|^2 \, h\, v\, h^\dag \ \vol_K \ = \ 0 \nonumber \ ,
\end{align}
for all matrices $v$ in $\su$. Any normalized function $s(h)$ that satisfies these identities, and hence also the complex conjugate identities, may be plugged into definition \eqref{DefinitionS} of $S(h)$ without affecting the overall conclusions of section 2.3.

Looking into the detailed calculation of the integrals above, one of the salient features of the original function
\beq
s_1 (h) \ := \  \, (h_{11} )^2 \, + \, (h_{21})^2 \, + \,  (h_{31} )^2   \ , 
\eeq
leading to the second integral identity, is its equivariance under the action of the embedded subgroup $\Utwo \to \SU$. Namely,
\beq \label{Equivariance}
s_1 \big(\,  h \cdot \iota(a) \, \big) \ = \ (\det a)^{-2} \, s_1 (h)  \ ,
\eeq
for all matrices $a \in \Utwo$, with the usual notation $\iota(a) = \diag \big( \det a^{-1}, \, a \big)$. In addition, for the first integral in \eqref{sIdentities} to vanish it was also helpful that $s_1(h)$ is symmetric in its three variables, since this produces a factor $v_{11} + v_{22} + v_{33}$ that is zero for all $v \in \su$. Therefore, one promising possibility to change the function $s_1(h)$ would be to multiply it by a totally symmetric real polynomial $p\big(|h_{11}|^2, \, |h_{21}|^2,\, |h_{31}|^2 \big)$, as this would not disturb the equivariance property \eqref{Equivariance}. A more detailed analysis of the integral calculations shows that this is indeed a viable option. One can rigorously prove that the new scalar function
\beq
 p\big(|h_{11}|^2, \, |h_{21}|^2,\, |h_{31}|^2 \big) \ s_1(h)
\eeq
also satisfies identities \eqref{sIdentities}, and hence leads to the correct fermionic representations in $M_4$ after normalization. In fact, more is true: given a second symmetric polynomial  in the same variables, $p_2\big(|h_{11}|^2, \, |h_{21}|^2,\, |h_{31}|^2 \big)$, one may generalize the transformation $S(h)$ of \eqref{DefinitionS} to
\beq \label{DefinitionS2}
S(h) \ := \ \bmatr \ p(h) \, \conj{s_1(h)}  \  &   \quad   \linebr   &  p_2(h) \, h \ \    \ematr  \ ,
\eeq 
and still the components of $\Psi^P (x, h)$ will project down to the fermionic representations found in the Standard Model. Hence a first conclusion seems to be that there are infinitely many vertical transformations $S(h)$ of $\Psi(x)$ that produce the correct representations in $M_4$, after fibre-integration along $K$, as long as we keep increasing the degree of the scalar polynomials involved. Heuristically, higher-degree polynomials on the entries of $h \in \SU$ will produce vertical functions $S(h)$ with more ``oscillations'' along $K$, and hence would presumably correspond to larger eigenvalues of the Laplacian and Dirac operators on $K$. In other words, would correspond to higher masses of the fermionic components of $\Psi (x)$ after projecting down to the four-dimensional Lagrangian.

What if we do not want to increase the degree of the polynomials involved? The original scalar function $s_1(h)$ has degree two in the entries $h_{k1}$ of $h$, and there exists a second symmetric polynomial of the same degree in these variables, namely
\beq
s_2 (h) \ := \   h_{11} h_{21}\, + \, h_{11} h_{31}\, +  h_{21} h_{31}   \ .
\eeq
Since $s_2(h)$ has the same equivariant behaviour as $s_1(h)$ under right-multiplication $h \to h\cdot \iota(a)$, satisfying \eqref{Equivariance}, it follows that $s_2 (h)$ also satisfies the second integral identity in \eqref{sIdentities}. However, the results in appendix A.4 show that
\bal \label{s2Identities}
\int_K \, \sbb_2 \, \big(\Lie_{v^\RRR} \, s_2\big) \ \vol_K \ &= \ \frac{2i}{3}\, \Imaginary (v_{12} + v_{13} + v_{23})  \, \int_K \, |s_2|^2 \ \vol_K  \linebr
\int_K\, |s_2|^2 \, h\, v\, h^\dag \ \vol_K \ &= \   \frac{1}{10}\, v_{11}  \bmatr 0 & 1 & 1 \\ 1 & 0 & 1 \\ 1 & 1 & 0 \ematr   \, \int_K \, |s_2|^2 \ \vol_K  \nonumber \linebr
\int_K\, |s_2|^2 \, h^\dag\, v\, h \ \vol_K \ &= \ \frac{i}{15}\, \Imaginary (v_{12} + v_{13} + v_{23}) \, \diag(-2,1,1) \, \int_K \, |s_2|^2 \ \vol_K   \nonumber \ ,
\end{align}
for all matrices $v \in \su$. This is very different from \eqref{sIdentities}, so the scalar function $s_2(h)$ on $K = \SU$ does not have the required behaviour along $K$ to produce the correct projections to $M_4$ of the spinor $\Psi^P (x, h)$.

This is not the end of the story for degree two polynomials, though. Persevering in the calculations of slightly more general integrals over $\SU$, one can show that linear combinations of both polynomials
\beq
s(h) \ := \   \alpha_1\, s_1(h) \, + \, \alpha_2\, s_2(h)   \ ,
\eeq
actually do satisfy identities \eqref{sIdentities} whenever the constants $\alpha_1$ and $\alpha_2$ are related by
\beq
|\alpha_2|^2 \ + \ 2 \, (\alpha_1\, \conj{\alpha_2} \, +  \conj{\alpha_1} \, \alpha_2 )  \ = \ 0  \ .
\eeq
Given an arbitrary $\alpha_1 \in \CC$, the general solution to this equation is just
\beq
\alpha_2 \ = \ 2 \, (1 + e^{i2\varphi}) \, \alpha_1
\eeq
for some phase $\varphi \in [ 0; \pi[$. Thus, for $\alpha_1 = 0$ we get $\alpha_2 = 0$ as well, and there is no non-trivial solution. For $\alpha_1 \neq 0$ we get a ``circle'' of solutions parametrized by $\varphi$:
\beq \label{DefinitionSPhi}
s_\varphi (h) \ := \   \alpha \, \big[ \, s_1(h) \, - \, 2 \, (1 + e^{i2\varphi})\, s_2(h) \, \big]   \ ,
\eeq
apart from the overall normalization factor $\alpha$. The simplest solution $s_1(h)$, discussed in the previous section, is recovered when $\varphi = \pi /2$.  

The conclusion is that the whole discussion of sections 2.2 and 2.3 remains valid if, instead of the choice $s = \sqrt{2}\, s_1$ used there, we choose as $s(h)$ any degree two polynomial belonging to the family $s_\varphi (h)$ defined by \eqref{DefinitionSPhi}.
Observe also that the formulae \eqref{A4I3} in appendix A.4 imply that the integrals of $|s_\varphi (h)|^2$ and $|s_\varphi (h)|^4$ are
\bal \label{sPhiIdentities}
\int_K\, |s_\varphi |^2 \ \vol_K \ &= \   \frac{1}{2}\, |\alpha|^2 \, \big(1+8\cos^2(\varphi) \big) \, (\Vol \, K)    \linebr
\int_K\, |s_\varphi |^4 \ \vol_K \ &= \   \frac{1}{15}\, |\alpha|^4 \, \big(5 \, + \,144 \cos^2(\varphi)\, + \,  256 \cos^4(\varphi) \big) \, (\Vol \, K)  \nonumber \ .
\end{align}
 The discussion in section 2.3 assumed a function $s$ normalized so that the fibre-integral of $|s|^2$ would be equal to $\Vol \, K$. Thus, the complex constant $\alpha$ in definition \eqref{DefinitionSPhi} should be picked to satisfy
\beq
|\alpha|^2 \ = \ \frac{2}{1+8\cos^2(\varphi)} \ .
\eeq 
The normalized functions $s_\varphi$ are then parameterized by two circles, not just one.

In the present section 2 we have identified relatively simple functions and vertical behaviours of the spinor $\Psi^P (x, h)$ that generate, after pairing and fibre-integration, the complete set of fermionic gauge representations appearing in the Standard Model. It would be desirable to find a more fundamental, {\it a priori} justification for the emergence of these particular vertical functions, other than the fact that they generate the experimentally observed gauge representations in four dimensions. Studying the properties of the spinors $\Psi^P (x, h)$ under the action of natural Dirac operators on $\SU$ could be a useful starting point.

\newpage

\section{Masses induced in four dimensions}

\subsection*{Mass in Kaluza-Klein theories} 
\addcontentsline{toc}{subsection}{Mass in Kaluza-Klein theories}

Traditional Kaluza-Klein theories regard the mass of particle fields as being due to the vibrations of the higher-dimensional fields along the internal space $K$, proposing that the energy associated to those movements will be perceived in four dimensions as the particle's mass. More precisely, if we endow $P = M_4 \times K$ with a ``vacuum'' product metric, the higher-dimensional Laplacian can be written as a sum of its lower-dimensional counterparts, $\Delta^P =  \Delta^M +  \Delta^K$. Therefore, if a scalar particle field behaves along $K$ as an eigenfunction of the internal operator $\Delta^K$, the kinetic term of the higher-dimensional Lagrangian will produce both kinetic and mass terms in the four-dimensional, Klein-Gordon Lagrangian:
\bal \label{DecompositionLaplaceTerms}
 \Delta^K \psi \ &= \ - \, \mu^2 \, \psi  \quad &\implies \qquad \ \  \ \left\langle \psi,\, \Delta^P \psi \right\rangle \ &= \  \left\langle \psi, \, \Delta^M \psi \right\rangle  \ -\ \mu^2 \left\langle \psi, \, \psi \right\rangle  \ .
\end{align}
For fermions the argument is similar. The higher-dimensional spinor space can be factorized as a tensor product $\Delta_{12} = \Delta_4 \otimes \Delta_8$. If $\Psi = \psi \otimes \chi$ is a spinor in this space, the higher-dimensional Dirac operator acts on it as
\[
 \dirac^P  (\psi \otimes \chi)  \ = \ (\dirac^M \psi) \otimes \chi \, + \, (\gamma^5 \psi) \otimes (\dirac^K \chi)  \ ,
\]
where $\gamma^5 = \diag(I_2, - I_2)$. 
In particular, if $\chi$ is an eigenspinor of $\dirac^K$ with a real eigenvalue $\mu$, then the classic Dirac kinetic term on the higher-dimensional manifold can be decomposed as
\bal \label{DecompositionDiracTerms}
i \, \big\langle \Gamma^0 \Psi,\, \dirac^P \Psi \big\rangle \ &= \  i\, \big\langle \gamma^0 \psi, \, \dirac^M \psi \big\rangle  \big\langle \chi , \chi \big\rangle  \ +\ i \mu \, \big\langle \gamma^0 \psi, \, \gamma^5 \psi \big\rangle \big\langle \chi , \chi \big\rangle   \nonumber \linebr
&= \ i\, \big\langle \gamma^0 \psi', \, \dirac^M \psi' \big\rangle  \ -\ \mu \, \big\langle \gamma^0 \psi', \,  \psi' \big\rangle \ . 
\end{align}
In the last equality, the spinor $\psi$ on Minkowski space was redefined through an isomorphism of $\Delta_4$ in order to obtain the four-dimensional kinetic and mass terms in their traditional form \cite[p. 22]{Duff}:
\[
\psi' \ := \ \sqrt{\frac{\langle \chi , \chi \rangle}{2}} \ \big( I_4   \, +\,  i  \gamma^5 \big) \, \psi \ = \ \sqrt{\langle \chi , \chi \rangle} \, \exp{(i \pi \gamma^5 / 4)}  \, \psi  \ .
\]

We now want to investigate the extent to which this attractive general picture can be applied to the $12$-dimensional spinor $\Psi^P(x, h)$ with its prescribed behaviour along $K$. Does the $h$-dependence prescribed by \eqref{VerticalTransformationSpinors} determine anything close to an eigenfunction of $\Delta^K$ or $\dirac^K$? In line with our discussion so far, we do not really need exact eigenfunctions: the right-hand side of equations \eqref{DecompositionLaplaceTerms} and \eqref{DecompositionDiracTerms} may have additional terms, as long as they integrate to zero when performing the integrals over $K$ that generate the four-dimensional Lagrangian.

What we aim to calculate, in fact, are the integrals of the pairings $\left\langle \Psi^P,\, \Delta^K \Psi^P \right\rangle$ and $\big\langle \Gamma^0 \Psi^P, \, (\gamma^5 \otimes \dirac^K) \Psi^P \big\rangle$ over the group $K=\SU$ equipped with a left-invariant metric similar to the metrics $g_\phi$ defined in \cite{Baptista}. Between these two pairings, the most relevant is the one involving the Dirac operator. Since its counterpart $\big\langle \Gamma^0 \Psi^P, \, \dirac^M \Psi^P \big\rangle$ produces the fermionic kinetic terms of the Standard Model Lagrangian, as studied in section 2.3, it would be great if the second pairing $\big\langle \Gamma^0 \Psi^P, \, (\gamma^5 \otimes \dirac^K) \Psi^P \big\rangle$  would produce the mass terms of the same Lagrangian, after fibre-integration over $K$. The algebra involved in the explicit calculation of the latter pairing seems to be relatively straightforward but more extended than what we can tackle here. Although we give a few preparatory details in the next short section, we will not carry out the full algebraic calculations here, so will not be able to offer a clear picture of the mass terms that appear in four dimensions after projecting $\big\langle \Gamma^0 \Psi^P, \, (\gamma^5 \otimes \dirac^K) \Psi^P \big\rangle$ down from $P$ through fibre-integration. This is a consequential point that deserves further investigation in the study of the present geometrical model.

The calculations are shorter in the case of the scalar Laplacian $\Delta^K$ over $K$, since it can be expressed purely in terms of Lie derivatives $\Lie_{v^\LLL}$, with no gamma matrices mixing the components of $\Psi^P$. We will therefore be able to calculate the explicit mass terms produced by the pairing $\big\langle \Psi^P, \, \Delta^K \Psi^P \big\rangle$ after projection to four dimensions. These calculations will occupy us in the latter parts of the present section.

\subsection*{Fermion masses from the internal Dirac operator} 
\addcontentsline{toc}{subsection}{Fermion masses from the internal Dirac operator}

By definition of left-invariant metric, if $\{ e_j \}$ is a $g$-orthonormal basis of $\su$, the corresponding left-invariant vector fields $\{ e_j^\LL \}$ will be everywhere orthonormal on $K$. Therefore, the Dirac operator on $(K, g)$ can be written on general grounds \cite{Bourguignon} as
\bal
\dirac^K \psi \ = \ \sum_j \, \Gamma^j \, \nabla^{\rm{\, spin}}_{v_j^\LLL} \psi  \ = \ \sum_j \, \Gamma^j  \left(    \Lie_{v_j^\LLL} \psi \, + \, \frac{1}{2} \sum_{k<l} \, g\big(\nabla_{v_j^\LLL} v^\LL_k , \, v_l^\LL \big) \, \Gamma^k \,  \Gamma^l \,  \psi \right) \ ,
\end{align}
where $\{ \Gamma^j \}$ is a set of gamma matrices corresponding to $\{ e_j \}$, and $\nabla$ is the original connection on the tangent bundle $TK$ that leads to the connection $\nabla^{\rm{\, spin}}$ on the spin bundle. When $\nabla$ is the Levi-Civita connection of a left-invariant metric, the Koszul formula allows us to express it in purely algebraic terms:
\beq
g\big(\nabla_{v_j^\LLL} v^\LL_k , \, v_l^\LL \big) \ = \ \frac{1}{2} \Big\{  g( [v_j, v_k] , \, v_l) \, - \, g([v_k, v_l] , \, v_j)  \, + \, g([v_l, v_j] , \, v_k)    \Big\} \ ,
\eeq
where we have used that the product $g(v_j^\LL, \, v_k^\LL) = g(v_j, \, v_k)$ is constant on $K$ and that $[v_j^\LL ,\,  v_k^\LL] = [v_j, \, v_k]^\LL$. Moreover, using the properties of gamma matrices and the fact that left-invariant vector fields have vanishing divergence on a unimodular Lie group equipped with a left-invariant metric,
\beq
\sum_j \, g\big(\nabla_{v_j^\LLL} v^\LL_k , \, v_j^\LL \big) \ = \ \divergence_g (v_k^\LL) \ = \ 0 \ ,
\eeq
the general formula for the Dirac operator on $(K, g)$ can be further simplified to \cite{Bar}
\beq \label{DiracOperator}
\dirac^K \psi \ = \ \sum_j \, \Gamma^j \left(    \Lie_{v_j^\LLL} \psi \right) \ + \ \sum_{j<k<l} \, \alpha_{jkl}\, \Gamma^j \, \Gamma^k \, \Gamma^l \, \psi \ .
\eeq
Here the totally anti-symmetric real coefficients $\alpha_{jkl}$ are defined by
\beq
\alpha_{jkl} \ := \ \frac{1}{4}\, \Big\{  g( [v_j, v_k] , \, v_l) \, + \, g([v_k, v_l] , \, v_j)  \, + \, g([v_l, v_j] , \, v_k)    \Big\} \ .
\eeq
Observe that when $g$ is a bi-invariant metric on $K$, the three terms in the preceding sum are equal to each other, so the formula for $\alpha_{jkl}$ is simpler. This simplification does not occur in the case of the left-invariant metrics $g_\phi$ and $\tilde{g}_\phi$ defined in sections 2 and 5 of \cite{Baptista}. Denoting by $v = v' + v''$ the standard decomposition of a vector in $\su \simeq \utwo \oplus \CC^2$, as in \cite{Baptista}, and inserting the definition of $g_\phi$ into the expression for $\alpha_{jkl}$, a little algebra shows that these coefficients may be written in terms of the Ad-invariant metric $\beta$ on $\su$ as
\bal
\alpha_{jkl} \ &= \ \frac{3}{4} \, \beta\big([v_j, v_k] , \, v_l\big) \ + \ \frac{1}{2} \, \beta\left( \phi, \, \big[[v_j', v_k'] , \, v_l''\big] \, +\, \big[[v_k', v_l'] , \, v_j'' \big]\, +\, \big[[v_l', v_j'] , \, v_k'' \big]  \right) \nonumber  \linebr
&= \ \frac{3}{4} \, \beta\big([v_j, v_k] , \, v_l\big) \ + \ \frac{1}{2} \, \sum_{\sigma}\, \beta\left( \, \phi, \, \big[[v_{\sigma(j)}', v_{\sigma(k)}'] , \, v_{\sigma(l)}''\big]  \big] \,  \right) \ ,
\end{align}
where the last sum is over the circular permutations of $\{j, k, l \}$. Therefore, the Dirac operator \eqref{DiracOperator} depends in two ways on the parameter $\phi$ of $g_\phi$: implicitly, through the choice of orthonormal basis $\{ e_j \}$ of $\su$; explicitly, in the formula for the coefficients $\alpha_{jkl}$.
The Dirac operator on $K$ will be simpler if the original connection $\nabla$ on the tangent bundle is taken to be the flat Schouten connection $\nabla^0$, instead of the usual Levi-Civita connection. This is a $g$-compatible connection with torsion $T^{\nabla^0} (u^\LL,\, v^\LL) = [u,v]^\LL$, and the coefficients $\alpha_{jkl}$ of the corresponding Dirac operator are identically zero. In Minkowski space $M_4$ these two connections coincide, since the Lie bracket of the abelian translations group is trivial and $\nabla^0$ is torsionless too.

In any case, the first step in the calculation of the projection of $\big\langle \Gamma^0 \Psi^P, \, (\gamma^5 \otimes \dirac^K) \Psi^P \big\rangle$ down to $M_4$ should probably be dealing with the derivative terms $\Gamma^j \, \Lie_{v_j^\LLL} \psi$ of the full $\dirac^K \psi$. More precisely, in analogy with the integrals \eqref{GaugeCouplings2} that lead to the fermionic kinetic terms in $M_4$, one may try to calculate the fibre-integrals
\beq \label{MassIntegrals1}
\int_K \ \sum_{l = 1}^8  \, \big\langle \, \big(\Gamma^0 \Gamma^{3+l} \Psi \big)^P , \  \Lie_{e_l^\LL} \big( \Psi^P \big) \, \big\rangle \ \vol_K \ ,
\eeq
and see what kind of mass terms they induce in four dimensions. Here $\{ e_l \}$ is a $g$-orthonormal basis of $\su$ but the $\Gamma^a$ are the gamma matrices of the 12-dimensional $M_4 \times \SU$ defined in \eqref{GammaMatrices}. So $\Gamma^{3+l}$ is the gamma matrix corresponding to $e_l$. This calculation is simplified by the work done in section 2.3, where it was shown that for general $8\times8$ matrices $\Psi_1 (x)$, $\Psi_2 (x)$ and for any vector $v$ in $\su$:
\beq \label{MassIntegrals2}
\int_K \ \big\langle \, \Psi_2^P , \  \Lie_{v^\LL} \big( \Psi_1^P \big) \, \big\rangle \ \vol_K  \ = \ \int_K \ \big\langle \, \Psi_2^P , \, ( \rho_{v}^\LL \, \Psi_1 )^P \, \big\rangle \ \vol_K \ .
\eeq
The algebraic action of $\rho_{v}^\LL$ on the components $\psi_\pm$ of $\Psi$ was explicitly written in \eqref{RhoAction}. In the case of interest we have $\Psi_2 = \Gamma^0 \Gamma^{3+l} \Psi$ and $\Psi_1 = \Psi$, so the basic task of the calculation will be to keep track of the components of the $8\times8$ matrices $\Gamma^0 \Gamma^{3+l} \Psi$ for $l=1, \ldots, 8$, in order to calculate the sum of all pairings $\big\langle \, \Gamma^0 \Gamma^{3+l} \Psi , \,  \rho_{v}^\LL \, \Psi  \, \big\rangle$.
The integrals over $K$ coming from the right-hand side of  \eqref{MassIntegrals2}, on the other hand, are easily computable. This is so because $\Psi_2$ and $\rho_{v}^\LL \, \Psi_1$ do not depend on the coordinate $h \in K$, and the prescription $\Psi \to \Psi^P$ that extends them to functions on $K$ is such that, when dealing with a product $\Tr\big[  (\Psi_2^P)^\dag \, \Psi_1^P \big]$, the only emerging integrals are those of the scalar functions $|s(h)|^2$ and $|s(h)|^4$, both of which have already been computed here. 

In the case of $\nabla$ being the Levi-Civita connection on $K$, the explicit calculation of the coefficients $\alpha_{jkl}$ should be a second extended algebraic task.

Finally, let us comment that even if the whole algebraic calculation would develop smoothly, its ideal result should not be too simple a set of mass terms on $M_4$, as we have not touched upon the subject of fermion generation mixing. In order to replicate the full mass structure of the Standard Model, one would ideally hope to find at least three distinct 12-dimensional spinor fields --- $\Psi_1^P (x, h)$, $\Psi_2^P (x, h)$ and  $\Psi_3^P (x, h)$ --- all of them having slightly different behaviour along the internal space $K$, though still projecting down to the correct gauge representations on $M_4$ (see the last part of section 2), and then calculate that only non-trivial linear combinations of the three spinors would behave as eigenfunctions of $\dirac^K$, after pairing and projection to $M_4$. This is an idealized scenario, of course, and should be taken as wishful speculation.

\subsection*{Masses from the internal Laplace operator} 
\addcontentsline{toc}{subsection}{Masses from the internal Laplace operator}

On a connected, unimodular Lie group, such as $K = \SU$, every left-invariant volume form is also right-invariant. Hence the natural volume form of a left-invariant Riemannian metric $g$ on $K$ will be fully bi-invariant, and both the right and left-invariant vector fields will have zero divergence on $(K, g)$. So if we take an orthonormal basis $\{ e_j \}$ of the Lie algebra and extend it to left-invariant vector fields $\{ e_j^\LL \}$, the scalar Laplacian on $(K, g)$ may be written, in terms of these orthonormal fields, simply as
\beq \label{Laplacian}
\Delta_g^K \, f \ = \ \sum_{j=1}^8 \, \Lie_{e_j^\LL} \big(\Lie_{e_j^\LL} \, f  \big)
\eeq
for any scalar function $f$ on $K$ \cite{DAtri}. This formula does not work with the right-invariant fields $\{ e_j^\RR \}$, as they generally are not orthonormal with respect to the left-invariant $g$.

Proceeding as in section 2, in order to calculate $\Delta_g \, \Psi^P$ more explicitly, we will decompose the 12-dimensional spinor $\Psi^P = [\, \psi_+^P \ \ \psi_-^P \, ]$ into several components that are not mixed by the Lie derivatives $ \Lie_{e_j^\LL}$, and hence by the Laplacian:
\beq
\psi_\pm^P \ = \ \bmatr \ a_\pm^P \  &   (c_\pm^P)^T   \linebr    b_\pm^P &   D_\pm^P \  \ematr  \ .
\eeq
Each of these components transforms along the internal space $K$ according to the rules \eqref{VerticalTransformationSpinors2}. Since the full pairing $\big\langle \,\Psi_2^P , \,  \Delta_g \, \Psi_1^P  \, \big\rangle$ is just a trace $\Tr \big[ (\Psi_2^P)^\dag  \,  \Delta_g \, \Psi_1^P  \big]$, it is clear that it will break into a sum of traces of the smaller components. Thus, the integral along $K$ of the full pairing breaks into a sum of fibre-integrals of the general form
\beq \label{MassIntegrals3}
\int_K \ \Tr \big[ (D_2^P)^\dag  \,  \Delta_g \, D_1^P  \big] \ \vol_K \ = \  \int_K \ \sum_{j=1}^8 \, \Tr \big[ (D_2^P)^\dag  \,  \Lie_{e_j^\LL}\,  \Lie_{e_j^\LL} \, D_1^P  \big] \ \vol_K \ ,
\eeq
and analogous integrals for the remaining components $a$, $b$ and $c$ of $\psi$. Using the explicit expressions for the $K$-dependence of $D^P(x, h)$ and the other components, as stated in \eqref{VerticalTransformationSpinors2}, we will now compute the Lie derivatives and integrals above.

\subsubsection*{Components $D_\pm (x,h)$} 
\addcontentsline{toc}{subsubsection}{Components $D_\pm (x,h)$}

Simplifying the notation of \eqref{VerticalTransformationSpinors2} for the transformation of $D_+$, write
\beq
D^P (h) \ = \ h\, D \, \conj{h} \ ,
\eeq
where it is implicit that the $3\times 3$ matrix $D$ may depend on the coordinate $x$ of $M_4$. The Lie derivatives of this function along any left-invariant vector fields $u^\LL$ and $v^\LL$ on $K$ are
\beq  \label{DoubleLieDerivativesD}
\big(\Lie_{u^\LLL} \, \Lie_{v^\LLL}\, D^P \big) \, (h) \ = \ h\, u\, v\, D \, \hb \ + \ h \, v\, D \, \hb \, \ub   \ + \ h \, u\, D \, \hb \, \vb \ + \  h \, D \, \hb \, \ub \, \vb \ ,
\eeq
where $u$ and $v$ are $3\times3$ matrices in $\su$. If we have two matrix functions $D_1^P$ and $D_2^P$, it is possible to construct the pairing
\beq  \label{PairingD}
\Tr \big[ (D_2^P)^\dag  \,  \Lie_{u^\LL}\,  \Lie_{v^\LL}  D_1^P  \big]  \, = \,  \Tr \big( D_2^\dag\, u\, v\, D_1 \, +  D_2^\dag\, v\, D_1 \, \hb \, \ub \, \hb^\dag   \, +  D_2^\dag\, u\, D_1 \, \hb \, \vb \, \hb^\dag \, +  D_2^\dag\, D_1 \, \hb \, \ub \, \vb \, \hb^\dag \big)  , \nonumber
\eeq
where the equality uses the cyclic properties of the trace. Integrating the right-hand side along the fibre $K$ leads to a significant simplification, since the integrals in appendix A.4 imply that
\bal
\int_K \, \hb \, \ub\, \hb^\dag \ \vol_K \ &= \  \int_K \, \hb \, \vb\, \hb^\dag \ \vol_K \ = \  0 \linebr
\int_K \, \hb \, \ub\, \vb\, \hb^\dag \ \vol_K \ &= \ \frac{1}{3} \, \Tr (\ub \, \vb) \, I_3 \ ,
\end{align}
for any traceless matrices $u$, $v$ in $\su$. So we get
\beq  
\int_K \, \Tr \big[ (D_2^P)^\dag  \,  \Lie_{u^\LL}\,  \Lie_{v^\LL}  D_1^P  \big] \ \vol_K \ = \  \left\{ \Tr \big( D_2^\dag\, u\, v\, D_1\big) \, + \, \frac{1}{3}\,\Tr( D_2^\dag\, D_1) \, \Tr(\ub \, \vb) \right\}   (\Vol\, K)  \ .\nonumber
\eeq
Using expression \eqref{Laplacian} for the Laplacian, we finally obtain
\beq  \label{IntegralPairingD}
\int_K \, \Tr \big[ (D_2^P)^\dag  \, \Delta_g^K  D_1^P  \big] \ \vol_K \ = \  \Tr \big( D_2^\dag \, \Omega^D_g \, D_1\big) \, (\Vol\, K)  \ ,
\eeq
where $\Omega^D_g$ is a $3\times3$ hermitian matrix that can be written in terms of the $g$-orthonormal basis $\{ e_j \}$ of $\su$ as
\beq \label{DefinitionOmegaD}
\Omega^D_g \ :=\ \sum_{j=1}^8 \, e_j \, e_j \, + \, \frac{1}{3} \, \Tr \big(  \conj{e_j} \, \conj{e_j}  \big)\, I_3 \ .
\eeq
An explicit expression for the matrix $\sum_{j} \, e_j \, e_j$ is given in appendix B, together with its eigenvalues and eigenvectors. All of these depend on the Higgs-like vector $\phi \in \CC^2$ that defines the metric $g_\phi$ on $K$, of course.

Choosing a basis of $\CC^3$ that diagonalizes $\Omega_g^D$ and taking the particular case where $D_2^P = D_1^P = D_+^P (x,h)$, the fibre-integral of the higher-dimensional scalar density is simply
\beq  
\int_K \, \Tr \big[ (D_+^P)^\dag  \, \Delta_g^K  D_+^P  \big] \ \vol_K \ = \  \Tr \Big\{ \big[D_+(x)\big]^\dag \, \diag(\mu_1, \mu_2, \mu_3)\, D_+(x)  \Big\} \, (\Vol\, K)  \ , \nonumber
\eeq
where $\mu_a$ are the eigenvalues of $\Omega_g^D$. In other words, after fibre-integration along $K$, the scalar density $\Tr \big[ (D_+^P)^\dag  \, \Delta_g^K  D_+^P  \big]$ on the product $P = M_4\times K$ generates Klein-Gordon mass terms for the fields $D_+(x)$ on $M_4$. The mass parameters are the eigenvalues $\mu_a$.

\subsubsection*{Components $b_\pm (x,h)$} 
\addcontentsline{toc}{subsubsection}{Components $b_\pm (x,h)$}

Simplifying the notation of \eqref{VerticalTransformationSpinors2}, the vertical behaviour along $K$ of the component $b_+^P(x,h)$ can be written as
\beq
b^P (h) \ = \ s_\varphi(h) \, h \, b \ , \nonumber
\eeq
where $s_\varphi : K \to \CC$ is any of the scalar functions defined in \eqref{DefinitionSPhi}. It is implicit that the vector $b \in \CC^3$ may depend on the coordinate $x$ of $M_4$. The Lie derivatives of $b^P$ along left-invariant vector fields $u^\LL$ and $v^\LL$ on $K$ are
\beq  \label{LieDerivativesb}
\big(\Lie_{u^\LLL} \, \Lie_{v^\LLL}\, b^P \big) \, (h) \ = \ (\Lie_{u^\LLL} \, \Lie_{v^\LLL}\, s_\varphi)\, h \, b \ + \ (\Lie_{v^\LLL}\, s_\varphi)\, h \, u\, b \ +  \ (\Lie_{u^\LLL}\, s_\varphi)\, h \, v\, b \ + \ 
s_\varphi \, h \, u \, v\, b \ ,  \nonumber
\eeq
where $u$ and $v$ are matrices in $\su$. If we have two vectorial functions $b_1^P$ and $b_2^P$, it is possible to take the hermitian product
\begin{multline} \label{auxPairingsb}
(b_2^P)^\dag   \big( \Lie_{u^\LLL}\,  \Lie_{v^\LLL}  b_1^P \big) \ = \   \sbb_{\varphi} \,  (\Lie_{u^\LLL} \, \Lie_{v^\LLL}\, s_{\varphi})\, b_2^\dag \, b_1 \ + \ \sbb_\varphi \, (\Lie_{v^\LLL}\, s_\varphi)\, b_2^\dag \, u\, b_1 \ \linebr +   \ \sbb_\varphi \, (\Lie_{u^\LLL}\, s_\varphi)\, b_2^\dag \, v\, b_1 \ + \ 
|s_\varphi|^2 \, b_2^\dag \, u \, v\, b_1 \ . 
\end{multline} 
The integrals calculated in \eqref{A4Int5} and \eqref{A4Int7} of appendix A.4 then say that
\bal
\int_K \, \sbb_\varphi \, (\Lie_{v^\LLL} \, s_\varphi ) \ \vol_K \ &= \ 2\, v_{11} \int_K |s_\varphi|^2 \ \vol_K \nonumber \linebr
\int_K \, \sbb_\varphi \, (\Lie_{u^\LLL} \, \Lie_{v^\LLL} \, s_\varphi ) \ \vol_K \ &= \ 2  \left[  u_{11}\, v_{11} \, + \,  \frac{1}{1+8\cos^2(\varphi)} \, (uv)_{11}  \right]  \int_K |s_\varphi|^2 \ \vol_K \nonumber
\end{align}
for any traceless matrices $u$, $v$ in $\su$. Applying these integral identities to the right-hand side of \eqref{auxPairingsb} we obtain
\begin{multline} \label{auxPairingsb2}
\int_K\, (b_2^P)^\dag   \big( \Lie_{u^\LLL}\,  \Lie_{v^\LLL}  b_1^P \big) \ \vol_K \ = \   b_2^\dag \, \Big\{    2  \left[  u_{11}\, v_{11} \, + \,  \frac{1}{1+8\cos^2(\varphi)} \, (uv)_{11}  \right]  I_3 \\ + \  2\, v_{11}\, u \ +   \  2\, u_{11}\, v \ +  \ u\, v  \Big\} \, b_1 \,  \int_K |s_\varphi|^2 \ \vol_K \ . \nonumber
\end{multline}
Coming back to expression \eqref{Laplacian} for the Laplacian, we finally get that
\beq  \label{IntegralPairingb}
\int_K \, (b_2^P)^\dag  \,  \big( \Delta_g^K  b_1^P \big)  \ \vol_K \ = \  \big( b_2^\dag \, \Omega^b_g \, b_1\big) \,  \int_K |s_\varphi|^2 \ \vol_K \ = \ \int_K \, (b_2^P)^\dag  \,  \big( \Omega_g^b \, b_1^P \big)  \ \vol_K  \ ,
\eeq
where $\Omega^b_g$ is a $3\times3$ hermitian matrix that can be written in terms of the $g$-orthonormal basis $\{ e_j \}$ of $\su$ as
\beq \label{DefinitionOmegaB}
\Omega^b_g \ :=\ \sum_{j=1}^8 \ e_j \, e_j \, + \,   4\, (e_j)_{11}\, e_j \, + \,  2  \left[  (e_j)_{11}^2 \, + \,  \frac{1}{1+8\cos^2(\varphi)} \, (e_j \, e_j)_{11}  \right]  I_3      \ .
\eeq
Expression \eqref{IntegralPairingb} says that, after pairing and fibre-integration over $K$, the action of the vertical Laplacian $\Delta_g^K$ on the functions $b^P$ corresponds to the simpler algebraic action of the matrix $\Omega^b_g$ on the vector $b \in \CC^3$. Taking the particular case where $b_2 = b_1 = b_+$ is an eigenvector of $\Omega^b_g$ with eigenvalue $\mu$, the fibre-integral of the vertical Laplacian kinetic term produces a simple Klein-Gordon mass term on $M_4$:
\beq  
\int_K \, (b_+^P)^\dag  \,  \big( \Delta_g^K  b_+^P \big)  \ \vol_K \ = \  \mu\, b_+^\dag \, b_+  \,  \int_K |s_\varphi|^2 \ \vol_K  \ .
\eeq
The matrix $\Omega^b_g$ and its eigenvalues depend on the Higgs-like vector $\phi \in \CC^2$ used in definition of the left-invariant metric $g_\phi$. The explicit dependence of $\Omega^b_g$ on $\phi$ may be derived from the formulae in the first part of appendix B.

\subsubsection*{Components $c_\pm (x,h)$} 
\addcontentsline{toc}{subsubsection}{Components $c_\pm (x,h)$}

The components $c_+^P(x,h)$ of $\psi_+^P$ behave along $K$ according to the rule \eqref{VerticalTransformationSpinors2}, which may be written in simplified notation as
\beq
c_+^P (x,h) \ = \ \conj{s_\varphi(h)} \, h^\dag \ c \ ,  \nonumber
\eeq
where $s_\varphi : K \to \CC$ is the scalar function defined in \eqref{DefinitionSPhi}. It is implicit that the vector $c \in \CC^3$ may depend on the coordinate $x$ of $M_4$. Given vectors $u$ and $v$ in $\su$, the Lie derivatives of $c^P$ along left-invariant vector fields $u^\LL$ and $v^\LL$ on $K$ are
\beq  \label{LieDerivativesc}
\big(\Lie_{u^\LLL} \, \Lie_{v^\LLL}\, c^P \big) \, (h) \ = \ (\Lie_{u^\LLL} \, \Lie_{v^\LLL}\, \sbb_\varphi)\, h^\dag \, c \ - \ (\Lie_{v^\LLL}\, \sbb_\varphi)\, u\, h^\dag \, c \ -  \ (\Lie_{u^\LLL}\, \sbb_\varphi)\, v\, h^\dag \, c \ + \ \sbb_\varphi \, v \, u\, h^\dag\, c \ ,  \nonumber
\eeq
where $u$ and $v$ are matrices in $\su$. If we have two vectorial functions $c_1^P$ and $c_2^P$, it is possible to take the hermitian product
\begin{multline} \label{auxPairingsc}
(c_2^P)^\dag   \big( \Lie_{u^\LLL}\,  \Lie_{v^\LLL}  c_1^P \big) \ = \   s_{\varphi} \,  (\Lie_{u^\LLL} \, \Lie_{v^\LLL}\, \sbb_{\varphi})\, c_2^\dag \, c_1 \ - \ s_\varphi \, (\Lie_{v^\LLL}\, \sbb_\varphi)\, c_2^\dag \, h\, u\, h^\dag \, c_1 \ \linebr -   \ s_\varphi \, (\Lie_{u^\LLL}\, \sbb_\varphi)\, c_2^\dag \, h\, v\, h^\dag \,  c_1 \ + \ 
|s_\varphi|^2 \, c_2^\dag \, h\, v \, u\, h^\dag \,  c_1 \ . 
\end{multline} 
The integral identities presented in \eqref{A4Integral1} and \eqref{A4Int6} of appendix A.4 then say that, for any traceless $u$ and $v$ in $\su$,
\bal
\int_K \, s_\varphi \, (\Lie_{v^\LLL} \, \sbb_\varphi ) \, h\, v\, h^\dag \ \vol_K \ &= 0 \nonumber \linebr
\int_K |s_\varphi|^2 \, \, h\, v\, u\, h^\dag \ \vol_K \ &= \  \frac{1}{3} \, \Tr(vu) \, I_3  \,  \int_K |s_\varphi|^2 \ \vol_K \nonumber \ .
\end{align}
This means that two terms in the right-hand side of \eqref{auxPairingsc} are killed off by fibre-integration over $K$, with a third term having a relatively simple form. The additional integral identity
\beq
\int_K \, s_\varphi \, (\Lie_{u^\LLL} \, \Lie_{v^\LLL} \, \sbb_\varphi ) \ \vol_K \ = \ 2  \left[  u_{11}\, v_{11} \, + \,  \frac{1}{1+8\cos^2(\varphi)} \, (vu)_{11}  \right]  \int_K |s_\varphi|^2 \ \vol_K   \nonumber
\eeq
from the same appendix, then allows us to write the integral of equation \eqref{auxPairingsc} in the simplified form
\begin{multline} \label{auxPairingsc2}
\int_K\, (c_2^P)^\dag   \big( \Lie_{u^\LLL}\,  \Lie_{v^\LLL}  c_1^P \big) \ \vol_K \ = \   c_2^\dag \, \bigg\{    2  \left[  u_{11}\, v_{11} \, + \,  \frac{1}{1+8\cos^2(\varphi)} \, (vu)_{11}  \right] \\ +  \  \frac{1}{3} \, \Tr(vu)  \bigg\} \, c_1 \,  \int_K |s_\varphi|^2 \ \vol_K \ . \nonumber
\end{multline}
Coming back to expression \eqref{Laplacian} for the Laplacian, we finally get that
\beq  \label{IntegralPairingc}
\int_K \, (c_2^P)^\dag  \,  \big( \Delta_g^K  c_1^P \big)  \ \vol_K \ = \  \big( c_2^\dag \, \Omega^c_g \, c_1\big) \,  \int_K |s_\varphi|^2 \ \vol_K \ = \ \int_K \, (c_2^P)^\dag  \,  \big( \Omega_g^c \,  c_1^P \big)  \ \vol_K  \ ,
\eeq
where $\Omega^c_g$ is a $3\times3$ hermitian matrix proportional to the identity $I_3$. In terms of a $g$-orthonormal basis $\{ e_j \}$ of $\su$, it may be written as
\beq \label{DefinitionOmegaC}
\Omega^c_g \ :=\ \sum_{j=1}^8 \, \left\{  2  \left[  (e_j)_{11}^2 \, + \,  \frac{1}{1+8\cos^2(\varphi)} \, (e_j \, e_j)_{11}  \right]  \ +\   \frac{1}{3} \, \Tr(e_je_j)  \right\} I_3   \ .
\eeq
Expression \eqref{IntegralPairingc} says that, after pairing and fibre-integration, the vertical Laplacian $\Delta_g^K$ acts on the vectors $c \in \CC^3$ as simple scalar multiplication by the real coefficient of $\Omega^c_g$. The middle expression in \eqref{IntegralPairingc} is a Klein-Gordon mass term for the field $c(x)$ on $M_4$.

\subsubsection*{Components $a_\pm (x,h)$} 
\addcontentsline{toc}{subsubsection}{Components $a_\pm (x,h)$}

The scalar component $a_+^P(x,h)$ of $\psi_+^P$ behave along $K$ according to the simple rule \eqref{VerticalTransformationSpinors2},
\beq
a^P (x,h) \ = \ |s_\varphi (h)|^2  \ a(x) \ ,  \nonumber
\eeq
where $s_\varphi : K \to \CC$ is the function defined in \eqref{DefinitionSPhi} and $a(x)$ is a complex number. Therefore, using standard properties of the Laplacian and an orthonormal basis  $\{ e_j \}$ of $\su$,
\bal  \label{IntegralPairinga}
\int_K \, (a^P)^\dag  \,  \big( \Delta_g^K  a^P \big)  \, \vol_K \, &= \,  \big| a^P \big|^2   \int_K |s_\varphi|^2  \, \big(  \Delta_g^K |s_\varphi|^2  \big) \, \vol_K  \,  = \, - \big| a^P \big|^2   \int_K \,   \big|\, \grad  |s_\varphi|^2 \, \big|^2  \, \vol_K      \nonumber \linebr 
&= \ - \big| a^P \big|^2  \int_K \,  \sum_j \left( \Lie_{e_j}  |s_\varphi|^2 \right)^2  \ \vol_K \ .
\end{align}
We do not offer here a more explicit formula for the last integral.

\newpage

\vspace{.3cm}

\begin{appendices}

\section{Integrals on $\SU$ }

\subsection{Integrals of general polynomials in $h_{k1}$ and $\hb_{j1}$}
\label{A1}

Let $K$ be the Lie group $\SU$ equipped with a bi-invariant volume form $\vol_K$. Denote by $h_{kj}$ the entries of a matrix $h$ in $\SU$ and consider the function $P: \SU \to \CC$ determined by a general monomial in the entries of the first column of $h$:
\beq \label{DefinitionPolynomial}
P(h) \ := \ (h_{11})^{k_1} \,  (h_{21})^{k_2} \,  (h_{31})^{k_2} \,  (\hb_{11})^{n_1} \,  (\hb_{21})^{n_2} \,  (\hb_{31})^{n_3} \ ,
\eeq
where the exponents $k_j$ and $n_j$ are all non-negative integers. The aim of this appendix is to prove that
\beq \label{IntegralPolynomial}
I [P] \ := \ \int_{h \in K} \, P(h) \ \vol_K \ = \ \delta_{k_1 n_1} \, \delta_{k_2 n_2} \, \delta_{k_3 n_3} \, \frac{2\, k_1! \, k_2!\, k_3!}{(2+ k_1+k_2+k_3)!} \, \big(\Vol \, K \big) \ .
\eeq
Observe that, by the symmetry of the problem, this result is valid also for monomials in the entries of any column of $h \in \SU$, other than the first column $h_{k1}$, or monomials in the entries of any line of $h$.

\vspace{.3cm}

\noindent
{\it{First part of the calculation}}

\noindent
We start by showing that the integral \eqref{IntegralPolynomial} vanishes if $k_j \neq n_j$ for some $j$ in $\{1,2,3 \}$. To this end, notice that the bi-invariance of the volume form $\vol_K$ implies that the integral is invariant under a change of  variable of integration of the form $h \mapsto \eta \, h$ or $h \mapsto h\, \eta$, for any fixed group element $\eta \in K$. 

Choosing $\eta$ of the form $\diag(e^{-2i \theta}, e^{i\theta}, e^{i\theta})$, we have that $(h\, \eta)_{k1} = e^{-2i \theta}\, h_{k1}$, and hence
\beq
P(h\, \eta) \ = \ e^{-2i \theta ( k_1 + k_2+ k_3 - n_1 -n_2 -n_3 )} \, P(h) \ .  \nonumber
\eeq
The invariance under change of variable then says that, for any phase $\theta$,
\[
I [P] \ = \ \int_{h \in K} \, P(h\, \eta) \ \vol_K \ = \ e^{-2i \theta ( k_1 + k_2+ k_3 - n_1 -n_2 -n_3 )} \ I [P] \ .
\]
So if the sum $k_1 + k_2+ k_3 - n_1 -n_2 -n_3$ is not zero, then $I [P]$ necessarily vanishes.

Now assume that $k_1 + k_2+ k_3 = n_1 + n_2 + n_3$. After pairing all the factors $h_{k1}$ with their complex conjugates in $P(h)$, as much as possible, the initial monomial can be written in the form
\beq \label{ReducedPolynomial}
P(h) \ = \    \big| \, h_{11}^{m_1}\   h_{21}^{m_2}\ h_{31}^{m_3} \,  \big|^2  \  h_{11}^{j_1}\   h_{21}^{j_2}\  \hb_{31}^{\, j_1+j_2}    \  
\eeq
for non-negative integers $m_i$ and $j_i$, or else in a similar form related to the one above by complex conjugation or by permutation of the entries $\{h_{11}, h_{21}, h_{31} \}$. To simplify the discussion, we will now assume that $P(h)$ has the form \eqref{ReducedPolynomial}, but the arguments used below work equally well for the complex conjugate monomial or for the monomials related to \eqref{ReducedPolynomial} by permutations of the entries. So there is no loss of generality.

Multiplying $h$ on the left by the element $\eta =\diag(e^{-2i \theta}, e^{i\theta}, e^{i\theta})$, it is clear from \eqref{ReducedPolynomial} that the monomial transforms as
\beq
P(\eta \, h) \ = \ e^{i ( -2j_1 + j_2 - j_1 - j_2) \theta} \, P(h)  \ = \   e^{ - 3 j_1 \theta} \, P(h)       \ .  \nonumber
\eeq
As before, the invariance of the integral $I [P]$ under the change of variable $h \mapsto \eta \, h$ then implies that the integral must vanish whenever $j_1 \neq 0$. In the remaining case where $j_1 = 0$, the monomial \eqref{ReducedPolynomial} takes the simplified form
\beq \label{ReducedPolynomial2}
P(h) \ = \    \big|\, h_{11}^{m_1}\   h_{21}^{m_2}\ h_{31}^{m_3} \,  \big|^2  \  h_{21}^{j_2}\  \hb_{31}^{j_2}    \  .
\eeq
Then choose the element $\tilde{\eta} = \diag(e^{\theta}, e^{-2i\theta}, e^{i\theta})$ in $\SU$ and multiply $h$ by it on the left. It follows from \eqref{ReducedPolynomial2} that
\beq
P(\tilde{\eta} \, h) \ = \ e^{i ( -2 j_2 - j_2) \theta} \, P(h)  \ = \   e^{ - 3 j_2 \theta} \, P(h)       \ .  \nonumber
\eeq
So the invariance of $I [P]$ by this operation implies that the integral also vanishes whenever $j_2 \neq 0$. Therefore, the integral $I [P]$ can be non-zero only when the monomial $P(h)$ is a simple norm $\big| h_{11}^{m_1}\   h_{21}^{m_2}\ h_{31}^{m_3}   \big|^2$, and this concludes the proof of the first part of the result.

\vspace{.3cm}

\noindent
{\it{Second part of the calculation}}

\noindent
The second part of the proof of formula \eqref{IntegralPolynomial} is to compute the integrals
\[
I [k_1, k_2, k_3] \ := \ \int_{h \in K} \,  \big| \, h_{11}^{k_1}\   h_{21}^{k_2}\  h_{31}^{k_3} \,  \big|^2 \ \vol_K   \ .
\]
Let $\vol_{\rm{Haar}}$ be the Haar volume form on $K$, i.e. the unique bi-invariant volume form with total volume normalized to 1 \cite{BD}. Since all bi-invariant volume forms are proportional to each other,
\[
I [k_1, k_2, k_3] \ = \  (\Vol \, K ) \, \int_{h \in K} \,  \big| \, h_{11}^{k_1}\   h_{21}^{k_2}\  h_{31}^{k_3} \,  \big|^2 \ \vol_{\rm{Haar}}  \ .
\]
The monomial $P(h) = \big| \, h_{11}^{k_1}\   h_{21}^{k_2}\  h_{31}^{k_3} \,  \big|^2$ is invariant under right-multiplication of $h$ by group elements of the form $\iota(a) = \diag(\det a^{-1} , a)$, for any $a \in \Utwo$, since this multiplies the entries $h_{k1}$ by a phase only. Thus, $P(h)$ descends to the quotient $\SU / \Utwo \simeq \CC P^2$. In other words, there exists a complex function $p$ on $\CC P^2$ such that $P = p \circ \pi$, where $\pi: \SU \to \CC P^2$ is the projection associated to the quotient. A well-know result about fibre-integration (see Proposition 5.16 of \cite{BD}) then says that
\beq
 \int_{h \in K} \,  P(h) \ \vol_{\rm{Haar}}  \ = \ \int_{z \in \CC P^2} p(z) \ \vol_{\rm{FS}} \ ,
\eeq
where $\vol_{\rm{FS}}$ is the unique volume form on $\CC P^2$ that is invariant under the $\SU$-action on this space and has total volume equal to 1. In other words, $\vol_{\rm{FS}}$ is the normalized volume form of the Fubini-Study metric on $\CC P^2$.

Using homogeneous coordinates on projective space, the projection $\pi: \SU \to \CC P^2$ has the simple form
\[
\pi (h) \ = \ [\, h_{11}, \, h_{21},  \, h_{31}\, ] \ .
\]
It follows that the function $p(z)$ on $\CC P^2$ corresponding to the monomial $P$ can be represented in homogeneous coordinates as
\beq
p( [z_1, z_2, z_3]) \ = \ \frac{\big| \, z_1^{k_1}\, z_2^{k_2}\, z_3^{k_3} \, \big|^2}{ \big( \, |z_1|^2 +  |z_2|^2 + |z_3|^2\, \big)^{k_1 + k_2 + k_3}} \ . \nonumber
\eeq
Indeed, the entries of any matrix $h$ on $\SU$ satisfy the identity $\sum_k |h_{k1}|^2 = 1$, so it is clear that $p \circ \pi (h) = P(h)$.

Now take the standard chart of projective space defined by the map $(z_1, z_2) \mapsto [z_1, z_2, 1]$  from $\CC^2$ to $\CC P^2$. In this chart the function $p$ looks like
\beq
\hat{p}(z_1, z_2) =  p( [z_1, z_2, 1]) \ = \ \frac{\big| \, z_1^{k_1}\, z_2^{k_2} \, \big|^2}{ \big( \, |z_1|^2 +  |z_2|^2 + 1\, \big)^{k_1 + k_2 + k_3}} \ . \nonumber
\eeq
Moreover, in this standard chart the normalized Fubini-Study volume form is just
\[
\vol_{\rm{FS}} \ = \ \frac{2}{ \pi^2 \, \big( \, 1 + |z_1|^2 +  |z_2|^2\, \big)^3 } \ \dd x_1 \wedge \dd y_1 \wedge \dd x_2 \wedge \dd y_2 \ ,
\]
where we have written $z_k = x_k +i y_k$. Thus, the integral of $p$ in the chart's coordinates is
\bal
\int_{z \in \CC P^2} p(z) \ \vol_{\rm{FS}} \ &= \ \frac{2}{ \pi^2} \, \int_{\CC^2} \,  \frac{\big| \, z_1^{k_1}\, z_2^{k_2} \, \big|^2}{ \big( \, 1 + |z_1|^2 +  |z_2|^2\, \big)^{3+ k_1 + k_2 + k_3}}  \ \dd x_1 \wedge \dd y_1 \wedge \dd x_2 \wedge \dd y_2 \nonumber \linebr
&= \ \frac{2}{ \pi^2} \,(2\pi)^2 \, \int_0^{+\infty} \int_0^{+\infty} \,  \frac{r_1^{2k_1 +1}\, r_2^{2k_2 + 1} }{ \big( \, 1 + r_1^2 +  r_2^2\, \big)^{3+ k_1 + k_2 + k_3}}  \ \dd r_1 \, \dd r_2  \nonumber \ ,
\end{align}
where $r_k := |z_k|$. Changing the variables of integration from $(r_1, r_2)$ to $(r_1, \theta)$, with the new coordinate defined by $\tan \theta := r_2 / \sqrt{1 + r_1^2}$, the preceding integral becomes
\[
8 \int_0^{+\infty} \int_0^{\pi /2} \,  \frac{r_1^{2k_1 +1}}{ \big( \, 1 + r_1^2 \, \big)^{2+ k_1 + k_3}} \, (\sin \theta)^{2k_2 +1}\, (\cos \theta)^{2k_1 + 2k_3 +3}    \ \dd \theta \, \dd r_1 \ .
\]
Defining a second change of variable by $r_1 = \tan \varphi$, the integral becomes
\[
8 \int_0^{\pi /2} \int_0^{\pi /2} \,  (\sin \varphi)^{2k_1 +1}\, (\cos \varphi)^{2k_3 + 1 }    \, (\sin \theta)^{2k_2 +1}\, (\cos \theta)^{2k_1 + 2k_3 +3}    \ \dd \theta \, \dd \varphi \ .
\]
But these integrals are related to the well-known beta function, which satisfies
\[
B(n, m) \ = \ 2 \int_0^{\pi /2} (\sin \theta)^{2n -1}\, (\cos \theta)^{2m-1}    \ \dd \theta \ = \ \frac{(m-1)! \, (n-1)!}{(m+n-1)!} \ , 
\]
for any positive integers $m$ and $n$. Thus, our main integral in just
\[ 
\int_{z \in \CC P^2} p(z) \ \vol_{\rm{FS}} \ = \ 2\, B(k_1 +1, \, k_3 +1) \, B(k_2 +1, \, k_1 + k_3 +2) \ = \  \frac{ 2\, k_1! \, k_2!\, k_3!}{(2+ k_1+k_2+k_3)!} \ .
\]
This concludes the second part of the calculation.

\subsection{Integrals of fourth order monomials $h_{kj} \, h_{mn}\, \hb_{k'j'} \, \hb_{m'n'} $}
\label{A2}

This  part of the appendix lists a few explicit integrals over $\mathrm{SU}(N)$ than can be calculated using general formulae found in \cite{Creutz}. The integrals are calculated either with respect to the Haar volume form $\vol_{\rm{Haar}}$, which has total volume equal to one; or with respect to a general bi-invariant top form $\vol_K$, which will be proportional to $\vol_{\rm{Haar}}$ and have total volume $\Vol \, K$.

Formula (23) in \cite{Creutz} says that, for any indices $k_i$ and $l_i$ in $\{1, \ldots, N \}$,
\beq
\int_{h \in \mathrm{SU}(N)} \, h_{k_1 l_1} \, h_{k_2 l_2}^\dag \ \vol_{\rm{Haar}} \ = \ \frac{1}{N} \  \delta_{k_1 l_2} \, \delta_{l_1 k_2} \ .
\eeq
Using this formula one can calculate that, for any bi-invariant volume form $\vol_K$ and any square matrix $v \in M_N(\CC)$,
\bal \label{Baux1}
\int_{h \in K} \, h\, v\, h^T \ \vol_K \ &= \ 0    \linebr
\int_{h \in K} \, h\, v\, h^\dag  \ \vol_K \ &= \ \frac{1}{N} \,  \Tr(v) \,  (\Vol\, K)  \, I_N    \nonumber  \linebr
\int_{h \in K} \, h\, v\, \hb  \ \vol_K \ &= \ \frac{1}{N} \, (\Vol\, K)  \, v^T  \nonumber  \ .
\end{align}
The first formula in \eqref{Baux1} is valid only when $N>2$. By general invariance properties under the change of variable of integration, as described in section 2 of \cite{Baptista}, these integrals remains unchanged if we substitute $h$ by $\hb$, $h^T$ or $h^\dag$ in the integrand function.
Now consider the integral of the fourth order monomial
\beq
I (i_1 \,  j_1  \, i_2 \,  j_2; \, k_1  \, l_1 \,  k_2 \,  l_2) \ := \ \int_{h \in \mathrm{SU}(N)} \, h_{i_1 j_1} \, h_{i_2 j_2} \, h_{k_1 l_1}^\dag \, h_{k_2 l_2}^\dag \ \vol_{\rm{Haar}}  \ .
\eeq
Formulae (25) and (26) of \cite{Creutz} say that, for any values of the indices\footnote{Formula (25) of \cite{Creutz} is missing the factor $\delta_{j_1 k_1}$ that appears in the first term on the right-hand side of \eqref{CreutzResult2}. It is a typo and the factor should be there.},
\begin{multline} \label{CreutzResult2}
N\, (N^2 -1)\, I (i_1 \,  j_1  \, i_2 \,  j_2; \, k_1  \, l_1 \,  k_2 \,  l_2) \ = \ N\, \big(  \delta_{i_1 l_1} \, \delta_{i_2 l_2} \,   \delta_{j_1 k_1} \, \delta_{j_2 k_2} \, + \,  \delta_{i_1 l_2} \, \delta_{i_2 l_1} \,   \delta_{j_1 k_2} \, \delta_{j_2 k_1}    \big) \linebr
- \  \big(  \delta_{i_1 l_1} \, \delta_{i_2 l_2} \,   \delta_{j_1 k_2} \, \delta_{j_2 k_1} \, + \,  \delta_{i_1 l_2} \, \delta_{i_2 l_1} \,   \delta_{j_1 k_1} \, \delta_{j_2 k_2}    \big) \, .
\end{multline}
Using the preceding formula, after some index manipulation one can calculate that, for any bi-invariant volume form $\vol_K$ and any square matrix $v \in M_N(\CC)$,
\bal \label{IntegralIdentitiesB2.2}
\int_{h \in K} \, h\, h\, v\, h^\dag \,  h^\dag \ \vol_K \ &= \ \int_{h \in K} \, h\, \hb\, v\, h \,  \hb \ \vol_K  \linebr
&= \ \frac{\Vol\, K}{N\, (N^2 -1)}\, \left[ N\, v \, + \, (N^2-2)\, \Tr(v) \, I_N  \right]  \nonumber  \\[0.9em]
\int_{h \in K} \, h^T\, h\, v\, (h^T\,  h)^\dag \ \vol_K \  &= \ \frac{\Vol\, K}{N+1}\, \left[ v^T \, + \, \Tr(v) \, I_N  \right]  \ .
\end{align}
The value of the integrals remains unchanged if we substitute $h$ by $\hb$, $h^T$ or $h^\dag$ in the integrand function.

\subsection{Integrals of functions of the form $q(h) \, h\, v \, h^\dag$}
\label{A3}

The aim of this appendix is to establish two identities that are useful in the calculations of the fibre-integrals studied in sections 2 and 3. They are identities for the integrals on $K$ of functions of the form $q(h) \, h\, v \, h^\dag$, where $q(h)$ is a scalar function on the group having certain invariance properties.

\vspace{-0.2cm}

\subsubsection*{Proposition.}

Let $K$ be the Lie group $\mathrm{SU} (N)$ equipped with a bi-invariant volume form $\vol_K$. Let $v$ be any square matrix on $M_N(\CC)$. Let $q: K \to \CC$ be a scalar function  invariant under left-multiplication on $K$ by the following group elements: 
\begin{enumerate}
\item The element $\theta_1 = \diag(-1, -1, 1, \ldots, 1)$ and its conjugates $\theta_k$ obtained by permuting the diagonal entries.
\item The block-diagonal element $\eta_1 = \diag (\eta, -1, 1, \ldots, 1 )$ and its conjugates $\eta_j$ obtained by permuting the diagonal blocks.
\end{enumerate}
Here we have denoted
\beq 
\eta \ := \ \bmatr  0 & 1  \\ 1 & 0  \ematr    \ . \nonumber
\eeq
Then the function $q(h)$ satisfies the following integral identities:
\beq \label{IntegralIdentitiesB3.1}
\int_{h\in K} \, q(h)\, h \, v\, h^\dag \ \vol_K \ = \ \int_{h\in K} \, q(h)\, h^\dag \, v\, h \ \vol_K \ = \ \frac{1}{N}\, \Tr(v) \, I_N \, \int_{h\in K} q(h)\, \vol_K \ .
\eeq
In particular, the first two integrals vanish when $v$ is traceless.

\vspace{.3cm}

\noindent
{\it{Proof.}}

\noindent
Consider the matrix-valued integral
\[
I_q (v) \ := \ \int_{h\in K} \, q(h)\, h \, v\, h^\dag \ \vol_K  \ .
\]
Since the volume form $\vol_K$ is bi-invariant, the integral is invariant under a change of variable of integration $h \mapsto h' h$ for any fixed element $h'$ in $\mathrm{SU} (N)$. If furthermore $q(h) = q(h' h)$ for the chosen element, then the integral satisfies the identity
\beq  \label{IdentityI1}
I_q (v) \ = \ h' \  I_q (v)\ (h')^\dag
\eeq
in the space of $N \times N$ complex matrices. Choosing $h'$ among the diagonal group elements $\theta_k$, it is clear that \eqref{IdentityI1} can be true for all $\theta_k$ only if $I_q (v)$ is a diagonal matrix.  Choosing $h'$ among the matrices $\eta_j$, identity \eqref{IdentityI1} further implies that the diagonal components of $I_q (v)$ must be equal to each other, that is, $I_q (v)$ must be proportional to the identity matrix $I_N$. But in this case we have that
\[
I_q (v)  =  \frac{1}{N}\, \Tr\big[I_q (v) \big] \, I_N  =  \frac{1}{N}\, I_N \int_{h\in K}  q(h)\,  \Tr( h \, v\, h^\dag) \, \vol_K \, = \,  \frac{1}{N}\, \Tr(v) \, I_N  \int_{h\in K}  q(h) \, \vol_K  ,
\]
as required. Now consider the second matrix-valued integral
\[
J_q (v) \ := \ \int_{h\in K} \, q(h)\, h^\dag \, v\, h \ \vol_K  \ .
\]
Using the bi-invariance of $\vol_K$, the same argument as before says that if $q(h) = q(h' h)$ for a fixed group element $h'$ and all $h$ in $\mathrm{SU} (N)$, then
\beq  \label{IdentityJ1}
J_q (v) \ = \  J_q \big[ \, (h')^\dag \, v \, h' \,\big] \ 
\eeq
in the space of $N \times N$ complex matrices. Since the integral identities \eqref{IntegralIdentitiesB3.1} are linear in the matrix $v$, it is enough to prove them for matrices with a single non-zero component. If the non-zero component of $v$ is off the main diagonal, then by choosing $h'$ to be the appropriate element $\theta_k$, one can achieve that $ (h')^\dag \, v \, h'   = -v $, and so identity \eqref{IdentityJ1} implies that $J_q (v)$ vanishes in this case, in agreement with \eqref{IntegralIdentitiesB3.1}. If the non-zero component of $v$ is in the main diagonal, then by choosing $h'$ to be the appropriate element among the $\eta_j$ or their products, one can achieve that $ (h')^\dag \, v \, h' $ has the same non-zero component as $v$ but in a different, arbitrary position within the main diagonal. Thus,
\bal
J_q \big[\, \diag(\alpha_1, \ldots, \alpha_N) \,\big] \ &= \ J_q \big[\,\diag(\alpha_1, 0, \dots, 0)\,\big] \,+ \  \cdots \ + \, J_q \big[\,\diag(0, \ldots,  0, \alpha_N)\,\big]  \nonumber \linebr
&= \ (\alpha_1 + \cdots + \alpha_N)\ J_q \big[\,\diag(1 , 0, \ldots,  0)\,\big] \ = \ \Tr(v) \ \frac{1}{N}\ J_q(I_N)  \nonumber  \linebr
&= \ \frac{1}{N} \ \Tr(v) \ \int_{h\in K} q(h)\, \vol_K \ ,
\end{align}
as required.

\subsection{Integrals involving the functions $s(h)$}
\label{A4}

Let $K$ be the Lie group $\SU$ equipped with a bi-invariant volume form $\vol_K$. Denote by $h_{kj}$ the entries of a matrix $h \in K$. In this appendix we will present the values of several integrals over $\SU$ involving the scalar functions
\bal
s_1 (h) \ &:= \   h_{11}^2\, + \, h_{21}^2\, +  h_{31}^2   \linebr
s_2 (h) \ &:= \   h_{11} h_{21}\, + \, h_{11} h_{31}\, +  h_{21} h_{31}    \nonumber   \linebr
s(h) \ &:= \  \alpha_1 \, s_1(h) \, +\, \alpha_2 \, s_2(h)    \nonumber \ , 
\end{align}
where $\alpha_1$ and $\alpha_2$ are any complex constants.

\subsubsection*{Integrals of $|s|^2$ and $|s|^4$}

The integrals of the functions $|s_a|^2$ and $|s_a|^4$ over $(K, \vol_K)$ can be calculated from the general formula \eqref{IntegralPolynomial} for the integrals of complex polynomials in the entries $h_{k1}$. The results are
\bal     \label{A4I1}
\int_K \, |s_1|^2  \ \vol_K \ &=  \ 2 \, \int_K \, |s_2|^2  \ \vol_K \ = \ \frac{1}{2} \, (\Vol \, K)  \linebr
\int_K \, s_1 \, \sbb_2  \ \vol_K \ &= \ 0     \nonumber    
\end{align}
for integrals quadratic in the functions $s_a$. For integrals involving four-fold products of those functions we have
\bal    \label{A4I2}
\int_K \, |s_1|^4  \ \vol_K \ &= \ \frac{1}{3} \, (\Vol \, K)  \linebr
\int_K \, |s_2|^4  \ \vol_K \ &= \ \int_K \, |s_1\, s_2|^2  \ \vol_K \ = \  \frac{1}{10} \, (\Vol \, K)      \nonumber \linebr
\int_K \, |s_2|^2 \, \sbb_1\, s_2  \ \vol_K \ &= \  \frac{1}{30} \, (\Vol \, K)     \nonumber   \linebr
\int_K \, |s_1|^2 \, \sbb_1\, s_2  \ \vol_K \ &= \  0     \nonumber \ .
\end{align}
From these results, one obtains that any linear combination $s = \alpha_1 s_1  + \alpha_2  s_2$ satisfies
\bal    \label{A4I3}
\int_K \, |s|^2  \ \vol_K \ &= \ \frac{1}{4} \,  \left( \, 2 \, |\alpha_1|^2 \, + \,    |\alpha_2|^2 \, \right)  (\Vol \, K)  \linebr
\int_K \, |s|^4  \ \vol_K \ &= \ \frac{1}{30}\,  \Big[ \, 10 \,|\alpha_1|^4 +   3 \, |\alpha_2|^4 + 18 \, |\alpha_1 \alpha_2 |^2 + 2 \,  |\alpha_2|^2 ( \alpha_1 \conj{\alpha_2} +  \conj{\alpha_1} \alpha_2 ) \, \Big] \, (\Vol \, K)  \ . \nonumber
\end{align}

\subsubsection*{Integrals of the form $ |s|^2  \, h\, v\, h^\dag$}

For any square matrix $v \in M_3 (\CC)$ one can show that
\bal  \label{A4Integral1}
\int_K \, |s|^2  \, h\, v\, h^\dag \ \vol_K \ &= \  B_1 \ \int_K \,  \, |s|^2  \ \vol_K   \linebr
\int_K \, |s|^2  \, h^\dag \, v\, h \ \vol_K \ &= \  B_2 \ \int_K \,  \, |s|^2  \ \vol_K   \ , \nonumber
\end{align}
where the matrices $B_1$ and $B_2$ are defined by
\bal   \label{A4Integral2}
B_1 \, &:= \,  \frac{1}{3}\, \Tr(v) \, I_3 \,+ \, \frac{1}{30} \,  (2\, v_{11} - v_{22}  - v_{33}  )   \, \frac{|\alpha_2|^2 \, + \,  2\, ( \alpha_1 \conj{\alpha_2} \, + \, \conj{\alpha_1} \alpha_2 )}{2 \, |\alpha_1|^2 \, + \,    |\alpha_2|^2} \,  \bmatr  0 & 1 & 1 \\ 1 & 0 &  1 \\ 1 & 1 & 0 \ematr    \nonumber  \\[0.4em]
B_2 \, &:= \,  \frac{1}{3}\, \Tr(v) \, I_3 \, + \, \frac{1}{30} \ \frac{|\alpha_2|^2 \, + \,  2\, ( \alpha_1 \conj{\alpha_2}  +  \conj{\alpha_1} \alpha_2 )}{2 \, |\alpha_1|^2 \, + \,    |\alpha_2|^2} \,  \left(  \sum_{j \neq k }\, v_{jk}     \right)   \diag(-2,1,1)   \ .
\end{align}
Observe how $B_2$ is a diagonal matrix, unlike $B_1$. Both these matrices are much simpler when the coefficients $\alpha_a$ are such that $|\alpha_2|^2  +   2\, ( \alpha_1 \conj{\alpha_2}  +  \conj{\alpha_1} \alpha_2 )$ vanishes, i.e. when the function $s(h)$ coincides with one of the functions $s_\varphi (h)$ defined in \eqref{DefinitionSPhi}. They are also simpler when $v$ is a traceless matrix.

These formulae are manifestly true when $v$ is proportional to the identity matrix $I_3$, for in this case $h\, v\, h^\dag = h^\dag \, v\, h = v$ and also $B_1 = B_2 = v$. Since any square matrix in $ M_3 (\CC)$ can be decomposed as sum $v =  \lambda I_3 + w + i u$, where $\lambda$ is a complex number and $u$ and $w$ are matrices in $\su$, it is clear that the formulae will be true in general if they hold for matrices in $\su$. The calculation for the case $v \in \su$ takes several pages. It can be done considering separately the integrals of the functions $s_1(h)$ and $s_2(h)$ and using the results in appendices A.1 and A.3. It is also useful to consider separately the case where $v$ belongs to the subspaces $\iota(\sutwo)$ or $\iota(\CC^2)$ of $\su$, since in this case many of the integrals involved can be shown to vanish by symmetry arguments. The calculation in the case where $v$ is proportional to $\diag(-2i, i, i)$, on the other hand, uses the results of appendix A.1. A relevant intermediate step is the calculation that
\bal   
\int_K \, \Real \big(   \alpha_1\, s_1 \, \conj{\alpha}_2 \, \sbb_2 \big)  \, h^\dag \, v\, h \ \vol_K \ &= \  \frac{1}{120} \, (\Vol\, K) \left(  \sum_{j \neq k }\, v_{jk} \right)  ( \alpha_1 \conj{\alpha_2}  +  \conj{\alpha_1} \alpha_2 ) \  \diag(-2,1,1)      \nonumber  \\[0.4em]
\int_K \Real \big(   \alpha_1\, s_1 \, \conj{\alpha}_2 \, \sbb_2 \big)  \, h\, v\, h^\dag \, \vol_K \, &= \, \frac{1}{120}  (\Vol\, K)  (2 v_{11} - v_{22}  - v_{33}  )    ( \alpha_1 \conj{\alpha_2}  +  \conj{\alpha_1} \alpha_2 ) \! \bmatr  0 & 1 & 1 \\ 1 & 0 &  1 \\ 1 & 1 & 0 \ematr   \nonumber   
\end{align}
These two integrals hold also in the general case $v \in M_3 (\CC)$.

\subsubsection*{Integrals involving Lie derivatives of $s(h)$}

Here we list the values of a group of integrals involving the Lie derivatives $\Lie_{v^\LLL} $ and $\Lie_{v^\RRR}$ of the functions $s_a(h)$ and $s(h)$ along left and right-invariant vector fields on $K = \SU$. These functions depend on the entries $h_{k1}$ of the matrix $h \in \SU$, so their Lie derivatives depend on the basic derivatives
\bal \label{A4Int1}
\Lie_{v^\LLL} \, \big( h_{kj} \big) \ &= \   \frac{\dd}{\dd t} \, \big[ h\, \exp(tv) \big]_{kj} \ |_{t=0} \ = \ (h\, v)_{kj}    \linebr
\Lie_{v^\RRR} \, \big( h_{kj} \big) \ &= \   \frac{\dd}{\dd t} \, \big[\exp(tv)\, h \big]_{kj} \ |_{t=0} \ = \ (v\, h)_{kj}  \nonumber \ ,
\end{align}
where $v$ is any matrix in $\su$. The calculation of the integrals involving the Lie derivatives of $s(h)$ makes extensive use of formula \eqref{IntegralPolynomial} for the integrals of complex polynomials in the entries of any column or row of $h \in \SU$.

For indices $a, b \in  \{ 1,2 \}$, the basic integrals involving Lie derivatives along left-invariant vector fields are
\beq \label{A4Int2}
\int_K \,  \sbb_a\, \big(\Lie_{v^\LLL} s_b \big) \ \vol_K \ = \ 2\, \delta_{ab} \, v_{11}\, \int_K \,  \, |s_a|^2  \ \vol_K  \ ,
\eeq
where no sum is intended over the repeated index $a$. The analogous integrals for derivatives along right-invariant vector fields are
\bal \label{A4Int3}
 \int_K \,  \sbb_1\, \big(\Lie_{v^\RRR} s_1 \big) \ \vol_K \ &=  \ 0  \linebr
 \int_K \,  \sbb_2\, \big(\Lie_{v^\RRR} s_2 \big) \ \vol_K \ &= \ \frac{1}{3}\, \sum_{j,\, k }\, v_{jk}  \, \int_K \, |s_2|^2 \ \vol_K  \nonumber  \linebr
 \int_K \,  \sbb_1\, \big(\Lie_{v^\RRR} s_2 \big) \ \vol_K \ &= \ \int_K \,  \sbb_2\, \big(\Lie_{v^\RRR} s_1 \big) \ \vol_K \ =\ \frac{1}{6}\, (\Vol\, K) \, \sum_{j,\, k }\, v_{jk}  \ .  \nonumber
\end{align}
Notice that for any matrix in $\su$ the sum of its entries is an imaginary number,
\beq   \label{A4Int4}
\sum_{j,\, k }\, v_{jk}  \ = \ \sum_{j \neq k}\, v_{jk}  \ = \ \sum_{j < k}\, v_{jk} - \conj{v_{jk}}  \ = \ 2\, i\, \Imaginary \big(v_{12} + v_{13} + v_{23} \big) \ .
\eeq
Combining the preceding integrals for the functions $s_1$ and $s_2$, one gets that linear combinations $s = \alpha_1 s_1  + \alpha_2  s_2$ satisfy
\bal   \label{A4Int5}
 \int_K \,  \sbb \, \big(\Lie_{v^\LLL} s \big) \ \vol_K \ &=  \ 2\, v_{11}\, \int_K \,  \, |s|^2  \ \vol_K  \ \linebr
 \int_K \,  \sbb \, \big(\Lie_{v^\RRR} s \big) \ \vol_K \ &=  \ \frac{ |\alpha_2|^2 \, + \,  2\, ( \alpha_1 \conj{\alpha_2} \, + \, \conj{\alpha_1} \alpha_2 )}{3\, \big(2 \, |\alpha_1|^2 \, + \,    |\alpha_2|^2 \big)} \ \sum_{j, \, k}\, v_{jk}  \ \int_K \, |s|^2 \ \vol_K  \ . \nonumber 
\end{align}
The integral in the second line vanishes for all right-invariant fields $v^\RRR$ whenever the coefficients $\alpha_a$ are such that $|\alpha_2|^2  +   2\, ( \alpha_1 \conj{\alpha_2}  +  \conj{\alpha_1} \alpha_2 )$ equals zero, i.e. when the function $s(h)$ coincides with one of the functions $s_\varphi (h)$ defined in \eqref{DefinitionSPhi}.

The scalar Laplacian on $K$ equipped with a left-invariant metric can be written as $\Delta s = \sum_k \,\Lie_{e_k^\LLL} \, \Lie_{e_k^\LLL}\,  s$, where $\{e_k \}$ is an orthonormal basis of the Lie algebra. The calculations of section 4.3 deal with those Laplacians and the following integral identities are used there. Firstly, for any $v \in \su$ and any square matrix $u \in M_3 (\CC)$, we have
\bal   \label{A4Int6}
\int_K \,  s_a\, \big(\Lie_{v^\LLL} \sbb_b \big) \, h\, u\, h^\dag \ \vol_K \ &= \ - \frac{2}{3} \, \delta_{ab} \, v_{11} \, \Tr(u) \,  \int_K \,  \, |s_a|^2  \ \vol_K  \linebr
\int_K \,  s\, \big(\Lie_{v^\LLL} \sbb \big) \, h\, u\, h^\dag \ \vol_K \ &= \ - \frac{2}{3} \, v_{11} \, \Tr(u) \,  \int_K \,  \, |s|^2  \ \vol_K  \ . \nonumber
\end{align}
Secondly, for any pair of matrices $u$, $v$ in $\su$, we have that
\bal   \label{A4Int7}
 \int_K \,  \sbb_1\, \big(\Lie_{u^\LLL} \Lie_{v^\LLL} s_2 \big) \ \vol_K \ &=  \ \int_K \,  \sbb_2\, \big(\Lie_{u^\LLL} \Lie_{v^\LLL} s_1 \big) \ \vol_K \ = \  0  \linebr
  \int_K \,  \sbb \, \big(\Lie_{u^\LLL} \Lie_{v^\LLL} s \big) \ \vol_K \ &=  \  \left[   2\, u_{11} \, v_{11} \, +\, \frac{4\, |\alpha_1|^2}{2 \, |\alpha_1|^2 \, + \,    |\alpha_2|^2 } \, (uv)_{11}   \right]    \int_K \, |s|^2 \ \vol_K \  .   \nonumber
\end{align}
A final group of integral identities involving the derivatives of the functions $s_a$ are
\bal   \label{A4Int8}
\int_K \, |s_1|^2 \, \sbb_2\, \big(\Lie_{v^\RRR} s_2 \big) \ \vol_K \ &= \  \frac{7}{180} \,  (\Vol \, K) \, \sum_{j,\, k }\, v_{jk}     \ = \ - \int_K \, |s_1|^2 \, s_2\, \big(\Lie_{v^\RRR} \sbb_2 \big) \ \vol_K  \nonumber   \linebr
\int_K \, |s_2|^2 \, \sbb_1\, \big(\Lie_{v^\RRR} s_1 \big) \ \vol_K \ &= \  \frac{1}{45} \,  (\Vol \, K) \, \sum_{j,\, k }\, v_{jk}      \ = \ - \int_K \, |s_2|^2 \, s_1\, \big(\Lie_{v^\RRR} \sbb_1 \big) \ \vol_K \ , 
\end{align}
for any matrix $v \in \su$.

\section{Laplacian of equivariant functions on $\SU$}

Let $\rho: \SU \times V \to V$ be a linear group representation and let $\dd \rho: \su  \times V \to V$ be the induced Lie algebra representation. For a fixed inner-product  $g$ on $\su$, pick an orthonormal basis $\{ e_j \}$ of this space and define an endomorphism $\Omega_g : V \to V$ by
\beq \label{DefinitionOmegaSmall}
\Omega_{g} \ := \ \sum_{j=1}^8 \, (\dd \rho)_{e_j} \,  (\dd \rho)_{e_j} \ .
\eeq
It is easy to check that $\omega_{g}$ does not depend on the choice of orthonormal basis. This operator on $V$ is the Casimir element associated to the product $g$ and the representation $\dd \rho$ of $\su$.  It emerges naturally when studying the action of the scalar Laplacian $\Delta^K_g$ on functions $\SU \to V$ that are equivariant with respect to the representation. Such functions are determined by their value at the identity element, since $f(h) = \rho_h \big[  f(I_3) \big]$. Their Lie derivatives with respect to left-invariant vector fields are simply
\bal
(\Lie_{v^\LLL} f) (h) \, &= \, \frac{\dd}{\dd t} \ f \big( h \, \exp(t v) \big) \ |_{t=0} \, =   \frac{\dd}{\dd t} \ \rho_h \circ  \rho_{\exp(t v)} \,  \big[f(I_3)\big] \ |_{t=0} \, = \, \rho_h \circ (\dd \rho)_v\, \big[f(I_3) \big]   \nonumber \linebr
(\Lie_{u^\LLL} \, \Lie_{v^\LLL}  f) (h) \, &=  \, \rho_h \circ (\dd \rho)_u \circ (\dd \rho)_v \, \big[f(I_3)\big] \nonumber \ .
\end{align}
If the product $g$ on $\su$ is extended to a left-invariant metric on the whole group $K= \SU$, then it follows from the general formula \eqref{Laplacian} for the scalar Laplacian that
\[
(\Delta^K_g  f) (h) \ =  \ \sum_j \ \rho_h \circ (\dd \rho)_{e_j} \circ (\dd \rho)_{e_j} \, \big[f(I_3)\big] \ = \ \rho_h \circ \Omega_g\, \big[f(I_3)\big] \ .
\]
This means that $\Delta^K_g  f$ is also $\rho$-equivariant on $\SU$ and that its value at the identity is just $\Omega_g\, \big[f(I_3)\big]$. In particular, if the vector $f(I_3)$ is an eigenvalue of the algebraic operator $\Omega_g$, then the function $f$ is an eigenfunction of $\Delta^K_g$ with the same eigenvalue.

\subsubsection*{Fundamental representation}

In the case where $\rho: \SU \times \CC^3 \to \CC^3$ is the fundamental representation, then
\beq \label{DefinitionOmegaSmall2}
\Omega_{g} \ = \ \sum_{j=1}^8 \, e_j \, e_j  \ 
\eeq
is the $3\times3$ hermitian matrix that appears in section 3.3. More precisely, it appears in formulae \eqref{DefinitionOmegaD}, \eqref{DefinitionOmegaB} and \eqref{DefinitionOmegaC} that define the matrices $\Omega^D_g$, $\Omega^b_g$ and $\Omega^c_g$, respectively. The latter matrices determine the action of the internal Laplacian $\Delta^K_g$ on the components of the higher-dimensional spinors $\Psi^P (x,h)$, after pairing and fibre-integration.

When the inner-product on $\su$ is the one defined in section 2 of \cite{Baptista}, denoted by $g_\phi$, the explicit orthonormal basis $\{ v_0, \ldots , v_3, \, w_1, \ldots, w_4  \}$ constructed there can be used to calculate that
\[
\Omega_{g_\phi} \ = \ \frac{-1}{3\,\lambda\, (1 - \mph^2)\, (1 - 4 \mph^2)} \, \bmatr   8 - 25 \mph^2 + 8 \mph^4   &     3\, (1 -4 \mph^2)\, \phi^\dag   \linebr 
 3\, (1 -4 \mph^2)\, \phi       &         (8 - 34 \mph^2 + 8 \mph^4) I_2   +  9\, \phi \phi^\dag        \ematr  \ . 
\]
This matrix is manifestly hermitian and only depends on the parameters $\lambda$ and $|\phi|^2$ that appear in the definition of $g_\phi$. In the special case where the deformation parameter vanishes, $\phi = 0$, the product $g_\phi$ reduces to the usual Ad-invariant product $\beta$ on $\su$ and the matrix $\Omega_{g_\phi}$ is proportional to the identity matrix,
\[
\Omega_{\beta} \ = \ - \, \frac{8}{3\,\lambda} \ I_3 \ .
\]
When $\phi \neq 0$, one can calculate that the matrix $\Omega_{g_\phi}$ has the following three independent eigenvectors in $\CC^3$:
\[
u_1 \ = \ \bmatr  0 \\ \conj{\phi_2}  \\  - \conj{\phi_1} \ematr    \qquad  \qquad  u_\pm \ = \ \bmatr  \pm \mph \\ \phi_1  \\  \phi_2 \ematr \ .
\]
Their respective eigenvalues are found to be
\bal
\mu_1 \ &= \ - \ \frac{8\, - \,34 \mph^2 \,+ \,8 \mph^4}{3\,\lambda\, (1 - \mph^2)\, (1 - 4 \mph^2)}   \\[0.6em]
\mu_\pm \ &= \ -\ \frac{8 \, - \, 25 \mph^2 \, +\, 8 \mph^4 \, \pm \,  3\, \mph \, (1-4 \mph^2)}{3\,\lambda\, (1 - \mph^2)\, (1 - 4 \mph^2)} \nonumber \ .
\end{align}
They reduce to the unique value $- 8 / (3\,\lambda)$ when the parameter $\phi$ vanishes, of course. The formulae of section 3.3 also include the scalars $\Tr(\Omega_{g})$ and $(\Omega_{g})_{11}$. When $g = g_\phi$, these can be easily read from the explicit formula of $\Omega_{g_\phi}$ given above. Other related scalars and matrices that appear in the same formulae of section 3.3 are
\bal
\sum_{j=1}^8 \, (e_j)_{11}^2  \ &= \  - \, \frac{2\, (1 - \mph^2)}{3\,\lambda\, (1 - 4 \mph^2)}   \linebr
\sum_{j=1}^8 \, (e_j)_{11} \, e_j \ &= \ \frac{1}{3\,\lambda\, (1 - 4 \mph^2)}  \bmatr  -2\, (1 - \mph^2)  &  3\, \phi^\dag  \linebr
3\, \phi    &    6 \, \phi \phi^\dag  +  (1 - 4 \mph^2) \, I_2 \ematr  \nonumber \ .
\end{align}
Both these identities were calculated using the explicit $g_\phi$-orthonormal basis of $\su$ described in \cite{Baptista}. Although the actual calculation of $\Omega_{g_\phi}$ is not presented here, we will finish this section by writing down some of the intermediary expressions that lead to the formula below \eqref{DefinitionOmegaSmall2}. Using the basis $\{ v_0, \ldots , v_3, \, w_1, \ldots, w_4  \}$ mentioned above, we get
\bal
\sum_{a=1}^3 \, v_a \, v_a \ &= \  \frac{-1}{2\,\lambda\, (1 - \mph^2)}   \bmatr   3 \, \mph^2      &     3 \, \phi^\dag   \nonumber    \linebr
3\, \phi    & (3 + 2 \mph^2) \, I_2  -   \phi \phi^\dag  \ematr   \linebr
\sum_{k=1}^4 \, w_k \, w_k \ &= \ \frac{-1}{ \lambda} \bmatr  2 & 0  \linebr  0  &  I_2 \ematr \ ,
\end{align}
and the longer expression
\[
v_0 \, v_0 \ = \ \frac{-1}{6\,\lambda\, (1 - \mph^2)\, (1 - 4 \mph^2)} \, \bmatr   4 + \mph^2 + 4 \mph^4   &    - 3\, (1 -4 \mph^2)\, \phi^\dag   \linebr 
 -3\, (1 -4 \mph^2)\, \phi       &         (1 - 4 \mph^2)^2\, I_2   +  3\, (7 -4 \mph^2)\, \phi \phi^\dag        \ematr  .
\]

\subsubsection*{Adjoint representation}

In the case where $\rho: \SU \times \su \to \su$ is the adjoint representation, the Casimir element associated to $g$ is the following endomorphism of $\su$:
\[
\Omega_g \ := \  \sum_{j=1}^8 \, \ad_{e_j} \, \ad_{e_j} \ = \  \sum_{j=1}^8 \, \big[ e_j, \, [e_j,\, \cdot \, ] \big]  \ .
\]
When the inner-product on $\su$ is the one defined in section 2 of \cite{Baptista}, denoted by $g_\phi$, an explicit calculation using the orthonormal basis of $\su$ defined there shows that
\[
\Omega_{g_\phi} \big[ \iota(\phi) \big] \ = \ -\, \frac{6 \, (1 - 2 \,\mph^2)}{\lambda\, (1 - \mph^2)\, (1 - 4\, \mph^2)}  \ \iota(\phi) \ ,
\]
where $\phi \in \CC^2$ is the deformation parameter of $g_\phi$ and $\iota: \utwo \oplus \CC^2 \to \su$ is the vector space isomorphism used in \cite{Baptista}. This exhibits the first eigenvector and eigenvalue of $\Omega_{g_\phi}$ for deformed metrics with $\phi \neq 0$. Four more eigenvectors can be written down using the matrices $w^1_\phi$  and $w^2_\phi$ defined in the first appendix of \cite{Baptista}. These matrices belong to the subspace $\iota[\sutwo]$ of $\su$ and an explicit calculation shows that
\bal
\Omega_{g_\phi} \Big(w^1_\phi \ \pm \ |\phi|^{-1}\, \big[ w^1_\phi,\, \iota(\phi) \big] \Big) \ &= \ - \, \mu_\pm \ \Big(w^1_\phi \ \pm \ |\phi|^{-1}\, \big[w^1_\phi,\, \iota(\phi) \big] \Big) \linebr
\Omega_{g_\phi} \Big(w^2_\phi \ \pm \ |\phi|^{-1}\, \big[ w^2_\phi,\, \iota(\phi) \big] \Big) \ &= \ - \, \mu_\pm \ \Big(w^2_\phi \ \pm \ |\phi|^{-1}\, \big[w^2_\phi,\, \iota(\phi) \big] \Big)  \nonumber \ ,
\end{align}
where the two degenerate eigenvalues $\mu_\pm$ are the real numbers
\beq
\mu_\pm \ := \ \frac{6 \, - \,  23\, \mph^2 \, + \, 8 \, \mph^4 \, \pm \, |\phi| \, ( 1 - 4\, \mph^2)}{\lambda\, (1 - \mph^2)\, (1 - 4\, \mph^2)} \ .
\eeq
We have exhibited five eigenvectors of $\Omega_{g_\phi}$ so far. Since this operator is self-adjoint with respect the Ad-invariant product $\beta$ on $\su$, we know that the three remaining eigenvectors will lie in the $\beta$-orthogonal complement to the subspace generated by the first five eigenvectors. This  $\beta$-orthogonal complement is spanned by the three matrices:
\[
e_1\ = \ \frac{1}{\sqrt{3}}\, \iota(iI_2) \qquad \qquad e_2\ = \ \mph^{-1} \, \iota(i\phi) \qquad \qquad e_3 \ = \ \iota\big[  2i\, |\phi|^{-2}\, \phi \phi^\dag - i\, I_2    \big] \ .
\]
The span of these matrices is necessarily an invariant subspace of $\su$ under the action of $\Omega_{g_\phi}$. In fact, in the basis $\{e_1, e_2, e_3 \}$ the operator can be identified with the $3\times 3$ matrix
\beq
\Omega_{g_\phi} \ = \  \frac{1}{\lambda\, (1 - \mph^2)\, (1 - 4\, \mph^2)} \, \hat{\Omega}_{g_\phi} \ ,
\eeq
where $\hat{\Omega}_{g_\phi} $ stands for the symmetric matrix
\beq
\hat{\Omega}_{g_\phi} \ := \ 
\bmatr    
3\, (-2 + 7 \mph^2 - 2 \mph^4)     &     \   3 \sqrt{3}\,  \mph \,(1 -2 \mph^2)  &  \ - \sqrt{3}\, \mph^2 \,(1+ 2 \mph^2)    \linebr
 3 \sqrt{3} \, \mph\, (1 -2 \mph^2)  & \ -2 \,(3 - 10 \mph^2 + 4 \mph^4 )    & \  \mph\, (10 \mph^2 -1)     \linebr
- \sqrt{3} \, \mph^2 \,(1+ 2 \mph^2)   & \  \mph\, (10 \mph^2 -1)     & - 6 + 23 \mph^2 - 2 \mph^4
 \ematr  \nonumber \ .
\eeq
We will not be able to offer an explicit expression for the eigenvalues and eigenvectors of $\hat{\Omega}_{g_\phi}$ as a function of $|\phi|^2$.

\newpage

\section{HDR to special relativity}

\subsubsection*{Null geodesics on the higher-dimensional spacetime}

In this secluded appendix we discuss a kinematic model for particles moving in the higher-dimensional spacetime $P = M_4 \times K$ equipped with a product metric $g_P = g_M \times g_K$. The metric on Minkowski space is taken with the usual form
\[
g_M \ = \ \dd x^1 \otimes \dd x^1  \, +\,  \dd x^2 \otimes \dd x^2 \, + \,  \dd x^3 \otimes \dd x^3 \, - \, c^2\,  \dd t \otimes \dd t \ .   
\]
The model's framework is that physical particles always follow {\it null} geodesics on $P$ and, as usual in Kaluza-Klein theories, a particle's mass and electromagnetic charge are quantities related to its motion along the internal space $K$. Thus, the null geodesic of a particle at spatial rest in Minkowski space must correspond to a vertical movement in internal space $K$ at full speed $c$. That would be the source of the energy of a particle at three-dimensional rest. Conversely, a particle moving at speed $c$ on Minkowski space corresponds to a horizontal null geodesic that cannot have any additional vertical component, and hence the particle has no mass or charge in the frame. While all vertical movement is perceived as mass in four dimensions, electromagnetic charge is related to the component of internal velocity along a specific Killing direction inside $K$. So there can be mass without charge but not the other way around. These classical rules incorporate the fact that a particle is massless if and only if it travels at speed $c$ in $M_4$, and that no charged massless particles have ever been observed. Having a unique finite speed $c$ for all fundamental particles is also an attractive feature, simpler than having a closed interval $[0; c]$ of possible speeds. The inescapable price is having to work with a higher-dimensional spacetime, of course.

Let $\gamma (s)$ be a null geodesic on $P$ and let $\gamma_M (s)$ and $\gamma_K (s)$ be its projections onto the factors $M_4$ and $K$, respectively. Since the metric on $P$ is the product metric, the curves $\gamma_M (s)$ and $\gamma_K (s)$ are geodesics on the respective spaces. Denote by $\dot{\gamma} (s) = \dot{\gamma}_M  + \dot{\gamma}_K$ the tangent vectors to $P$ obtained by derivation of $\gamma (s)$ with respect to the parameter $s$.  The fact that the geodesic is null  is equivalent to
\beq
g_K (\dot{\gamma}_K, \, \dot{\gamma}_K) \ = \ - \, g_M (\dot{\gamma}_M, \, \dot{\gamma}_M) \ .
\eeq
Since the left-hand side is non-negative for any Riemannian metric on $K$, the projection $\gamma_M (s)$ must be a time-like or null geodesic on Minkowski space. The quantity $g_K (\dot{\gamma}_K, \, \dot{\gamma}_K)$ then coincides with the square of $c$ times  $\frac{\dd \tau}{\dd s}$, the rate of change with respect to $s$ of the proper time $\tau$ associated to the particle in free fall along the geodesic $\gamma_M (s)$.

Let us describe this in more detail. Since the vector field $\frac{\partial}{\partial t}$ is Killing on $M_4$, the internal product $g_M (\dot{\gamma}_M, \, \frac{\partial}{\partial t}) =  - c^2 \,(\dd t) (\dot{\gamma}_M)$ is constant and independent of the parameter $s$. In particular, after an affine reparameterization if necessary, we may pick the parameter $s$ of the geodesic so that $(\dd t) (\dot{\gamma}_M)$ is equal to 1 everywhere along the curve. It is possible to do this for every null geodesic $\gamma(s)$ except the trivial one. It is equivalent to choosing the time coordinate $t$ as the parameter of the geodesic. With this choice, the quantities $(\dd x^k) (\dot{\gamma}_M)$ can be identified with the particle's physical velocities $v_M^k$ along the $x^k$-direction, for $k= 1, 2, 3$, in the given coordinate system on $M_4$. So the particle's speed along Minkowski space in this frame is
\[
 v_M \ :=\  \sqrt{ (v_M^1)^2 \, +\, (v_M^2)^2 \, +\, (v_M^3)^2 }\ = \ \sqrt{\, \ g_M\left(\frac{\dd {\gamma}_M}{\dd t} ,\, \frac{\dd {\gamma}_M}{\dd t} \right)  \ + \ c^2 } \ .
 \] 
The particle's speed along the internal space $K$, as seen in the same frame with time coordinate $t$, is given by
\[
 v_K \ :=\ \sqrt{\, g_K\left(\frac{\dd {\gamma}_K}{\dd t} ,\, \frac{\dd {\gamma}_K}{\dd t} \right) } \ .
 \] 
The condition that the geodesic $\gamma(t)$ is null can then be rephrased as
\beq \label{CelestialSphere}
 \ v_M^{\, 2} \ + \ v_K ^{\, 2}  \ = \  c^2  \ .
\eeq
Thus, the particle's speed along $M_4$ always lies between zero and $c$, for any null geodesic on $P$ and any Minkowski coordinates. In a coordinate system on $M_4$ where the particle is at rest, i.e. a coordinate system with $v_M = 0$, the time coordinate $t = \tau$ must be such that the speed of the particle along the internal space has the value $v_K = c$. In a coordinate system with $v_M > 0$, the time coordinate $t$ is such that the internal speed is $\sqrt{c^2 - v_M^{\, 2}}$. The Riemannian distance travelled by the particle along the internal space $K$, as the geodesic's parameter evolves from $s_1$ to $s_2$, is related to the particle's proper time by
\[
\int_{s_1}^{s_2} \sqrt{\, g_K\left(\dot{\gamma}_K ,\, \dot{\gamma}_K \right) }  \ \dd s \ = \ \int_{s_1}^{s_2} \sqrt{-  g_M\left(\dot{\gamma}_M ,\, \dot{\gamma}_M  \right) }  \ \dd s \ = \    \int_{\tau(s_1)}^{\tau(s_2)} c \ \dd \tau \ = \ c\ \big[ \tau(s_1)\, -\, \tau(s_2) \big] \ .
\]
Thus, in this model the particle's proper time is a direct measure of the distance travelled in internal space. Expression \eqref{CelestialSphere} also tells us that the set of null geodesics starting from a fixed point in $P$ and moving forward in time, up to affine reparameterizations, defines a ``celestial'' sphere inside a space of velocities of dimension $\dim M_4\, - \, 1 \, +\,  \dim K$. In other words, it is a set parameterized by a topological sphere of dimension $2 + \dim K$.

Since a geodesic on $P$ can be reconstructed from its components on $M_4$ and $K$, let us spend a few lines discussing the geodesics $\gamma_K (s)$ on internal space. When the group $K$ is equipped with a general left-invariant metric $g_K$, the problem of finding its geodesics is not a simple one. It is well-known that the second order geodesic equation can be replaced by two natural, first-order differential equations: the Euler-Arnold equation for a curve $w(s)$ in the Lie algebra $\kk$ of $K$; and a linear differential equation to reconstruct the geodesic $\gamma_K(s)$ on the group from the solution $w(s)$ on $\kk$. In the case of a matrix group, these two equations can be written as \cite{Arnold, Modin}
\bal \label{EulerArnold}
g_K (\dot{w}, \, u) \ &= \  g_K \big(\, w, \, [ w, \, u ] \, \big)  \qquad  {\rm{for\ all \ }} u \in \kk  \nonumber  \linebr
\dot{\gamma_K} \ &= \ \gamma_K(s) \ w(s) \ .
\end{align}
Although these equations are difficult to solve in general, simple solutions can be written when the metric $g_K$ has left-invariant Killing fields. If $w \in \kk$ is a vector such that $\Lie_{w^\LLL} g_K = 0$, then the formula for this derivative presented in section 2.2 of \cite{Baptista} says that $g_K ( w, \, [ w, \, u ]  )$ vanishes for all $u \in \kk$. So the constant function $w(s) = w$ and the curve $\gamma(s) = h\, \exp(s\, w)$ are solutions of \eqref{EulerArnold}, and hence define a geodesic on $K$ starting at the point $h$. In the special case of a bi-invariant metric $g_K$, the vector fields $w^\LL$ are Killing for all $w \in \kk$, so this procedure generates all the geodesics on $K$. In the case where  $K = \SU$ is equipped with the left-invariant metrics $g_\phi$, or with the more general variants $\tilde{g}_\phi$ defined in \cite{Baptista}, the isometry group is $\rm{U} (1) \times \SU$. Up to normalization, only the electromagnetic vector $e_\phi \in \su$ defines a left-invariant Killing field $e_\phi^\LL$ on the group manifold.\footnote{The electromagnetic vector inside $\su$ is denoted by $\gamma_\phi$ in \cite{Baptista} and in the other sections of the present study. In this section it is denoted by $e_\phi$ to avoid confusion with the notation $\gamma$ for the geodesics.} Thus, the closed curves $s \mapsto h\, \exp(s\, e_\phi)$ are the only explicit geodesics of the metric $g_\phi$ constructed in this way.

Now consider again a general geodesic $\gamma(s)$ on the higher-dimensional $P$. As usual, if $X$ is a Killing vector field for the metric $g_P$, we have that
\[
\frac{\dd}{ \dd s} \ g_P(\dot{\gamma}, \, X) \ |_{\gamma(s)} \ = \ g_P( \nabla_{\dot{\gamma}} \dot{\gamma}, \, X) \ + \ g_P(\dot{\gamma}, \, \nabla_{\dot{\gamma}}  X) \ = \ 0 .
\]
So the internal product $g_P(\dot{\gamma}, \, X)$ is constant along the geodesic $\gamma(s)$. For a product metric $g_P = g_M \times g_K$ this means that, besides the constants of motion associated to the Killing fields of Minkowski space, for every geodesic there are additional constants of motion related to the Killing fields of the internal metric $g_K$. In the case where  $K = \SU$ is equipped the left-invariant metrics $g_\phi$ of \cite{Baptista}, we have nine constants of geodesic motion associated to the isometry group $\rm{U} (1) \times \SU$. These are the products of $\dot{\gamma}_K$ with the right-invariant vector fields, $\ g_\phi(\dot{\gamma}_K, \, v^\RR)$, for any $v \in \su$, and the product $g_\phi(\dot{\gamma}_K, \, e_\phi^\LL)$ with the left-invariant field generated by the electromagnetic vector $e_\phi \in \su$. In the next paragraph we will verify that, as usual in Kaluza-Klein theories, the constant of motion $g_\phi(\dot{\gamma}_K, \, e_\phi^\LL)$ can be identified with the particle's charge to mass ratio.

\subsubsection*{Turning on the gauge fields}

As in section 3 of \cite{Baptista}, let the higher-dimensional spacetime $P= M_4 \times K$ be equipped with a metric $g_P$ determined by: a metric $g_M$ on $M_4$; a fixed metric $g_K$ on the fibres $K$; one-forms $A_\LL$ and $A_\RR$ on $M_4$ with values in the Lie algebra of $K$. By construction, the map $\pi: P \to M_4$ is a Riemannian submersion. Any curve $\gamma(s)$ on $P$ can still be reconstructed from its projections $\gamma_M(s)$ and $\gamma_K(s)$. However, the decomposition $\dot{\gamma} (s) = \dot{\gamma}_M  + \dot{\gamma}_K$ of the tangent vector is no longer orthogonal with respect to $g_P$. Instead, the orthogonal decomposition defined by the metric $g_P$ is now $\dot{\gamma} (s) = \dot{\gamma}^\VV + \dot{\gamma}^\HH$. The two decompositions are related by
\bal
\dot{\gamma}^\VV \ &= \   \dot{\gamma}_K   \,  - \,  A^j_\LL (\dot{\gamma}_M) \, e_j^\LL    \, + \,  A^j_\RR (\dot{\gamma}_M) \, e_j^\RR      \linebr
\dot{\gamma}^\HH \ &= \   \dot{\gamma}_M  \, +\,  A^j_\LL (\dot{\gamma}_M) \, e_j^\LL    \, - \,  A^j_\RR (\dot{\gamma}_M) \, e_j^\RR  \nonumber \ ,
\end{align}
where $\{ e_j\}$ is a basis of the Lie algebra of $K$. Writing the geodesic equation on $P$ as $\nabla^P_{\dot{\gamma}} \dot{\gamma}$, where $\nabla^P$ is the Levi-Civita connection associated to $g_P$, one can ask how this equation looks like once decomposed into its horizontal and vertical components. This decomposition can be obtained from a more general calculation in \cite{ONeill} about parallel transport in Riemannian submersions. Adapting the notation, it implies that for any vertical vector $V$ and horizontal vector $X$ on $TP$,
\bal \label{DecompositionGeodesicEquation}
g_P \big(\nabla^P_{\dot{\gamma}} \dot{\gamma} , \, V \big) \ &= \   g_P \big( \nabla^P_{\dot{\gamma}} \, \dot{\gamma}^\VV , \, V \big)    \, - \,   g_P \big( S_V \, \dot{\gamma}^\VV, \, \dot{\gamma}^\HH \big)       \linebr
g_P \big(\nabla^P_{\dot{\gamma}} \dot{\gamma} , \, X \big) \ &= \   g_M \big( \nabla^M_{\dot{\gamma}_M} \, \dot{\gamma}_M ,\, \pi_\ast X \big)   \, +\,  2 \, g_P \big(  \FF_X \, \dot{\gamma}^\HH,  \,  \dot{\gamma}^\VV \big) \, +\, g_P \big( S_{\dot{\gamma}^\VV} \, \dot{\gamma}^\VV, \, X \big)     \nonumber \ .
\end{align}
The notation here is as in section 3 of \cite{Baptista}. So the tensor $S: \VV \times \VV \to \HH$ is the second fundamental form of the fibres $K$; the tensor $\FF: \HH \times \HH \to \VV$ is essentially the curvature of the gauge fields; the covariant derivative $\nabla^M_{\dot{\gamma}_M} \, \dot{\gamma}_M $ is a vector field on the curve $\gamma_M(s)$ over $M_4$.

Let $\gamma_M(s)$ be an arbitrary curve on $M_4$ and let $\gamma (s)$ be a horizontal lift of that curve to $P$. Then $\dot{\gamma}^\VV = 0$ by definition. Inspecting \eqref{DecompositionGeodesicEquation}, it is clear that in these conditions the product $g_P \big( \nabla^P_{\dot{\gamma}} \dot{\gamma} , \, V \big)$ vanishes for all vertical $V$, while the product $g_P \big(\nabla^P_{\dot{\gamma}} \dot{\gamma} , \, X \big)$ vanishes precisely if $g_P \big( \nabla^M_{\dot{\gamma}_M} \, \dot{\gamma}_M ,\,  \pi_\ast X \big)$ is zero. 
Thus, we recognize that $\nabla^P_{\dot{\gamma}} \dot{\gamma} = 0$ is equivalent to $\nabla^M_{\dot{\gamma}_M} \, \dot{\gamma}_M = 0$. A horizontal curve $\gamma(s)$ is a geodesic on $P$ if and only if its projection $\gamma_M(s)$ is a geodesic on $M$. If the horizontal curve $\gamma(s)$ is null on $P$, then we have that $0= g_P(\dot{\gamma}^\HH, \dot{\gamma}^\HH) = g_M(\dot{\gamma}_M, \dot{\gamma}_M)$, and the curve $\gamma_M(s)$ is null on $M$ too. Thus, in this model all  particles that travel at speed $c$ on $M_4$ are following horizontal null geodesics on $P$ that project down to null geodesics on $M_4$. In particular, the motion in $M_4$ of particles that travel at speed $c$ is not directly affected by the presence of the gauge fields $A_\LL$ and $A_\RR$, they follow geodesics determined by $g_M$ solely. In other words, although the metric $g_P$ and the geodesic motion on $P$ are affected by the presence of gauge fields, the projection of this motion down to $M_4$ is independent of the gauge fields in the case of particles travelling at speed $c$ on $M_4$. This will not be the case for particles travelling at lower speeds on $M_4$, since the corresponding geodesics on $P$ have non-zero vertical components $\dot{\gamma}^\VV$ that, according to \eqref{DecompositionGeodesicEquation}, couple to the gauge fields through the non-vanishing tensors $\FF$ and $S$.

Let $\gamma(s)$ be a general geodesic on $P$ satisfying $\nabla^P_{\dot{\gamma}} \dot{\gamma} = 0$. Then the first equation in \eqref{DecompositionGeodesicEquation} implies that the norm of the vertical component $\dot{\gamma}^\VV$ evolves according to
\beq   \label{NormVerticalComponent}
\frac{\dd}{\dd s} \ g_P \big(\dot{\gamma}^\VV, \, \dot{\gamma}^\VV   \big) \ = \ 2\ g_P \big(  \nabla^P_{\dot{\gamma}} \, \dot{\gamma}^\VV, \, \dot{\gamma}^\VV   \big) \ = \ 2\ g_P \big( S_{ \dot{\gamma}^\VV } \, \dot{\gamma}^\VV , \,  \dot{\gamma}^\HH \big) \ .
\eeq
If the geodesic $\gamma(s)$ is null on $P$, then we also have that
\[
g_P \big(\dot{\gamma}^\VV, \, \dot{\gamma}^\VV   \big) =  - g_P \big(\dot{\gamma}^\HH, \, \dot{\gamma}^\HH   \big) = - g_M \big(\dot{\gamma}_M, \, \dot{\gamma}_M   \big) \ . 
\]
So the norm in $M_4$ of the tangent to the projected curve $\gamma_M(s)$ may not be constant as the parameter $s$ varies. The (zero) norm of the full tangent vector $\dot{\gamma}$ is always preserved along the geodesic, but the norm of the individual components $\dot{\gamma}^\VV$ and $\dot{\gamma}^\HH$ may change, with opposite signs, in regions of Minkowski space where the tensor $S$ is non-zero, i.e. in regions where the fibres of $P$ are not totally geodesic. In those regions the rate of change of proper time $\frac{\dd \tau}{ \dd s}$ may not be constant along $\gamma_M(s)$.

Applying the formula for the tensor $\FF$ given in section 3 of \cite{Baptista}, the term $ \FF_X \, \dot{\gamma}^\HH$ can be written more explicitly in terms of the curvature forms of the gauge fields,
\bal
2\,  \FF_X \, \dot{\gamma}^\HH \ &= \ F_{A_\LLL}^j \big(\pi_\ast X, \, \pi_\ast \dot{\gamma}^\HH \big) \, e_j^\LL \ - \ F_{A_\RRR}^j \big(\pi_\ast X, \, \pi_\ast \dot{\gamma}^\HH \big) \, e_j^\RR  \ .
\end{align}
Since the projection $ \pi_\ast \dot{\gamma}^\HH $ is just $\dot{\gamma}_M$, it follows from the second equation in \eqref{DecompositionGeodesicEquation} that the curve $\gamma_M(s)$ on Minkowski space obtained by projecting down a geodesic $\gamma(s)$ on $P$ satisfies the equation of motion
\[
g_M \big(\nabla^M_{\dot{\gamma}_M}\, \dot{\gamma}_M , \, Y \big) \ = \ g_P (e_j^\RR, \, \dot{\gamma}) \,  F_{A_\RRR}^j \big(Y, \, \dot{\gamma}_M \big)  \ - \ g_P (e_j^\LL, \, \dot{\gamma}) \, F_{A_\LLL}^j \big(Y, \, \dot{\gamma}_M \big) \ - \ g_M \big(\pi_\ast\,  S_{\dot{\gamma}^\VV} \, \dot{\gamma}^\VV, \, Y \big)
\]
for all vectors $Y \in TM$. Unless the fibre-metric $g_K$ is bi-invariant, this equation for the curve $\gamma_M(s)$ on $M_4$ is still strongly coupled to the vertical component of the geodesic $\gamma(s)$, through factors like $g_P (e_j^\LL, \, \dot{\gamma})$ and the vector $S_{\dot{\gamma}^\VV} \, \dot{\gamma}^\VV$.

The decomposition \eqref{DecompositionGeodesicEquation} of the geodesic equation on $P$ simplifies considerably in regions of Minkowski space where the tensor $S$ vanishes. This happens, for example, in regions where the left-invariant fibre metric $g_K$ is constant across the fibres and where the only non-vanishing gauge fields are the massless ones. More precisely, let us assume that we are in a region $\mathcal{U}$ of $M_4$ where the background metric $g_P$ satisfies the following two conditions:
\begin{enumerate}
\item[$\bullet$]  The inner-product $g_P(u^\LL, \, v^\LL)$ is a constant function on $\pi^{-1}(\mathcal{U})$ for any left-invariant vector fields $u^\LL$ and $v^\LL$ on $K$, regarded as fields on $P$.
\item[$\bullet$] The one-form $A_\RR$ is arbitrary but $A_\LL$ is such that the vertical fields $A_\LL^j (X) \, e^\LL_j$ are Killing for the fibre metric $g_K$, for all tangent vectors $X \in T \mathcal{U}$.
\end{enumerate}
In this case, it follows from the formula of section 5.3 of \cite{Baptista}:
\[
2\, g_P ( \, S_{u^\LLL} v^\LL, \,  X^\HH) \ = \  - \, \Lie_X \, \big[ \,  g_P (u^\LL, v^\LL) \, \big]  \ - \  A_\LL^k (X) \  (\Lie_{e_k^\LLL}\, g_K)(u^\LL.v^\LL)    \ , 
\]
that the tensor $S$ --- the second fundamental form of the fibres --- vanishes identically over the region $\mathcal{U}$, and so the fibres are totally geodesic. If we take a null geodesic $\gamma(s)$ on $P$ and project it down to Minkowski space, $\gamma_M = \pi \circ \gamma$, it is a consequence of \eqref{NormVerticalComponent} that the norms of the horizontal and vertical components of $\dot{\gamma}$, namely $g_M \big(\dot{\gamma}_M, \, \dot{\gamma}_M   \big)$ and $g_P \big(\dot{\gamma}^\VV, \, \dot{\gamma}^\VV   \big)$, will both be constant along $\gamma(s)$. Moreover, the decomposition \eqref{DecompositionGeodesicEquation} is simpler in this region of $P$. It implies that for a geodesic $\gamma(s)$, any vertical vector $V \in TP$ and any $Y$ tangent to $M_4$, we have 
 \bal \label{DecompositionGeodesicEquation2}
g_P( \nabla_{\dot{\gamma}} \, \dot{\gamma}^\VV , \, V) \ & = \  0 \    \linebr
g_M \big( \nabla^M_{\dot{\gamma}_M} \, \dot{\gamma}_M ,\, Y \big)   \ & = \      g_P (e_j^\RR, \, \dot{\gamma}) \, F_{A_\RRR}^j \big(Y, \, \dot{\gamma}_M \big) \, - \,  g_P (e_j^\LL, \, \dot{\gamma}) \,  F_{A_\LLL}^j \big(Y, \, \dot{\gamma}_M \big)      \nonumber \ .
\end{align}
Thus, the projection on Minkowski space of the higher-dimensional geodesic is a curve $\gamma_M (s)$ satisfying an equation of motion on $M_4$ similar to the Lorentz-force law, only with more gauge fields involved. The inner-products $g_P (e_j^\RR, \, \dot{\gamma})$ and $g_P (e_j^\LL, \, \dot{\gamma})$ play the role of ``charges'', coupling the geodesic equation for $\gamma_M$ with the curvature of the background gauge fields. We will now investigate the extent to which these inner-products are constant along the geodesic $\gamma(s)$, so that \eqref{DecompositionGeodesicEquation2} truly resembles a Lorentz-force equation of motion.

Let $V$ be a vector field on $K$, regarded as a field on $P$ that is constant along the $M_4$-directions. Then along a geodesic $\gamma (s)$ we have that
\begin{multline} \label{DecompositionProductGeodesic}
\frac{\dd}{\dd s} \ g_P \big(  V,  \dot{\gamma} \big) \, = \, g_P \big( \nabla_{\dot{\gamma}} \,  V, \, \dot{\gamma} \big) \ = \ 
g_P \big( \nabla_{\dot{\gamma}^\HH} \,  V,  \, \dot{\gamma}^\HH \big) \, +\, g_P \big( \nabla_{\dot{\gamma}^\VV} \,  V,  \, \dot{\gamma}^\HH \big) \,  \linebr +  \, g_P\, \big( \nabla_{\dot{\gamma}^\HH} \,  V, \, \dot{\gamma}^\VV \big) \, + \, g_P \big( \nabla_{\dot{\gamma}^\VV} \,  V,  \, \dot{\gamma}^\VV \big)  \, .
\end{multline}
The first term on the right-hand side always vanishes. Indeed, using that $\nabla$ is the Levi-Civita connection, that $V$ is vertical and hence orthogonal to $\dot{\gamma}^\HH$, and the definition of the tensor $\FF_X Y$ and its anti-symmetry, one calculates that
\[
g_P \left( \nabla_{\dot{\gamma}^\HH} \,  V,  \, \dot{\gamma}^\HH \right) \, = \, \Lie_{\dot{\gamma}^\HH} \left[ g_P \left(  V,  \, \dot{\gamma}^\HH \right)  \right] \, - \, g_P \left(  V,  \, \nabla_{\dot{\gamma}^\HH}\, \dot{\gamma}^\HH \right) \, = \, - \, g_P \left(  V,  \, \FF_{\dot{\gamma}^\HH}\, \dot{\gamma}^\HH \right)  = \ 0 \, .
\]
The second term on the right-hand side of \eqref{DecompositionProductGeodesic} can be written in terms of the second fundamental form of the fibres, which vanishes on $\pi^{-1} (\mathcal{U})$ as already discussed,
\[
g_P \left( \nabla_{\dot{\gamma}^\VV} \,  V,  \, \dot{\gamma}^\HH \right) \ = \ g_P \left( S_{\dot{\gamma}^\VV} \,  V,  \, \dot{\gamma}^\HH \right) \ =  \ 0 \ .
\]
The last term on the right-hand side of \eqref{DecompositionProductGeodesic} only involves vertical fields, so well-known properties of Riemannian submersions \cite[section 9.C]{Besse} say that it can be calculated using the fibres' Levi-Civita connection $\nabla^K$. In particular, if $V$ is a Killing field of the fibre metric $g_K$, we have
 \[
g_P \left( \nabla_{\dot{\gamma}^\VV} \,  V,  \, \dot{\gamma}^\VV \right) \ = \ g_K \left( \nabla^K_{\dot{\gamma}^\VV} \,  V,  \, \dot{\gamma}^\VV \right) \ =  \ 0 \ .
\]
To calculate the third term on the right-hand side of \eqref{DecompositionProductGeodesic}, let us take care and consider an extension of $\dot{\gamma}$ to a local vector field $\hat{\gamma}$ defined on an open set of $P$ around the curve $\gamma(s)$. By assumption, the geodesic $\gamma(s)$ is non-trivial and null, so the projection $\gamma_M = \pi \circ \gamma$ is locally injective and the extension $\hat{\gamma}$ can be taken of the form
\[
\hat{\gamma} (x,h) \ = \ \hat{\gamma}^\HH (x,h) \ + \ \hat{\gamma}^\VV (x,h)  \ = \ \sum_{\mu = 0}^3 \, a^\mu(x) \ X_\mu^\HH \, |_{(x,h)} \ + \ \sum_j \, b^j(x)\ e_j^\LL\, |_{h} \ . 
\]
Here $(x,h)$ are the coordinates on $M_4 \times K$; the coefficients $a^\mu(x)$ and $b^j(x)$ are real functions on an open set of $M_4$; we have picked a basis $\{ e_j \}$ for the Lie algebra $\kk$ and a basis $\{ X_\mu \}$ for the tangent space $TM$; the symbol $X^\HH_\mu$ denotes the lift of $X_\mu$ to a horizontal, basic vector field on $P$. Using the properties of the Levi-Civita connection on $P$, we then have that
\bal
g_P\, \big( \nabla_{\hat{\gamma}^\HH} \,  V, \, \hat{\gamma}^\VV \big) \ &= \ g_P\, \big( \nabla_{V} \, \hat{\gamma}^\HH \,+ \, \big[\hat{\gamma}^\HH, \, V \big] \, , \, \hat{\gamma}^\VV \big) \linebr 
&= \ - \,  g_P\big(\hat{\gamma}^\HH, \nabla_{V} \, \hat{\gamma}^\VV \big)  \, + \,  g_P\, \big( \big[\hat{\gamma}^\HH, \, V \big] \, , \, \hat{\gamma}^\VV \big)  \nonumber  \linebr
&= \ - \,  g_P\big(\hat{\gamma}^\HH, S_{V} \, \hat{\gamma}^\VV \big)  \, + \,  g_P\, \big( \big[\hat{\gamma}^\HH, \, V \big] \, , \, \hat{\gamma}^\VV \big)  \nonumber \linebr
&= \ a^\mu \, b^j\  g_P \big( \,\big[X^\HH_\mu, \, V \big]  , \, e_j^\LL \, \big)  \nonumber \ .
\end{align}
Using the explicit expression of the basic lift $X^\HH_\mu$, as appears in the definition of the higher-dimensional metric $g_P$, we can also write
\[
\big[X^\HH_\mu, \, V \big] \ = \  \big[X_\mu, \, V \big] \, + \, A_\LL^i(X_\mu) \, \big[ e_i^\LL , \, V \big] \, - \, A_\RR^i(X_\mu)  \, \big[ e_i^\RR , \, V \big]   \ .
\]
But $V$ is a vector field on $K$ while $X_\mu$ is a field on $M_4$, so the bracket $\big[X_\mu, \, V \big]$ necessarily vanishes on $P$. Moreover, by definition of the extension $\hat{\gamma}$, we have that   $b^j \, e_j^\LL = \dot{\gamma}^\VV$ and that $a^\mu \,A_{\LL / \RR}^i(X_\mu) =  A_{\LL / \RR}^i( \dot{\gamma}_M)$ over the curve $\gamma(s)$. Thus, bringing all the calculations together, we obtain that for any Killing field $V$ on the internal space $K$, regarded as a vector field on $P$, the inner-product with the tangent $\dot{\gamma}$ to the geodesic evolves as
\beq
\frac{\dd}{\dd s} \ g_P \big(  V,  \, \dot{\gamma} \big) \ = \  A_\LL^i(\dot{\gamma}_M) \  g_P \big( \big[ e_i^\LL , \, V \big], \, \dot{\gamma} \big) \ - \ A_\RR^i(\dot{\gamma}_M)  \ g_P \big( \big[ e_i^\RR , \, V \big], \, \dot{\gamma} \big)  \ 
\eeq
in the region $\pi^{-1} (\mathcal{U})$ where the fibres of $P$ are totally geodesic. In the special case where the Killing field $V$ is a purely left-invariant or right-invariant one, then making use of the general relations between the Lie bracket of invariant vector fields and the bracket on the Lie algebra,
\[
[u^\LL, \, v^\LL ] \ = \ [u, \, v ]^\LL_\kk  \qquad \qquad  [u^\RR, \, v^\RR ] \ = \ - [u, \, v ]^\RR_\mathfrak{k}   \qquad \qquad   [u^\LL, \, v^\RR ] \ = \ 0 \ ,
\]
we obtain the simpler evolution equations
\bal \label{EvolutionCharges}
\frac{\dd}{\dd s} \ g_P \big(  v^\LL, \, \dot{\gamma} \big) \ &= \ g_P \left( \big[ A_\LL(\dot{\gamma}_M), \, v \big]^\LL , \, \dot{\gamma} \right)  \linebr
\frac{\dd}{\dd s} \ g_P \big(  v^\RR, \, \dot{\gamma} \big) \ &= \   g_P \left( \big[ A_\RR(\dot{\gamma}_M), \, v \big]^\RR , \, \dot{\gamma} \right)  \nonumber \ .
\end{align}
Again, bear in mind that these equations are valid only when $v^\LL$ or $v^\RR$ are Killing for $g_K$ and in regions of Minkowski space with totally geodesic fibres.

To finish this appendix, let us now come back to the case where $K= \SU$ is equipped with the left-invariant fibre metric $g_K = g_\phi$ with ${\mathrm U}(1) \times \SU$
 isometry, as studied in \cite{Baptista}. Suppose that the Higgs-like field $\phi$ is constant over a region of Minkwoski space where the one-form $A_\RR$ is arbitrary but $A_\LL = A_\LL^e \,  e_\phi$ has values in the line spanned by the electromagnetic vector $e_\phi $ inside $\su$. This means that the strong-force and electromagnetic gauge fields are arbitrary but all the weak-force gauge fields vanish. Since the left-invariant extension $e_\phi^\LL$ is Killing for $g_\phi$, and all right-invariant fields are Killing as well, we are in the conditions where the tensor $S$ vanishes and the simpler equations \eqref{DecompositionGeodesicEquation2} and \eqref{EvolutionCharges} are valid. In particular, for any geodesic $\gamma(s)$ on $P$ we have that
\[
\frac{\dd}{\dd s} \ g_P \big(  e_\phi^\LL, \, \dot{\gamma} \big) \ = \  A_\LL^{e}(\dot{\gamma}_M) \ g_P \left( \big[ e_\phi , \, e_\phi \big]^\LL , \, \dot{\gamma} \right) \ = \ 0 \ .
\]
This means that in the second equation of \eqref{DecompositionGeodesicEquation2}, which now reads
\[
g_M \big( \nabla^M_{\dot{\gamma}_M} \, \dot{\gamma}_M ,\, Y \big)   \ = \  - \,  g_P (e_\phi^\LL, \, \dot{\gamma}) \,  F_{A_\LLL^e} \big(Y, \, \dot{\gamma}_M \big)   \ + \  \sum_j \  g_P (e_j^\RR, \, \dot{\gamma}) \  F_{A_\RRR}^j \big(Y, \, \dot{\gamma}_M \big)  \ , 
\]
the coefficient $g_P \big(  e_\phi^\LL, \, \dot{\gamma} \big)$ of the electromagnetic field-strength $F_{A_\LLL^e}$ is constant along $\gamma(s)$. As usual in five-dimensional Kaluza-Klein theories, comparing with the Lorentz-force equation of motion of a moving charge in $M_4$, one recognizes that the constant $-g_P \big(  e_\phi^\LL, \, \dot{\gamma} \big)$ can be identified with the charge to mass ratio $q/m$ of the particle in free fall along the higher-dimensional geodesic $\gamma(s)$.

\end{appendices}

\newpage


\vspace{1cm}

\end{document}